%% file: main.tex
\documentclass[aps,prb,twocolumn,nofootinbib]{revtex4-2}

\usepackage[a-1b]{pdfx}
\input{defs}

\hypersetup{
colorlinks = true,
linkbordercolor = {white}
}

\newcommand{\cVec}{\cV\mathrm{ec}}
\newcommand{\eVec}{\eV\mathrm{ec}}
\newcommand{\cRep}{\cR\mathrm{ep}}

\def\arraystretch{1.3} 

\newcommand{\ket}[1]{\lvert #1 \rangle}

\begin{document}
\title{Phases with non-invertible symmetries in 1+1D \\
-- symmetry protected topological orders as duality automorphisms
}

\author{\"Omer M. Aksoy}
\affiliation{Department of Physics, Massachusetts Institute of Technology,
Cambridge, Massachusetts 02139, USA}

\author{Xiao-Gang Wen}
\affiliation{Department of Physics, Massachusetts Institute of Technology,
Cambridge, Massachusetts 02139, USA}

\begin{abstract}

We explore 1+1 dimensional (1+1D) gapped phases in systems with non-invertible
symmetries, focusing on symmetry-protected topological (SPT) phases (defined as
gapped phases with non-degenerate ground states), as well as 
SPT orders (defined as the differences between 
gapped/gapless phases with identical bulk excitations spectrum).  For
group-like symmetries, distinct SPT phases share identical bulk excitations and
always differ by SPT orders.  However, for certain non-invertible symmetries,
we discover novel SPT phases that have different bulk excitations and thus
do not differ by SPT orders.  Additionally, we also study the spontaneous
symmetry-breaking (SSB) phases of non-invertible symmetries.
Unlike group-like symmetries, non-invertible
symmetries lack the concept of subgroups, which complicates the definition of SSB phases as well as their identification.
This challenge can be addressed by employing the symmetry-topological-order
(symTO) framework for the symmetry.  The Lagrangian condensable algebras and
automorphisms of the symTO facilitate the classification of gapped phases in
systems with such symmetries, enabling the analysis of both SPT and SSB phases (including those that differ by SPT orders).  
Finally, we apply this methodology to investigate gapless phases in symmetric systems and to study
gapless phases differing by SPT orders.

\end{abstract}

\maketitle

\setcounter{tocdepth}{1}
{\small \tableofcontents }

\section{Introduction}
\label{intro}

Given a group $G$, what are the gapped phases that can be realized as the
ground states of $G$-symmetric Hamiltonians?  Landau-Ginzburg (LG)
paradigm answers this question in terms of the spontaneous symmetry breaking
(SSB) patterns of group $G$~\cite{L3726,L3745}.  An SSB pattern is described by
a conjugacy class $[H]$ of the subgroups of $G$, \ie
\begin{align}
[H]=
\{g H g^{-1} \, |\, H \subset G,\ g\in G\},
\end{align}
such that each (degenerate) ground state is symmetric under a subgroup in the
conjugacy class $[H]$.  We then say that the $G$ symmetry is spontaneously
broken down to the subgroup $H$.  For example, the permutation group of order six
\begin{subequations}
\begin{align}
S_3 = \{
(), (1,2), (2,3), (1,3), (1,2,3), (1,3,2)
\},
\end{align}
has four such conjugacy classes of subgroups:
\begin{align}
& 
[\Z_1]
\equiv
\big[\{()\}\big],
\nonumber\\
& 
[\Z_2]
=
\big[\{(),(1,2)\}\big]
=
\big[\{(),(2,3)\}\big]
=
\big[\{(),(3,1)\}\big],
\nonumber\\
& 
[\Z_3]
\equiv
\big[\{(),(1,2,3),(1,3,2)\}\big],
\nonumber\\
& 
[S_3]
=
\big[\{
(), (1,2), (2,3), (1,3), (1,2,3), (1,3,2)
\}\big].
\end{align}
\end{subequations} 
Thus, the $S_3$-symmetric systems have four SSB patterns 
(described by conjugacy classes of subgroups)
with unbroken
symmetries $\Z_1$, $\Z_2$, $\Z_3$, and $S_3$.  Given an SSB pattern, the
symmetry breaking ground states are distinguished by the non-vanishing
expectation value of an $H$-symmetric local order parameter that transform
non-trivially under the broken symmetries.  

While the LG paradigm is successful in characterizing many SSB phase,
it has been realized that states with identical SSB pattern can still belong to
different phases.  First, it has been shown that  phases with the same SSB
pattern can differ by their different topological orders \cite{W8987,W9039}, \ie
different patterns of long-range entanglement \cite{CGW1038}.  Second, even
without long-range entanglement, in the presence of symmetry, the states with
identical SSB patterns can still belong to different phases
\cite{GW0931,CLW1141,CGW1107,SPC1139}, that differ by symmetry protected
topological (SPT) orders.  The SPT orders protected by a symmetry group $G$ can
be classified by group cohomology of $G$ \cite{CGL1314}, or more generally, by
group cohomology of $G\times SO(\infty)$ (with proper restrictions)
\cite{W1477}. They can also be classified by cobordism theory
\cite{K1459,FH160406527}.

However, symmetries in quantum systems can go beyond those described by groups.
Those generalized symmetries can be a combination of group-like symmetry,
higher-form and higher-group symmetry
\cite{NOc0605316,NOc0702377,KT13094721,GW14125148} (described by higher-group),
anomalous group-like symmetry \cite{H8035,CGL1314,W1313,KT14030617}, anomalous
higher symmetry
\cite{KT13094721,GW14125148,TK151102929,T171209542,P180201139,DT180210104,BH180309336,ZW180809394,WW181211968,WW181211967,GW181211959,WW181211955,W181202517},
non-invertible 0-symmetry (in 1+1D)
\cite{PZh0011021,CSh0107001,FSh0204148,FSh0607247,FS09095013,DR11070495,BT170402330,CY180204445,TW191202817,KZ191213168,I210315588,Q200509072},
non-invertible higher symmetry (also called algebraic higher symmetry)
\cite{KZ200308898,KZ200514178,HV201200009,KZ211101141,CS211101139,BT220406564},
and/or non-invertible gravitational anomaly
\cite{KW1458,FV14095723,M14107442,KZ150201690,KZ170200673,JW190513279}.
Non-invertible gravitational anomaly include anomaly-free/anomalous
non-invertible higher symmetry (described by fusion higher category)
\cite{JW191213492,KZ200308898,KZ200514178,FT220907471}.  To describe all those
different symmetries in a unified framework, a holographic theory was developed
\cite{KZ150201690,JW191213492,KZ200308898,KZ200514178,AS211202092,MT220710712,FT220907471}
(see Section \ref{rev} for a short review).

What are the SPT phases for those generalized symmetries?  What are the SSB
patterns for the generalized symmetries that are not described by group and have
no concept of subgroup?  It turns out that one needs to use new approaches to
address  these physical questions.  Indeed, several different methods have been
introduced to study various gapped and gapless phases with  
generalized symmetries beyond group structure~\cite{LZ230107899,LZ230704788,FA231209272,BW240300905,SS240401369,BT240505302,CW240505331,PA240612962,PA240918113,BS241019040B,WS241215024,MG241220546}.

In this paper, we will use the holographic theory of symmetry to study 1+1D SPT
phases and SSB phases with non-invertible symmetries. The holographic theory
allows us to compute all 1+1D gapped phases (which include both SPT and SSB
phases) for systems with non-invertible symmetries.  
After obtaining those gapped phases, we can use ground state degeneracy (GSD)
on a ring to further characterize them.  However, GSD cannot
distinguish all different SSB patterns.  For example, a group $G$ may have multiple
subgorups of the same order. In this case, different SSB
pattern where $G$ is broken down to any of these subgroups may have the same GSD.  
The goal of this work is to describe distinct SSB patterns and SPT phases
for non-invertible symmetries including those with the same GSD.  

As we shall discuss, the holographic theory of symmetry suggests that 
the notion of an SPT order also needs to be generalized.
There are two definitions of SPT order or phase in 1+1D:
\begin{enumerate}
\item
The distinct gapped phases with GSD = 1 on a ring correspond to distinct
\emph{SPT phases} \cite{CGL1314}. This definition only applies to gapped phases with
non-degenerate ground state. We cannot use it to define gapless SPT phases and
SSB patterns that also support (distinct) SPT phases in each of their degenerate ground states.

\item
The distinct phases with identical bulk excitations are said to differ by
\emph{SPT orders} \cite{KZ200514178}. This definition is based on equivalence relations.  
Two phases with identical bulk excitations are said to be in the same equivalence class,
in which the distinct equivalent phases differ by SPT orders.  This definition can be used to define
gapless SPT phases~\cite{Scaffidi170501557,Jiang18,Parker18,Thorngren21,Verresen21,wen2023classification} and symmetry breaking SPT phases.

\end{enumerate}

\renewcommand*{\arraystretch}{1.6}
\begin{table*}[t] 
\caption{ 1+1D gapped phases for some generalized symmetries } \label{Phases} 
\centerline{
\begin{tabular}{ |c|l|}
\hline 
Symmetry & Gapped Phases (Lagrangian condensable algebras)\\ 
\hline 
$S_3$ & 
\{[\textbf{1}]$_{1}$,[\textbf{1}]$_{3}$\},
\{[\textbf{1}]$_{2}$,[\textbf{1}]$_{6}$\}  \\
\hline 
$\cRep_{S_3}$ & 
\{[\textbf{1}]$_{1}$,[\textbf{1}]$_{3}$\},
\{[\textbf{1}]$_{2}$,[\textbf{1}]$_{3}$\}  \\
\hline 
$A_4$ & 
\{[\textbf{2}]$_{1}$,[\textbf{1}]$_{4}$\},
\{[\textbf{2}]$_{3}$,[\textbf{1}]$_{12}$\},
\{[\textbf{1}]$_{6}$\}  \\
\hline 
$\cRep_{A_4}$ & 
\{[\textbf{2}]$_{1}$,[\textbf{1}]$_{4}$\},
\{[\textbf{1}]$_{2}$\},
\{[\textbf{2}]$_{3}$,[\textbf{1}]$_{4}$\}  \\
\hline 
$\tl\cR^{\eD(A_4)}_{\cA_{1,1}}$ &
\{[3$\times$\textbf{1}]$_{2}$,\},
\{[\textbf{1}]$_{6}$,\},
\{[3$\times$\textbf{1}]$_{6}$\}  \\
\hline 
$S_3\hskip-1mm \times \hskip-1mm \Z_3$ &
\{[\textbf{1}]$_{1}$,2$\times$[\textbf{1}]$_{3}$,[\textbf{1}]$_{9}$\},
\{[\textbf{1}]$_{2}$,[\textbf{1}]$_{6}$\},
\{[\textbf{1}]$_{2}$,2$\times$[\textbf{1}]$_{6}$,[\textbf{1}]$_{18}$\}  \\
\hline 
$\cRep_{S_3\times \Z_3}$ &
\{[\textbf{1}]$_{1}$,[\textbf{1}]$_{3}$\},
\{[\textbf{1}]$_{1}$,2$\times$[\textbf{1}]$_{3}$,[\textbf{1}]$_{9}$\},
\{[\textbf{1}]$_{2}$,[\textbf{1}]$_{3}$,[\textbf{1}]$_{6}$,[\textbf{1}]$_{9}$\} \\
\hline 
$\tl\cR^{\eD(S_3\times \Z_3)}_{\cA_{3,2}}$ &
\{[\textbf{1},\textbf{1}]$_{1}$,[\textbf{1},\textbf{1}]$_{3}$\},
\{[\textbf{1},\textbf{1}]$_{2}$,[\textbf{1},\textbf{1}]$_{6}$\},
\{[\textbf{1}]$_{6}$,[\textbf{1}]$_{10}$\}  \\
\hline 
$D_8$ &
\{[\textbf{2}]$_{1}$,[\textbf{1},\textbf{1}]$_{4}$\},
\{[\textbf{1}]$_{2}$,[\textbf{2},\textbf{2}]$_{2}$,[\textbf{1}]$_{8}$\},
\{[\textbf{1}]$_{4}$\} \\
\hline 
$\cRep_{D_8}$ &
\{[3$\times$\textbf{1}]$_{1}$,[3$\times$\textbf{1}]$_{4}$\},
\{[\textbf{1}]$_{2}$\},
\{[3$\times$\textbf{1}]$_{2}$,[\textbf{1}]$_{5}$\} \\
\hline 
$\tl\cR^{\eD(D_8)}_{\cA_{1,1}}$ &
\{[4$\times$\textbf{1}]$_{2}$\},
\{[6$\times$\textbf{1}]$_{4}$\},
\{[\textbf{1}]$_{8}$\}\\
\hline 
$H_8$ &
\{[\textbf{1}]$_{1}$,[\textbf{1}]$_{2}$,[\textbf{1}]$_{4}$,[\textbf{1}]$_{5}$\},
\{[\textbf{1}]$_{2}$,[\textbf{1}]$_{4}$\} \\
\hline 
$\tl\cR^{\eM_\mathrm{Haag}}_{\cA_{1,1}}$ &
\{[\textbf{1}]$_{4}$\}, \{[\textbf{1},\textbf{1}]$_{4}$\} \\
\hline 
$\tl\cR^{\eM_\mathrm{Haag}}_{\cA_{2,1}}$ &
\{[\textbf{1}]$_{2}$,[\textbf{1}]$_{6}$\}, \{[\textbf{1}]$_{4}$\}
\\
\hline 
$\tl\cR^{\eM_\mathrm{tHaag}}$ &
\{[\textbf{1}]$_{6}$\} \\
\hline 
\end{tabular}
}
\end{table*}

For group-like symmetries, the above two definitions are equivalent to one another.  
In this work, we show by way of example that, for some non-invertible symmetries, the above two
definitions do not imply each other.  For that reason, we shall use two different terms:
\emph{SPT phases} and \emph{SPT orders} to describe the two concepts.  We
stress that SPT orders actually describe the difference of two phases.  

Also there are three different ways to define what ``identical bulk excitations'' means,
which leads to three different notions of SPT orders: 
\begin{enumerate} 
\item
Two phases have \emph{identical bulk excitations} if there are two states, one
in each phases, such that the two state have identical energy spectrum on a
ring and the corresponding excitations have identical symmetry charge.  This
leads to the notion of \emph{SPT order}, which corresponds to the usual SPT
order described by group cocycles.

\item Two phases have \emph{identical bulk excitations} if there are two
states, one in each phases, such that the two states have identical energy
spectrum on a ring.  However, the corresponding excitations may have permuted
symmetry charges.  This leads to the notion of \emph{permute-charge SPT (pSPT)
order} or \emph{pseudo SPT (pSPT) order}.

\item Two phases have \emph{identical bulk excitations} if there are two
states, one in each phases, such that the two states have identical energy
spectrum on a ring within a symmetric sub-Hilbert space. However, the two
states may have different total spectra.  This leads to the notion of
\emph{automorphism SPT (aSPT) order}.  

\end{enumerate}

The name aSPT order comes from the following result.  If there is a
symmetry-preserving duality transformation that maps a model with a generalized
symmetry to another model with the same generalized symmetry, then, the ground
states of the two models support identical excitation spectrum in the appropriate  sub-Hilbert space, 
and differ by an aSPT order according to the above definition.  Later we will see that
symmetry-preserving duality transformations correspond to the automorphisms of
the symmetry-topological-order (symTO).  So we can compute symmetry-preserving
duality transformations via the automorphisms of the symTO.  To stress this
relation, we refer to symmetry-preserving duality transformation as
fixed-symmetry duality (FSD) automorphisms. We also refer to aSPT order as
FSD automorphism.  Two phases connected by a FSD automorphism will be referred
to as two phases differ by an aSPT order.

Similarly, two phases connected by a fixed-charge duality (FCD) automorphism
correspond to two phases differ by an SPT order.  \Rf{KZ200514178} used such FCD
automorphisms to classify SPT orders in any dimensions for general
non-invertible higher symmetries  (which was called algebraic higher
symetries).  Two phases connected by a fixed-algebra duality (FAD) automorphism
correspond to two phases differ by a pSPT order. Every FCD automorphism is also
an FAD automorphism in which the permutation of charges is trivial.
In turn,  every FAD automorpghism is an FSD autormorphism in which 
the duality holds in the entire global Hilber space. See Section \ref{autos} for details.

We remark that following the hierarchy of automorphisms, 
an aSPT order.  The essense of the aSPT orders (as well as SPT-
and pSPT-orders) is symmetry-preserving duality automorphisms: \frmbox{Two
phases are said to differ by an aSPT order if for each state in one phase, there
exist another state in the other phase, such that the two states are related by
a symmetry-preserving duality transformation (\ie FSD automorphism).} Thus a
study of aSPT order is a study of symmetry-preserving duality tansformation
\cite{LZ230107899}. Importantly, we use the term \emph{duality} 
to refer to both invertible maps between two Hilbert spaces where the excitation spectrum is completely
preserved and non-invertible maps that become invertible only in a particular sub-Hilbert space.
An example of the former is a unitary transformation, while an example of the latter is the Kramers-Wannier (KW)
transformation.

Duality automorphisms (\ie FSD automorphism), as maps between different
Hamiltonians of the same symmetry, can be composed and form a group.  As a
result, the sets of SPT orders, pSPT orders, and aSPT orders form groups (or
more precisely, form torsors).  In contrast, the set of SPT phases may not form
a group.


\subsection{Summary of Results}

\begin{table*}[t] 
\caption{ 28 distinct gapped phases for each of the six symmetries in the
Morita-equivalence class of $\eD(S_3\times S_3)$-symTO.  All  six symmetries are
anomaly-free. Only $S_3\times S_3$ and $\tl \cR^{\eD(S_3\times
S_3)}_{\cA_{2,1}}$ symmetries have non-trivial SPT order. The other four
symmetries have more-than-one SPT phases, that are not connected by
SPT-orders.} 
\label{PhasesS3xS3}
\centerline{
\begin{tabular}{ |c|>{\raggedright\arraybackslash}p{6in}|}
\hline 
Symmetry & Gapped Phases (Lagrangian condensable algebras)\\ 
\hline 
$S_3\hskip-1mm \times \hskip-1mm S_3$ & 
\{[\textbf{2}]$_{1}$,[\textbf{2},\textbf{2}]$_{3}$,[\textbf{2}]$_{9}$\},
\{[\textbf{1}]$_{2}$,[\textbf{1}]$_{6}$\},
\{[\textbf{1}]$_{2}$,[\textbf{1},\textbf{1}]$_{6}$,[\textbf{1}]$_{18}$\},
\{[\textbf{1},\textbf{1}]$_{2}$,2$\times$[\textbf{1},\textbf{1}]$_{6}$,[\textbf{1},\textbf{1}]$_{18}$\},
\{[\textbf{1}]$_{4}$,[\textbf{1}]$_{12}$\},
\{[\textbf{1}]$_{4}$,[\textbf{1},\textbf{1}]$_{12}$,[\textbf{1}]$_{36}$\}
\\
\hline 
$\cRep_{S_3 \times  S_3}$ & 
\{[\textbf{1}]$_{1}$,[\textbf{1}]$_{3}$\},
\{[\textbf{1}]$_{1}$,[\textbf{1},\textbf{1}]$_{3}$,[\textbf{1}]$_{9}$\},
\{[\textbf{1}]$_{1}$,[\textbf{1},\textbf{1}]$_{3}$,[\textbf{1}]$_{4}$,[\textbf{1}]$_{6}$,[\textbf{1},\textbf{1}]$_{6}$,[\textbf{1}]$_{9}$\},
\{[\textbf{1}]$_{2}$,[\textbf{1}]$_{3}$\},
\{[\textbf{1}]$_{2}$,[\textbf{1},\textbf{1}]$_{3}$,[\textbf{1}]$_{6}$\},
\{[\textbf{1},\textbf{1}]$_{2}$,[\textbf{1},\textbf{1}]$_{3}$,[\textbf{1},\textbf{1}]$_{6}$,[\textbf{1},\textbf{1}]$_{9}$\}
\\
\hline 
$\cRep_{S_3}\hskip -1mm \times\hskip -1mm S_3$ & 
\{[\textbf{1}]$_{1}$,[\textbf{1}]$_{3}$\},
\{[\textbf{1}]$_{1}$,2$\times$[\textbf{1}]$_{3}$,[\textbf{1}]$_{9}$\},
\{[\textbf{1}]$_{1}$,2$\times$[\textbf{1}]$_{3}$,[\textbf{1}]$_{4}$,[\textbf{1}]$_{6}$,[\textbf{1}]$_{9}$,[\textbf{1}]$_{12}$, [\textbf{1}]$_{18}$\},
\{[\textbf{1}]$_{2}$,[\textbf{1}]$_{6}$\},
\{[\textbf{1}]$_{2}$,2$\times$[\textbf{1}]$_{6}$,[\textbf{1}]$_{18}$\},
\{[\textbf{1},\textbf{1}]$_{2}$,[\textbf{1},\textbf{1}]$_{3}$,[\textbf{1},\textbf{1}]$_{6}$,[\textbf{1},\textbf{1}]$_{9}$\}
\\
\hline 
$\tl \cR^{\eD(S_3\times S_3)}_{\cA_{2,1}}$ &
\{[\textbf{2},\textbf{2}]$_{1}$,[\textbf{2},\textbf{2}]$_{3}$\},
\{[4$\times$\textbf{1}]$_{2}$,[4$\times$\textbf{1}]$_{6}$\},
\{[\textbf{1},\textbf{1}]$_{2}$,[\textbf{1},\textbf{1}]$_{6}$\},
\{[\textbf{1},\textbf{1}]$_{4}$,[\textbf{1},\textbf{1}]$_{12}$\},
\{[\textbf{1}]$_{6}$,[\textbf{1}]$_{10}$\},
\{[\textbf{1}]$_{12}$,[\textbf{1}]$_{20}$\}
\\
\hline 
$\tl \cR^{\eD(S_3\times S_3)}_{\cA_{5,1}}$ &
\{[4$\times$\textbf{1}]$_{1}$,[4$\times$\textbf{1}]$_{3}$\},
\{[\textbf{1},\textbf{1}]$_{2}$,[\textbf{1},\textbf{1}]$_{6}$\},
\{[4$\times$\textbf{1}]$_{2}$,[4$\times$\textbf{1}]$_{3}$\},
\{[\textbf{1},\textbf{1}]$_{4}$,[\textbf{1},\textbf{1}]$_{6}$\},
\{[\textbf{1}]$_{6}$,[\textbf{1}]$_{10}$\},
\{[\textbf{1}]$_{6}$,[\textbf{1}]$_{8}$\}
\\
\hline  
$\tl \cR^{\eD(S_3\times S_3)}_{\cA_{6,1}}$ &
\{[\textbf{1},\textbf{1}]$_{1}$,2$\times$[\textbf{1},\textbf{1}]$_{3}$,[\textbf{1},\textbf{1}]$_{9}$\},
\{[\textbf{1}]$_{2}$,[\textbf{1}]$_{6}$\},
\{[\textbf{1}]$_{2}$,[\textbf{1},\textbf{1}]$_{6}$,[\textbf{1}]$_{18}$\},
\{[\textbf{1},\textbf{1}]$_{2}$,[4$\times$\textbf{1}]$_{3}$,[\textbf{1},\textbf{1}]$_{6}$\},
\{[\textbf{1}]$_{4}$,[\textbf{1}]$_{6}$\},
\{[\textbf{1}]$_{4}$,[\textbf{1},\textbf{1}]$_{6}$,[\textbf{1}]$_{12}$\}
\\
\hline 
\end{tabular}
}
\end{table*}

Tables \ref{Phases} and  \ref{PhasesS3xS3} summarize distinct 1+1D gapped
phases for some generalized symmetries.  The phases are grouped into
SPT-classes (\ie the phases that differ by SPT orders).  Each bold-face number
correspond to a SPT-class, and the number indicates the number of gapped phases
in that SPT-class.  A bold-face number larger than 1 indicates the presence of
non-trivial SPT order.  The SPT-classes are then grouped into pSPT-classes (\ie
the phases that differ by pSPT orders).  Each $[\cdots]_\text{GSD}$ represents
a pSPT-class and GSD is the ground state degeneracy of the phases in the
pSPT-class.  If GSD $>$ 1, then the phases in the pSPT-class all have SSB.  The
pSPT-classes are further grouped into aSPT-classes, which is denoted by curly brackets $\{\cdots\}$.  
All phases in an aSPT-class are connected by FSD automorphisms, and they have identical spectrum
within the symmetric sub-Hilbert space despite they may have different SSB
patterns and GSDs. We use the shorthand notation, say, 3$\times$\textbf{1}
to denote $\textbf{1}, \textbf{1}, \textbf{1}$. Here,  $H_8$ is the  
non-invertible symmetry described by a Hopf algebra.  

From the table \ref{Phases}, we see that $A_4$ symmetry has two gapped
symmetric phases (with GSD = 1) that differ by SPT orders (\textbf{2} in
[\textbf{2}]$_{1}$).  $A_4$ symmetry also has two gapped SSB phases (with
GSD = 3 and $A_4\to \Z_2\times \Z_2$ symmetry breaking pattern) that differ by
SPT orders (\textbf{2} in [\textbf{2}]$_{3}$).  We call these two phases as
SSB-SPT phases.  

Furthermore, there are three phases in the aSPT-class
\{[\textbf{2}]$_{1}$,[\textbf{1}]$_{4}$\} connected by FSD 
automorphisms.  Those three phases have identical energy spectrum in their respective
symmetric sub-Hilbert spaces, despite having different SSB patterns (two
phases are $A_4$-symmetric, while the third phase has $(A_4\to \Z_3)$-SSB).  We
remark that the $(A_4\to \Z_3)$-SSB phase has domain-wall excitations, while the
two symmetric phases have only $A_4$-charged excitations.  Despite this
difference, the three phases still have identical excitation spectrum in the
symmetric sub-Hilbert space, since the domain-wall excitations in the symmetry
breaking phase have the same energy as certain $A_4$-charged excitations in the
symmetric phases.

We note that, in a phase diagram, the boundary of a phase is formed by
continuous and first order phase transitions to other phases.  Two phases
connected by an FSD automorphism (which include two phases differ by SPT
and pSPT order as special cases) have the same boundary, formed by the
same set of continuous and first order phase transitions.  

We also see that $\cRep_{A_4}$ non-invertible symmetry has two gapped symmetric
phases that differ by SPT orders (\textbf{2} in [\textbf{2}]$_{1}$).
$\cRep_{A_4}$ symmetry also has two gapped SSB phases that differ by SPT orders
(\textbf{2} in [\textbf{2}]$_{3}$).  Since $\cRep_{A_4}$ is non-invertible, we
cannot obtain the $\cRep_{A_4}$ SPT orders using group cohomology.  New methods
of duality automorphisms of symTO are developed to compute SPT orders, pSPT
orders, and FSD automorphisms for non-invertible symmetry.

For $\cRep_{S_3\times \Z_3}$ non-invertible symmetry, we find that it has two
SPT-phases, but the two SPT-phases belong to two different aSPT-classes,
\{[\textbf{1}]$_{1}$,[\textbf{1}]$_{3}$\} and
\{[\textbf{1}]$_{1}$,2$\times$[\textbf{1}]$_{3}$,[\textbf{1}]$_{9}$\}.  In
other words, the two SPT-phases are not connected by any FSD automorphism, and
as a result, they have different bulk excitation spectra and can be
distinguished via bulk measurement \cite{W2024Talk}.  In fact, the excitations
in the two SPT phases form a $\cVec_{S_3\times \Z_3}$ fusion category and
$Z_3\times\Z_3$ Tambara-Yamagami $TY(\Z_3\times\Z_3)$ fusion category 
as calculated in \Rf{MG241220546}~\footnote{ For a
group-like symmetry $G$, the symmetry defects in any SPT phase follow the
fusion rules of the category $\cVec_G$ while the symmetry charges follows the
fusion rules of the category $\cRep_G$.  }.   The above result implies that SPT
phases with non-invertible symmetry may have different fusion rules of charges
described by distinct fusion categories.  Strikingly, this means that \frmbox{
the same non-invertible symmetry can have more than one distinct set of
symmetry charges forming distinct fusion categories.} This is another new
result of this paper.

We also find many FSD automorphisms, and their corresponding duality
symmetries, for the $\cRep_{S_3\times \Z_3}$ non-invertible symmetry (as well
as for other symmetries). For example, the symmetric phase and the SSB phase in
\{[\textbf{1}]$_{1}$,[\textbf{1}]$_{3}$\} are connected by a duality
automorphism.  This duality automorphism also implies a duality symmetry that
connects the two phases.  The duality symmetries of the aSPT classes of $S_3$
symmetry, \{[\textbf{1}]$_{1}$,[\textbf{1}]$_{3}$\} and
\{[\textbf{1}]$_{2}$,[\textbf{1}]$_{6}$\}, have been studied and realized
explicitly on lattice model in \cite{CW240505331}.

The above results are obtained using a holographic theory of symmetry, which 
we briefly review in the following.

\section{A short review of holographic theory of symmetry} \label{rev}

Given a $d$-dimensional model $\underline{\cC}$, which can be a non-lattice
model such as a quantum field theory, the holographic theory of symmetry
describes the decomposition of universal symmetry properties of
$\underline{\cC}$ from its non-universal dynamical data. This is achieved by
identifying the low-energy properties of the $d$-dimensional model
$\underline{\cC}$ with that of a boundary of a topological order in one-higher
dimension.

\begin{figure}[t]
\hskip -1.7in
\includegraphics[scale=0.6]{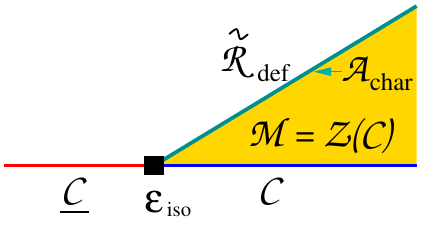}
\caption{ An isomorphic holographic decomposition $\eps_\text{iso}$ for a model
$\underline{\cC}$ (or a quantum field theory).  $\tl\cR_\text{def}$ is a fusion
higher category describing the gapped excitations on a gapped boundary of the
bulk topological order $\eM$.  $\tl\cR_\text{def}$ describes the (emergent)
symmetry in $\underline{\cC}$.  We will refer to such a symmetry as
$\tl\cR_\text{def}$-symmetry. In fact, $\tl\cR_\text{def}$ describes the fusion
of the symmetry defects.  Also, $\eZ(\cC)$ is the symTO $\eM$ describing the
Morita-equivalence class of symmetry $\tl\cR_\text{def}$.  The boundary
$\tl\cR_\text{def}$ is induced by a  Lagrangian condensable algebra
$\cA_\text{char}$.  The anyons in $\cA_\text{char}$ is defined as the charge of
the symmetry $\tl\cR_\text{def}$.  } \label{CCmorph} 
\end{figure}

More precisely, in  \Rf{KZ150201690}, an isomorphic holographic decomposition
$\eps_\text{iso}$ was proposed to reveal hidden gravitational anomaly in a
model $\underline{\cC}$. \footnote{Here, we adopt the notations used in
\Rf{KZ200514178} to make it easy to compare with the results in
\Rf{KZ200514178}.}  This decompostion is given by an isomorphism
$\eps_\text{iso}$, which describes the low-energy equivalence of the model
$\underline{\cC}$ and the composite model $\cC\boxtimes_\eM \tl\cR_\text{def}$
(see Fig.\ \ref{CCmorph}).  Here, the composite model is formed by a
$d+1$-dimensional bulk topological order $\eM$, and its two boundaries
$\tl\cR_\text{def}$ and $\cC$, which we call top and bottom boundaries,
respectively.  

The top boundary $\tl\cR_\text{def}$ is  gapped and induced by a  Lagrangian
condensable algebra $\cA_\text{char}$ of $\eM$.
Correspondingly, there is a fusion higher category, which we also label by
$\tl\cR_\text{def}$, describing the gapped excitations, on the top boundary
$\tl\cR_\text{def}$. As we shall explain, the top boundary  $\tl\cR_\text{def}$
describes the  symmetry of the equivalent $d$-dimensional model
$\underline{\cC}$.

The bottom boundary $\cC$ may or may not be gapped.  With the assumption that
the bulk topological order $\eM$ and the top boundary $\tl\cR_\text{def}$ have
large energy gaps, the bottom boundary $\cC$ describes the dynamics of
excitations below these large energy gaps.  The low-energy equivalence means
that the two models $\underline{\cC}$ and $\cC\boxtimes_\eM \tl\cR_\text{def}$
are equivalent below the gaps of $\eM$ and $\tl\cR_\text{def}$, \ie they have
the identical energy spectra below the gaps.  The low-energy equivalence also
implies that the model $\underline{\cC}$ and model $\cC$ are locally
indistinguishable at low energies.  More precisely, all the local operators in
model $\cC$ have corresponding local operators in model $\underline{\cC}$ with
the same long-distance behavior in their correlation functions. 

As a stand-alone model, the bottom boundary $\cC$ has a gravitational
anomaly\footnote{Here, gravitational anomaly is defined as the obstruction for
the model to have a lattice regularization in the same dimension. With this
definition, the gravitational anomaly can be non-invertible in the sense that
one cannot construct a lattice regularization by stacking the model with
another one. The invertible case then corresponds to the traditional
gravitational anomaly which is  defined via obstruction to diffeomorphism
invariance.}, since its bulk is a non-trivial topological order \cite{KW1458}.
The isomorphic holographic decomposition $\eps_\text{iso}$ indicates that the
anomaly-free model  $\underline{\cC}$ is local low-energy equivalent to the
anomalous model $\cC$. Thus the isomorphic holographic decomposition
$\eps_\text{iso}$ reveals the hidden gravitational anomaly in the model
$\underline{\cC}$.  Knowing the hidden gravitational anomaly is very helpful
since anomalies constrain low-energy dynamics.  In other words,  knowing the
anomaly-free model $\underline{\cC}$ has a  hidden gravitational anomaly, its
low energy properties are constrained just like the anomalous model $\cC$.

It has been realized that a symmetry (up to Morita-equivalence\footnote{Two
symmetries are  Morita-equivalent if two symmetric systems are identical after
restricted to their corresponding symmetric sub-Hilbert space, \ie, there is a
duality between the systems with these two symmetries when restricted to
appropriate sub-Hilbert spaces.   For example, in 1+1D, $\Z_2\times
\Z_2$-symmetry with the mixed anomaly is Morita-equivalent to
$\Z_4$-symmetry~\cite{CW220303596,ZL220601222}.  The Morita-equivalence was
called holo-equivalence in \Rf{KZ200514178}.}) is actually a (non-invertible)
gravitational anomaly \cite{JW190513279,JW191213492,KZ200514178}.  This
identification is vary natural since both symmetry and gravitational anomaly
constrain low-energy dynamics.  Non-invertible gravitational anomalies are
capable to describe group-like symmetry, anomalous symmetry, higher-form
symmetry, non-invertible symmetry, and their combinations, in a unified way.
Since gravitational anomaly corresponds to topological order in one higher
dimension, we may say that generalized symmetry (up to Morita-equivalence) is
described by topological order in one higher dimension.  Such a bulk
topological order is named as the symmetry topological order (symTO\footnote{In
some early papers, symTO was called \emph{categorical
symmetry}.})\cite{JW191213492,KZ200514178} or symmetry topological field theory
(symTFT) \cite{AS211202092,FT220907471} \footnote{More precisely, symTO can be
viewed as symTFT plus UV regulation. As a result, the automorphisms of symTFT
(symmetries of symTFT) are not symmetries of symTO. The boundaries of symTO do
not have those automorphism symmetry in general.}.

Under the symmetry-topological-order (sym/TO) correspondence, topological
orders with gappable boundaries in one higher-dimension are in one-to-one
correspondence with the Morita-equivalence classes of finite generalized
symmetries.  Physically, the boundary of topological order exactly simulate
symmetric systems in the limit that the bulk gap approaches infinity.

To describe and to further distinguish different symmetries within the same
Morita-equivalence class, we need to include more information, in addition to
the bulk topological order $\eM$ (\ie the symTO).  This additional information
happens to be the fusion higher category $\tl\cR_\text{def}$ that describes the
excitations on the gapped boundary $\tl\cR_\text{def}$ in Fig.\  \ref{CCmorph}
\cite{TW191202817,KZ200514178,FT220907471}.  Since fusion higher category
$\tl\cR_\text{def}$ also determines the non-degenerate braided fusion higher
category $\eM$ that describes the excitations in the bulk topological order
\cite{KW1458,KZ150201690}, we  say that the \emph{fusion higher category
$\tl\cR_\text{def}$  determines the generalized symmetry}.  As a result, in
this paper, we will refer to a symmetry as $\tl\cR_\text{def}$-symmetry, which
can be group-like, anomalous, higher-form, non-invertible, or their
combinations.

It turns out that the fusion higher category $\tl\cR_\text{def}$ describes the
fusion and the braiding properties of the \emph{symmetry defects} of the
corresponding symmetry.  The Lagrangian condensable algebra $\cA_\text{char}$
that produces the gapped $\tl\cR_\text{def}$-boundary is formed by the
\emph{symmetry charges}, which are particles and/or extended excitations with
trivial self and mutual statistics above the trivial symmetric ground state.  

Those \emph{symmetry charges} also form a fusion higher category denoted by
$\cR$. This fusion higher category too is induced by a Lagrangian condensable
algebra $\tl\cA_\text{def}$, which is now formed by the \emph{symmetry defects}
\cite{KZ200514178}.  

We like to remark that for an anomalous algebraic higher symmetry
$\tl\cR_\text{def}$, the symmetry defects have non-trivial braiding, and they
cannot form a Lagrangian condensable algebra.  In this case,
$\tl\cA_\text{def}$ and $\cR_\text{char}$ do not exist.  In other words, there
is an obstruction to define fusion higher category that describe symmetry
charges for an anomalous symmetry.  This is expected, since physically symmetry
charges are defined as excitations above gapped symmetric non-degenerate ground
state, while an anomalous symmetry does not admit gapped, symmetric, and
non-degenerate ground state.

\section{Gapped phases for systems with generalized symmetry -- a
general discussion}

For systems with group-like symmetries, there can be multiple SPT phases, which
have gapped non-generate and symmetric ground states with trivial toplogical
order that cannot be smoothly deformed to each other without going through
phase transition or symmetry breaking~\cite{GW0931,CLW1141,CGL1314}.  Such SPT
phases are known to be decribed by the group cohomology of the symmetry group.
What are the SPT phases for generalized symmetries?

First, generalized symmetries can be anomalous, while the SPT phases exist only
for anomaly-free symmetries.  But how to define anomaly-free symmetry?
\Rfs{TW191202817,KZ200514178} propose that a symmetry is anomaly-free if it
allows non-degenerate gapped ground state on closed spaces of any homotopy
type.  Using the point of view of sym/TO correspondence, a symmetry that is
described by a fusion higher category $\tl\cR_\text{def}$ is anomaly-free if
and only if there exist another fusion higher category $\cR_\text{char}$, that
satisfies $\eZ(\cR_\text{char})= \eZ(\tl\cR_\text{def})=\eM$ (\ie
$\cR_\text{char}$ describes the excitations on the gapped lower boundary in
Fig.\  \ref{CCmorph}), and that $\tl\cR_\text{def}  \boxtimes_{\eM}
\cR_\text{char} = \eVec$ (\ie the stacking of $\tl\cR_\text{def}$ and
$\cR_\text{char}$ through the common bulk $\eM$ gives rise to a trivial
topological order -- a short-range entangled state)
\cite{TW191202817,KZ200514178}.

How does one compute SPT phases protected by a generalized
$\tl\cR_\text{def}$-symmetry?  More generally, what are the possible gapped
phases of systems with $\tl\cR_\text{def}$-symmetry?  Using sym/TO
correspondence, we can compute gapped phases by computing gapped boundary
phases of the symTO $\eM$ (\ie computing gapped boundary $\cC$ in Fig.~\ref{CCmorph}).  

A 2+1D topological order with gappable boundary contains one or more sets
anyons, that can condense together.  Anyons in such a set have trivial self and
mutual statistics.  Or more precisely, anyons that can condense together are
described by a direct sum of anyons: $\cA = a\oplus b \oplus \cdots$ where
$a,b,\cdots$ are the anyons with trivial self and mutual statistics.  The
direct sum of anyons, $\cA$, describes a vector space.  One also need to
supplement a multiplication rules for the fusion channels of these anyons.
Such a structure is referred to as a condensable algebra, that describes a set
anyons, that can condense together (for a review, see \cite{K13078244}).

If a condensable algebra $\cA$ contains enough anyons, then condensing those
anyons drives the topological order to a trivial order described by short-range
entangled states.  We will refer to such a condensable algebra as Lagrangian
condensable algebra.  For the simple cases that we will discuss in this paper,
it will be enough to specify direct sums of anyons $\cA$ to uniquely determine
the Lagrangian condensable algebras.

\begin{figure}[t]
\hskip -1.7in
\includegraphics[scale=0.6]{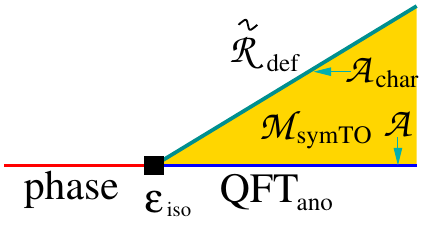}
\caption{ In the sym/TO correspondence, for systems with a symmetry described
by a fusion higher category $\tl R_\text{def}$, the gapped phases are
one-to-one classified by Lagrangian condensable algebras $\cA$ of the symTO
$\eM =\eZ(\tl R_\text{def})$.  The gapless phases are characterized (\ie
many-to-one classified) by non-Lagrangian condensable algebras $\cA$ of the
symTO $\eM$.  In the symmetric sub-Hilbert space (\ie the charge neutral
sector), the low energy properties of the phase are described by an anomalous
quantum field theory $QFT_\text{ano}$.  } \label{phaseQFTmorph} \end{figure}

When condensation of a Lagrangian condensable algebra $\cA$ drive a topological
order to trivial order, the same Lagrangian condensable algebra $\cA$ creates a
gapped boundary of the topological order, since such a boundary can be viewed
as an interface between the topological order and trivial order.  Therefore,
according to \Rf{K13078244,CZ190312334}, the gapped boundary phases of a
topological order can all be obtained from its Lagrangian condensable algebras.
This means that the gapped phases for systems with a symmetry are classified by
the Lagrangian condensable algebra of the symTO (see Fig.\ \ref{phaseQFTmorph}).  The direct sums of anyons
$\cA$ for all possible Lagrangian condensable algebras can be computed from
the bulk topological order $\eM$  (For a summary of calculation, see Appendix
of \Rf{NW230809670}).  In other words, this provides a way to compute all
possible gapped phases for a system $\underline{\cC}$ using the bulk symTO.

Once all gapped phases are enumerated, we can further characterize them in
terms of their ground state degeneracy (GSD). GSD can be computed from the
Lagrangian condensable algebras $\cA$ that produce the gapped phases at the
bottom boundary $\cC$ and the Lagrangian condensable algebra $\cA_\text{char}$
that produce the fusion higher category $\cR_\text{def}$ 
at the top boundary~\cite{CW221214432}.  


\section{Three kinds of duality automorphisms and SPT orders}
\label{autos}

After obtaining the gapped phases using the sym/TO correspondence, we want to
characterize those gapped phases in more details. Following
\Rfs{KZ200308898,KZ200514178}, we organize gapped phases by using the
\emph{automorphisms} of the symTO $\eM$ , \ie invertible maps $\mathfrak{A}\in
\mathrm{Aut}(\eM)$ from  $\eM$ to itself.  These are the so-called braided
auto-equivalences whose action does not affect the gauge invariant quantities
such as quantum dimensions, topological spins, or fusion rules~\cite{BW14104540}.  As we will see
below, these automorphisms correspond to dualities in physics.
We will, hence, refer to these automorphisms as \emph{duality automorphisms}.

In what follows, we will introduce an hierarchy of duality automorphisms that
are differentiated by how they act on the anyons $\cA_\text{char}$
corresponding to symmetry charges.  Namely, we are going to define three kinds
of duality automorphisms: \emph{fixed-charge duality (FCD)},
\emph{fixed-algebra duality (FAD)}, and \emph{fixed-symmetry duality (FSD)}
automorphisms.  We will explain how Lagrangian condensable algebras can be
organized  by those duality automorphisms, which produce corresponding SPT
orders and SSB patterns.  

For example, we will define an \emph{SPT-class} of gapped phases whose
corresponding Lagrangian condensable algebras are related by FCD
automorphisms. As we will argue, for group-like symmetries the group of FCD
automorphisms classify all SPT phases, while this is not necessarily so for
non-invertible symmetries. 

\subsection{Fixed-charge duality (FCD) automorphisms}
\label{charauto}

\emph{Fixed-charge duality (FCD) automorphisms} are invertible maps that 
do not act on the symmetry charges in the condensable algebra $\cA_\text{char}$, \ie
for all $a\in \cA_\text{char}$ the FCD automorphism $\mathfrak{A}_\text{FCD}$ act as
\begin{align}
\mathfrak{A}_\text{FCD}(a) = a.	
\end{align}
Hence, for the model $\underline{\cC}$, they correspond to unitary operators
that maps any local charged operator  to another local charged operator with
the same charge.  

We use FCD automorphisms as equivalence relations, such that the phases in the
same equivalence classes have the same SSB patterns.\footnote{However, 
phases in different SPT-classes do not necessarily have different SSB patterns,
since phases in different SPT-classes may have non-degenerate ground states
which do not break any symmetry (\ie have the same SSB pattern).} Those phases
are said to have different SPT-orders, by definition.

To better understand what an FCD automorphism is, let us consider the example
of a  \emph{globally-symmetric finite-depth quantum circuit (FDQC)}.  An FDQC
is formed by finite layers of unitary transformations each of which is formed
by local unitary operators acting on non-overlapping finite intervals on the
lattice.  Such an FDQC is called locally symmetric if each local unitary
operator commutes with the symmetry generators. Similarly, an FDQC is called
globally symmetric if the entire quantum circuit commutes with the symmetry
generators. Note that while the former implies the latter, the converse is not
necessarily true. 

A locally-symmetric FDQC maps states in the same gapped
phase into each other.  In fact, the equivalence classes induced by
such FDQCs can be used to define gapped phases
for systems with a symmetry.  In contrast, a globally-symmetric
FDQC can map one gapped phase to a different gapped phase.  
In particular, for group-like symmetries, it is known that globally-symmetric FDQCs,
the so-called \emph{entanglers}, can map
trivial symmetric product state to non-trivial SPT states.  

Since a globally-symmetric FDQC maps local operators to
local operators, they map local charged operators to local charged operators
that carry the same charge. Hence,  
the two phases connected by a globally-symmetric
FDQC have identical correlations for the
corresponding local operators.  Thus we cannot distinguish the two phases from
the bulk properties alone.  We need to use boundary properties to distinguish
the two phases. This bulk-indistinguishably is a characteristic property of SPT
orders. In this paper, we are going to regard the  bulk-indistinguishably as the
defining property of SPT order.

Motivated by these properties of FDQCs, we use FCD automorphisms of the symTO
as equilvalence relations  to group the gapped phases of symmetric systems into
equivalence classes, called \emph{SPT-classes}:  phases in same SPT-class are
connected by  FCD automorphisms, and phases in different SPT-classes are not
connected by FCD automorphisms.

In general, the phases in the same SPT-class only differ by SPT orders.
Therefore, if an SPT-class describing SSB phases contains more than one phase,
then we have distinct SSB phases that are distinguished by the additional SPT
orders they support.  We shall call these SSB-SPT phases.  While the phases in
the same SPT-class always have the same SSB pattern, \ie the same condensation
pattern of symmetry charges $\cA_\text{char}$, the converse is not true.  It
turns out that SPT-class is a more refined characterization than SSB pattern in
the sense that the phases in different SPT-classes may still have the same SSB
pattern (such as no symmetry breaking).

All SPT orders are described by FCD automorphisms
\cite{KZ200308898,KZ200514178}, which characterize the differences of the
phases in the same SPT-class.  For group-like symmetries, 1+1D gapped SPT
phases (\ie gapped phase with GSD = 1) all belong to the same SPT-class.  Those
SPT phases differ by SPT orders (\ie are connected by FCD automorphisms).
However, for non-invertible symmetries, there can exist SPT phases that belong
to different SPT-classes, which, thus, are not connected by FCD automorphisms.
Therefore, those  SPT phases are not characterized by their different SPT
orders (described by FCD automorphisms), nor by their different SSB patterns
since the symmetry is not broken.  They are characterized by their different
SPT-classes.

Using the holographic theory of symmetry, in this paper, we computed gapped SPT
phases, gapped SSB phases, and gapped SSB-SPT phases for some non-invertible
symmetries (as well as some group-like symmetries).  The results are presented
in Tables \ref{Phases} and \ref{PhasesS3xS3}.

From the table, we see that $A_4$, $\cRep_{A_4}$, $D_8$ symmetries have
non-trivial SPT orders.  $A_4$, $\cRep_{A_4}$, $D_8$ symmetries also have
non-trivial SSB-SPT orders  (\ie there are different gapped SSB phases that
only differ by SPT orders).  The gapped phases that only differ by SPT orders
cannot be distinguished by local bulk measurements.  One need use boundary
properties to distinguish them.

From the table, we can also observe that, for group-like symmetries, the gapped
phases with GSD = 1 form a single SPT-class. These phases are classified as
symmetric SPT phases. However, in the case of non-invertible symmetries such as
$\cRep_{S_3}\times \Z_3$ and $\cRep_{D_8}$, the GSD = 1 phases belong to
several distinct SPT-classes. These are examples of gapped SPT phases, whose
differences are beyond SPT orders.

The concept of SPT order was originally developed to describe gapped phases.
However, it has since been extended to apply to gapless phases as well~\cite{Scaffidi170501557,Jiang18,Parker18,Thorngren21,Verresen21,wen2023classification}. If we
use FCD automorphisms to characterize SPT order and define SPT-classes, then the
notion of SPT order can be naturally generalized to include gapless phases.
When two gapless phases belong to the same SPT-class (\ie, are connected by
FCD automorphisms), their differences are described by SPT order. This allows us
to classify gapless phases according to their SPT-classes and SPT orders. This
is another key result obtained in this paper.

From the Table \ref{gaplessPhases}, we see that $A_4$, $\cRep_{A_4}$, $D_8$
symmetries have non-trivial gapless SPT orders.  $A_4$, $D_8$ symmetries also
have non-trivial gapless SSB-SPT orders.  

\subsection{Fixed-algebra duality (FAD) automorphisms}
\label{symauto}

In the previous section, we used FCD automorphisms generated by globally-symmetric
FDQCs to characterize and organize phases of symmetric
systems.  This leads to the notion of SPT orders and SSB patterns.  
We can now further characterize  phases of symmetric systems, by using what we call
\emph{fixed-algebra duality (FAD) automorphisms} which are generated by
quantum circuits where the condition of being globally symmetric is relaxed.

While a globally-symmetric FQDC maps a local
charged operator to another local charged operator with the same charge,
by definition, a FAD automorphism maps a local charged operator to another local
charged operator which may carry a different charge.  Thus the quantum circuits
that generate FAD automorphisms may not be symmetric (\ie may not commute with
symmetry automorphisms).  In other
words, such quantum circuits corresponding to FAD automorphisms 
permute the symmetry charges, but keep the set of symmetry charges invariant. 

Or more precisely, the set of symmetry charges form a condensable algebra of
the symTO, and the FAD automorphisms keep the condensable algebra invariant, \ie
\begin{align}
\mathfrak{A}_\text{FAD}(A_\text{char}) \cong A_\text{char}.
\end{align}
Hence, we used the term ``fixed-algebra automorphism''.
Therefore, every FCD automorphism is also an FAD automorphism with trivial permutation of the
charges. The FAD automorphisms generated by quantum circuits include automorphisms of
symmetry itself.  For example, a $\Z_2\times \Z_2$ symmetry has an  automorphism that
exchange the two $\Z_2$ symmetries, which is an FAD autormorphism in the corresponding symTO,
$\Z_2\times\Z_2$ quantum double. 
 In contrast,this is not an FCD automorphism since charges of $\Z_2\times\Z_2$ symmetry are acted on.

We use the FAD automorphisms to organıze the phases of symmetric systems into
the pSPT-classes: the phases in a pSPT-class are connected by the
FAD automorphisms,  which include the automorphisms of the global symmetry.  
Phases related by FAD automorphisms have the same GSD, \ie a pSPT-class consists of 
phases with the same GSD while they may support distinct SPT-orders.

\subsection{Fixed-symmetry duality (FSD) automorphisms}
\label{symTOauto}

\begin{figure}[t]
\hskip -3.2in
\includegraphics[scale=0.5]{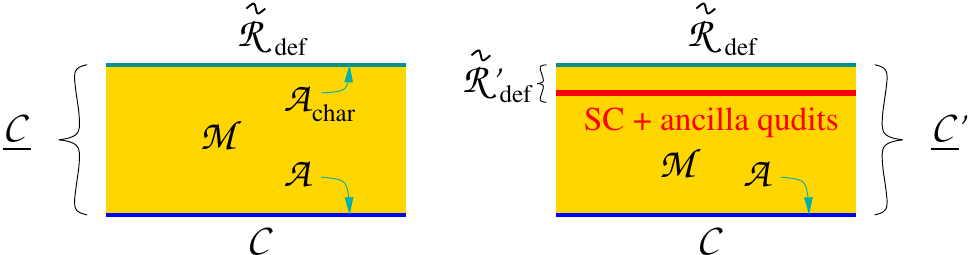}
\caption{ Adding ancilla qudits and applying sequential circuit (SC) to the
bulk topological order $\eM$ generate a duality transformation: $
\underline{\cC} \to \underline{\cC}' $.  $ \underline{\cC} $ and $
\underline{\cC}' $ can have different symmetries described by
$\tl\cR_\text{def}$ and $\tl\cR'_\text{def}$. But they have identical energy
spectrum within the their corresponding symmetric sub-Hilbert space (\ie the
corresponding charge neutral sectors).  A duality transformation becomes a FSD
automorphism if $\tl\cR'_\text{def}=\tl\cR_\text{def}$.  } \label{dualM} 
\end{figure}

The two previously discussed quantum circuits generate duality transformations.
The most general duality transformations correspond to changes of the top boundary in
Fig.\ \ref{CCmorph}: $\tl\cR_\text{def} \to \tl\cR'_\text{def}$.  This can be
achieved by adding ancilla qudits and applying sequential quantum circuits
formed by local symmetric unitary operators (see Fig.\  \ref{dualM})
\cite{LV200811187,VC240906647}.  From the point of view of the sym/TO
correspondence, the duality transformations are actually properties of
symmetry: 
\[
\boxed{\text{No symmetry, no duality.}}
\]
Between any pair of Morita-equivalent symmetries 
(which moy or may not be the same symmetry), 
there are one or more duality transformations.

In this work, we are going to focus on the
duality transformations that keep the symmetry $\tl\cR_\text{def}$ unchanged.  
We conjecture that such in general such duality transformations can be realized by sequential
quantum circuits without ancilla qudits, and correspond to automorphisms of the
symTO $\eM$.  We will refer to those duality transformations as fixed-symmetry
duality (FSD) automorphisms since the global symemtry structure is unchanged.

A FSD automorphism, as a map between Hamiltonians of the same symmetry, is an
linear unitary transformation (on a particular sub-Hilbert space), that maps a local symmetric operator to another
local symmetric operator.  It can be realized by sequential quantum circuits
formed by local symmetric unitary operators:
\begin{align} 
\label{SCact}
H_{\underline{\cC}} = \sum_i O_i\ \ &\to \ \  
H_{\underline{\cC}'} = \sum_i O_i' 
\nonumber\\ 
\text{where }\ \  O_i' &= U_{SC} O_i U_{SC}^\dag .  
\end{align} 
Here it is important that we restrict ourselves to a particular sub-Hilbert space (typically a definite symmetry eigensector).
In Section \ref{ZNZM}, we give explicit examples of quantum circuits
that realize the FSD automorphisms (as well as FCD and FAD automorphisms).

In the holographic picture of Fig.\ \ref{dualM}, an FSD automorphism is realized
as a sequential circuit acting on the bulk topological order $\eM$.  Note that
the local operators acting on the lower boundary $\cC$ are local symmetric
operators and they automatically commute with the sequential quantum circuits
in Fig.\ \ref{dualM}.

It is then clear that the two systems related by the FSD automorphism has
identical low energy spectrum in the symmetric sub-Hilbert space, which is
described by the lower boundary $\cC$.  But the ground states of the two
systems may belong to two different phases, despite they share identical
excitations in their symmetric sub-Hilbert spaces.  

From Eq.~\eqref{SCact}, we see that
duality automorphisms (\ie FSD automorphism) can lead to a duality symmetry, if
a duality automorphism maps a Hamiltonian into itself
\cite{WS170302426,CW221214432,BS231217322,CW240505331}:  
\begin{align} 
\label{SCsym}
H_{\underline{\cC}} = \sum_i O_i\ \ &\to \ \  
H_{\underline{\cC}} = \sum_i O_i 
\nonumber\\ 
\text{where }\ \  O_i &= U_{SC} O_i U_{SC}^\dag .  
\end{align} 
Such duality symmetries can be anomalous depending on the global
symmetry structure, \ie symmetric states in 1+1D must be gapless unless the symmetry is (spontaenously) broken. 

In Tables \ref{Phases}, \ref{PhasesS3xS3}, and \ref{gaplessPhases}, we list the
gapped and gapless phases for various generalized symmetries. These phases are
grouped into aSPT-classes.  Phases in the same aSPT-class are connected by FSD
automorphisms.  Within each aSPT-class, the phases are further divided into
pSPT-classes.  Phases in the same pSPT-class are connected by FAD automorphisms.
Finally, the phases within each pSPT-class are subdivided into SPT-classes.
Phases in the same  SPT-class are connected by FCD automorphisms.

\subsection{The physical meaning of the three kinds of duality automorphisms}

For two systems related by a duality automorphism, the algebra of their local
symmetric operators must be isomorphic to one another~\cite{CN09070733,CN11032776}. 
Then the local symmetric operators that are mapped  to each other have identitcal correlations 
on the symmetric states. In this sense, two such dual states are \emph{locally
equivalent in charge neutral sector}.  This local equivalence of charge neutral
sector is the essence of duality automorphisms.  The phases in the same
aSPT-class are locally equivalent in their charge neutral sectors.

For example, consider excitations on a gapped state in a local region in space with
zero total symmetry charge.  Such excitations are described by a fusion
category.  For two gapped states related by a FSD automorphism, the fusion
categories describing the excitations must be the same.

For two gapped states related by a FAD automorphism, again the fusion
categories describing those excitations are the same.  However, the
assignment of the objects in the fusion category to the symmetry charges may
differ by a permutation of of the latter between the two states.

Similarly, for two gapped states related by a FCD automorphism, the fusion
categories describing those excitations are the same.  Furthermore, the
assignment of the objects in the fusion category to the symmetry charges is
also the same in both states.  The FSD and FAD automorphisms may not have this
property.

The above understanding of the three types of automorphisms can also be extended to
gapless states. First, it's important to note that a system with symmetry can
be described by multiple partition functions. Specifically, for each sub-Hilbert space
characterized by a particular symmetry charge and a specific symmetry-twisted
boundary condition, there is a corresponding partition function. We will refer
to these sub-Hilbert spaces as ``symmetry-charge symmetry-twist (SCST) sectors.''

If two gapped or gapless states are related by a FSD automorphism, then the
partition function of the first system in a given SCST
sector will be identical to the partition function of the second systme in the
corresponding SCST sector. The relationship between
these sectors is determined by the FSD automorphism. In other words, the two
systems are indistinguishable, except for a permutation of their SCST sectors.

In particular, the partition functions of two systems related by FSD autormorphisms
coincide with one another when the
untwisted periodic conditions and the neutral charge sector are chosen.  If the two
gapless states are described by two conformal field theories (CFT) and are
related by an FSD automorphism, then the primary fields generated by local
symmetric operators will be the same for the two CFTs.  In other words, the
partition functions $Z_\onebb(\tau,\bar \tau)$ (see Appendix \ref{S3Zs}) on a
ring in the charge-neutral symmetry-untwisted sector are the same for the two
CFTs.  Furthermore, the correlation functions of all local
symmetric operators related by FSD automorphisms are identical on a ring with untwisted periodic conditions.


If two gapped or gapless states are related by an FCD automorphism, then
their corresponding partition functions on a ring with an untwisted periodic condition are
identical in any symmetry-charge sector. Additionally, the correlation
functions of all corresponding local symmetric and non-symmetric operators are
also identical on such a ring. In other words, the two states are
indistinguishable as long as the periodic boundary conditions remain untwisted. This
implies that an SPT order cannot be detected through bulk measurements.

Similarly, if two gapped or gapless states are connected by an
FAD automorphism, then on a ring with an untwisted periodic condition,
the partition function of the first state in any symmetry-charge sector will be
identical to the partition function of the second state in the corresponding
symmetry-charge sector. The relationship between the symmetry-charge and its
corresponding symmetry-charge is determined by the FAD automorphism.
(If the FAD automorphism is an FCD automorphism, then the
symmetry-charge sectors will be identical.) In
particular, the partition functions of both states on a ring with an untwisted
periodic condition are the same. Furthermore, the correlation functions of all
local operators that are dual to each other are identical on such a ring, even though they 
may carry distinct symmetry charges. (If the
FAD automorphism is an FCD automorphism, the dual
local operators will carry identical symmetry charges.) In other words, the two
states are indistinguishable up to a permutation of symmetry charges, as long
as the periodic condition is not twisted. This highlights the physical
significance of FAD automorphisms.

\section{1+1D gapped phases with finite Abelian symmetry}
\label{ZNZM}

As a warm up, let us illustrate how symTO point of view 
allows us to organize the gapped phases of 
local Hamiltonians with finite Abelian symmetries in 1+1D.
For simplicity, we consider the case $G=\Z^a_2\times \Z^b_2$~\footnote{
The uperscripts $a$ and $b$ are used to distinguish the two subgroups. 
Accordingly, the diagonal subgroup is denoted by $\Z^{ab}_2$.}
which can be straightforwardly generalized to any finite Abelian group.
For any finite symmetry $G$ in 1+1D, the corresponding symTO is the 2+1D
topological order $\mathcal{D}(G)$ -- the quantum double of $G$.
The quantum double $\mathcal{D}(\Z^a_2\times\Z^b_2)$ is an Abelian topological
order with  $2^2\times 2^2=16$ anyons.
The set of all anyons are generated by fusing the four generators
that we label as $e_a$, $e_b$, $m_a$ and $m_b$. These four anyons satisfy 
\begin{align}
e^{2}_a = m^{2}_a =
e^{2}_b = m^{2}_b = \onebb,
\end{align} 
and any other anyon can be obtained as the fusion product
$e^{\al_1}_a m^{\al_2}_a e^{\bt_1}_b\,m^{\bt_2}_b$,
where $\al_1,\,\al_2,\,\bt_1,\,\bt_2 \in \Z_2$.
Physically, the anyons $e_a$ and $e_b$ can be thought of as 
charges of $\Z_2$ subgroups while the anyons
$m_a$ and $m_b$ as their symmetry fluxes, respectively.

For the remaining of this section, we  choose the top boundary such that
the corresponding fusion category is 
$\tl\cR =\cVec_{\Z_2 \times \Z_2}$ which is obtained by condensing all symmetry charges,
\ie we choose our top boundary to be induced by the Lagrangian condensable algebra
\begin{align}
\label{eq:Z2xZ2 symm boundary}
\cA_\text{char}
=
\onebb \oplus e_a \oplus e_b \oplus e_a\,e_b
\end{align}
where all the charges of $\Z^a_2\times\Z^b_2$ symmetry are condensed.
This means that the $\Z^a_2\times\Z^b_2$ symmetry of the model is generated by the
anyon lines $m^\alpha_a\,m^\bt_b$ that correspond to various symmetry fluxes. 

Let us use an 1+1D model with $\Z^a_2\times\Z^b_2$ symmetry to illustrate the
above three types of duality automorphisms: FCD automorphisms,
FAD automorphisms, and FSD automorphisms.  
The corresponding symTO, $\Z^a_2\times\Z^b_2$ quantum double
$\cD(\Z^a_2\times\Z^b_2)$, has six Lagrangian condensable algebras. 
These are 
\begin{align}
\label{eq:LA for Z2xZ2}
\begin{split}
&
\cA_{1,1} = \onebb \oplus e_a \oplus e_b \oplus e_a\,e_b,
\\	
&
\cA_{1,2} = \onebb \oplus m_a \oplus e_b \oplus m_a\,e_b,
\\
&
\cA_{1,3} = \onebb \oplus e_a \oplus m_b \oplus e_a\,m_b,
\\
&
\cA_{1,4} = \onebb \oplus m_a \oplus m_b \oplus m_a\,m_b,
\\
&
\cA_{1,5} = \onebb \oplus e_a\,e_b \oplus m_a\,m_b \oplus e_a\,e_b\,m_a\,m_b,
\\
&
\cA_{1,6} = \onebb \oplus e_a\,m_b \oplus m_a\,e_b \oplus e_a\,e_b\,m_a\,m_b.
\end{split}
\end{align}
With the choice of $\cA_\text{char}$ in Eq.~\eqref{eq:Z2xZ2 symm boundary} as 
the top boundary, the Lagrangian algebra $\cA_{1,1} \equiv \cA_\text{char}$ describes 
gapped phase where global $\Z^a_2\times\Z^a_2$ symmetry is completely broken down to 
$Z_1$. Similarly, the Lagrangian algebras $\cA_{1,2}$, $\cA_{1,3}$, $\cA_{1,5}$ describe three
SSB patterns where  $\Z^a_2\times\Z^b_2$ symmetry is broken down to the $\Z^a_2$, $\Z^b_2$, and
$\Z^{ab}_2$ subgroups, respectively. 
Finally, the remaining two Lagrangian algebras  $\cA_{1,4}$ and $\cA_{1,6}$ describes two SPT orders that
preserve the global $\Z^a_2\times \Z^b_2$ symmetry. 

Let us now describe various duality automorphisms.
To do so, let us construct simple exactly solvable Hamiltonians 
corresponding to each of these gapped phases. 
We consider a one-dimensional lattice of $L$ sites 
where each site $i$ supports
two flavors of qubit operators $\{X_i,Z_i,\tl X_i. \tl Z_i\}$
that satisfy the algebra 
\begin{align}
&
X_i\,
Z_j
=
(-1)^{\delta_{i,j}}\,
Z_j\,
X_i,
\quad
X_i^{2}
=
Z_i^{2}
=
1,
\\
&
\tl X_i\,
\tl Z_j
=
(-1)^{\delta_{i,j}}\,
\tl Z_j\,
\tl X_i,
\quad
\tl X_i^{2}
=
\tl Z_i^{2}
=
1.
\end{align}
The global 
$\Z^a_2\times \Z^b_2$ symmetry is generated by the unitary operators
$U_a=\prod_i X_i$ and $U_b=\prod_i \tl X_i$.
Then, the six phases are realized by the following six fixed-point Hamiltonians:
\begin{subequations}
\label{eq:fixed-point Ham for Z2xZ2}
\begin{align}
&
H_{1,1}
=
-
\sum_{i=1}^L
\left(
Z_i\,Z_{i+1}
+
\tl Z_i\,\tl Z_{i+1}
\right),
\\
&
H_{1,2}
=
-
\sum_{i=1}^L
\left(
X_i
+
\tl Z_i\,\tl Z_{i+1}
\right),
\\
&
H_{1,3}
=
-
\sum_{i=1}^L
\left(
Z_i\,Z_{i+1}
+
\tl X_i\,
\right),
\\
&
H_{1,4}
=
-
\sum_{i=1}^L
\left(
X_i\,
+
\tl X_i\,
\right),
\\
&
H_{1,5}
=
-
\sum_{i=1}^L
\left(
Z_i\,\tl Z_i\,
Z_{i+1}\, \tl Z_{i+1}
+
X_i\,
\tl X_i\,
\right),
\\
&
H_{1,6}
=
-
\sum_{i=1}^L
\left(
Z_i\,
\tl X_i\,
Z_{i+1}
+
\tl Z_i\,
X_{i+1}\,
\tl Z_{i+1}
\right).
\end{align}
\end{subequations}
In what follows, we will construct the three kinds of
duality automorphisms explicitly with the emphasis on their physical meaning.

\subsection{FCD automorphisms}
An FCD automorphism for $\Z^a_2\times\Z^b_2$ symmetry is generated by a
finite-depth quantum circuit 
\begin{align}
E = \prod_i E_i,
\quad
E_i = \frac{1}{2}
(1+Z_{i}\,Z_{i+1}) + \tl Z_i \,  \frac{1}{2}(1-Z_{i}\,Z_{i+1}),
\end{align}
which implements the transformation
\begin{align}
\begin{split}
&
E\, Z_i \, E^\dag = Z_i,  
\qquad\qquad\,\,\,\,
E\, \tl Z_i \, E^\dag = \tl Z_i,
\\
&
E\, X_i \, E^\dag = \tl Z_{i-1}\,X_i\, \tl Z_i, \quad E\, \tl X_i \, E^\dag =   Z_{i}\,\tl X_i\, Z_{i+1}.
\end{split}
\end{align}
It is straightforward to verify that $E$ commutes with both symmetry operators 
$U_a$ and $U_b$, while the local unitaries $E_i$ do not. 
Hence, the unitary $E$ is a globally symmetric but locally non-symmetric FDQC. 
Physically, this FDQC is the so-called entangler that exchanges the trivial and non-trivial
SPT orders with $\Z^a_2\times\Z^b_2$ symmetry.

Comparing the action of FDQC $E$ on the fixed-point Hamiltonians 
\eqref{eq:fixed-point Ham for Z2xZ2} and the corresponding Lagrangian algebras
\eqref{eq:LA for Z2xZ2}, we identify the FCD automorphism~\footnote{Note that Hamiltonians
$H_{1,2}$, $H_{1,3}$, and $H_{1,5}$ are not invariant under the FDQC  $E$, while their respective
gapped ground states are. }
\begin{align}
\mathfrak{A}_E: 
\left(e_a, \, e_b, \, m_a,\, m_b\right)
\mapsto
\left(e_a, \, e_b, \, m_a\,e_b,\, m_b\, e_a\right).
\end{align}
Note that this automorphism does not act on any of the anyons condensed at the 
top symmetry boundary $\cA_\text{char}\equiv \cA_{1,1}$. Hence, it is indeed an 
FCD automorphism.  Furthermore,
all Lagrangian algebras \eqref{eq:LA for Z2xZ2} are invariant under 
$\mathfrak{A}_U$, except $\cA_{1,4}$ and $\cA_{1,6}$ which are the two 
SPT orders. Hence, there are 5 SPT-classes with $\Z^a_2\times\Z^b_2$ symmetry
\begin{align*}
(\cA_{1,1})_4,\, (\cA_{1,2})_2, \, (\cA_{1,3})_2, \, (\cA_{1,5})_2,
\, (\cA_{1,4},\, \cA_{1,6})_1
\end{align*}
which are denoted by round brackets with the subcripts denoting the respectrive
GSD of each SPT-class. In fact, with the choice of the top boundary
\eqref{eq:Z2xZ2 symm boundary}, $\mathfrak{A}_E$ is the only FCD automorphism
for the symTO $\cD(\Z^a_2\times\Z^a_2)$.  Accordingly, there are only two SPT
orders with $\Z^a_2\times Z^b_2$ symmetry.  These SPT orders form the group
$\Z_2$ under stacking which can be understood as the group generated by the FCD
automorphism $\mathfrak{A}_E$ which squares to the identity autormorphism.

\subsection{FAD automorphisms}

There are 12 FAD automorphisms that leave the top boundary 
$\cA_\text{char}$ \eqref{eq:Z2xZ2 symm boundary}
invariant. By definition, every FCD automorphism is also a FAD automorphism. Hence, there are 
6 FAD automorphisms that act on the $e_a$ and $e_b$ anyons while keeping the 
Lagrangian algebra $\cA_\text{char}$ invariant. At the level of lattice models, these 6 automorphisms are generated
by the following two unitary transformations
\begin{subequations}
\begin{align}
&
S = \prod_i S_i, 
\\
&
S_i = \frac{1}{2}\left(1 + Z_i\,\tl Z_i + X_i\,\tl X_i - Z_i\,\tl Z_i\,X_i\,\tl X_i\right),
\end{align}
and
\begin{align}
\begin{split}
&
R = \prod_i R_i, 
\\
&
R_i = \frac{1}{2}
\left(
1 + \tl Z_i + X_i - X_i\,\tl Z_i
\right)\,
S_i
\end{split},
\end{align}
which satisfy
\begin{align}
R^3 = 1, \qquad S^2=1, \qquad S\,R = R^\dag\,S.
\end{align}
\end{subequations}
These two unitary transformations are FDQCs that obey the multiplication rules 
of the non-Abelian group $S_3$. 
They implement the transformations
\begin{subequations}
\begin{align}
\begin{split}
&
S\, Z_i\, S^\dag = \tl Z_i,
\quad\,\,\,
S\, \tl Z_i\, S^\dag =  Z_i,
\\
&
S\, X_i\, S^\dag = \tl X_i,
\quad
S\, \tl X_i\, S^\dag = X_i,
\end{split}
\end{align}
and
\begin{align}
\begin{split}
&
R\, Z_i\, R^\dag = \tl Z_i,
\quad\,\,\,
R\, \tl Z_i\, R^\dag =  Z_i\, \tl Z_i,
\\
&
R\, X_i\, R^\dag = X_i\, \tl X_i,
\quad
R\, \tl X_i\, R^\dag = X_i.
\end{split}
\end{align}
\end{subequations}
Importantly, these two FDQCs do not commute with the $\Z^a_2\times\Z^b_2$ 
symmetry generators. Instead, they implement the group isomorphisms of $\Z^a_2\times\Z^b_2$, 
\ie they form a representation of the automorphism group of $\Z^a_2\times\Z^b_2$ 
which is precisely $S_3$.  Hence, these FDQCs permute the charges and fluxes of the 
$\Z^a_2\times\Z^b_2$ symmetry.

Comparing the actions of $S$ and $R$ on the fixed-point Hamiltonians 
\eqref{eq:fixed-point Ham for Z2xZ2} and the corresponding Lagrangian algebras
\eqref{eq:LA for Z2xZ2}, we identify the two FAD automorphisms to be
\begin{subequations}
\begin{align}
&
\mathfrak{A}_S: 
\left(e_a, \, e_b, \, m_a,\, m_b\right)
\mapsto
\left(e_b, \, e_a, \, m_b,\, m_a\right),
\\
&
\mathfrak{A}_R: 
\left(e_a, \, e_b, \, m_a,\, m_b\right)
\mapsto
\left(e_b, \, e_a\,e_b,  \, m_a\,m_b,\, m_a\right).
\end{align}
\end{subequations}
It is straightforward to verify that the condensable algebra,
$\cA_\text{char}\equiv \cA_{1,1}$ that induces the top boundary, is invariant
under both FAD automorphisms.  We can organize the six Lagrangian algebras
\eqref{eq:LA for Z2xZ2} into 3 pSPT-classes 
\begin{align*}
[(\cA_{1,1})]_4,\, 
[(\cA_{1,2}), \, (\cA_{1,3}), \, (\cA_{1,5})]_2, \, 
[(\cA_{1,4},\, \cA_{1,6})]_1
\end{align*}
which are denoted by brackets with subscripts denoting the respective GSD of each pSPT-class. 

\subsection{FSD automorphisms}

The FSD automorphisms 
for $\Z^a_2\times\Z^b_2$ symmetry are generated by three
non-invertible operators $D_a$, $D_b$, and $D_{ab}$ with
\begin{subequations}
\label{eq:KW duality transformations}
\begin{align}
&
D_a =\frac{1+\prod_i X_i}{2}\ee^{\ii \frac{\pi}{4} X_L}\prod_{i=L-1}^{1} \ee^{\ii \frac{\pi}{4} X_i} \ee^{\ii \frac{\pi}{4} Z_iZ_{i+1}},
\\
&
D_b =
\frac{1+\prod_i \tl X_i}{2}\ee^{\ii \frac{\pi}{4} \tl X_L}\prod_{i=L-1}^{1}\ee^{\ii \frac{\pi}{4} \tl X_i} \ee^{\ii \frac{\pi}{4} \tl Z_i\tl Z_{i+1}},
\\
&
D _{ab} =R^\dag \frac{1+\prod_i X_i\tl X_i}{2}
\ee^{\ii \frac{\pi}{4} X_L\tl X_L}\prod_{i=L-1}^{1} \ee^{\ii \frac{\pi}{4} X_i\tl X_i} \ee^{\ii \frac{\pi}{4} \tl Z_i \tl Z_{i+1}}
\nonumber
\\
&\quad\,\,\,
=
D_a\,R^\dag.
\end{align}
\end{subequations}
Note that these non-invertible operators are composition of sequential quantum circuits (SQC)
with projectors to the $U_a=1$, $U_b=1$, or  $U_a\,U_b=1$ subspaces.
The SQCs (i) consist of local unitaries that are symmetric under $\Z_2\times\Z_2$ symmetry
and (ii) have depth linear in system size, hence, the name sequential.
These threee non-invertible opertators implement the Kramer-Wannier duality transformations~\cite{SS230702534}
corresponding to gauging the $\Z^a_2$, $\Z^b_2$, and $\Z^{ab}_2 $ subgroups, respectively. 
We remark that the projectors in the non-invertible operators \eqref{eq:KW duality transformations} 
are trivialized in the $\Z^a_2\times\Z^b_2$-symmetric sub-Hilbert space. In this subspace,
these nin-invertible operators reduce to the unitary SQCs which are precisely the FSD automorphisms 
described in Sec.\ \ref{symTOauto} (recall Eq.\ \eqref{SCact}).

The non-invertible duality transformations $D_a$, $ D_b$, and $D_{ab}$  
map the local $\Z^a_2\times \Z^b_2$-symmetric
operators to the same set of local symmetric operators
\begin{align}
\begin{split}
&
D_a X_i = Z_iZ_{i+1} D_a, \ \ \ D_a Z_iZ_{i+1} = X_{i+1} D_a,
\\
&
D_b \tl X_i = \tl Z_i\tl Z_{i+1} D_b, \ \ \  D_b \tl Z_i\tl Z_{i+1} = \tl X_{i+1} D_b,
\\
&
D_{ab} X_i = \tl X_i D_{ab}, \ \ \  D_{ab} \tl X_i = \tl X_i Z_i Z_{i+1} D_{ab}, 
\\
&
D_{ab} Z_i Z_{i+1} =  \tl Z_i\tl Z_{i+1}  X_{i+1}  D_{ab},
\ \ \
D_{ab} \tl Z_i\tl Z_{i+1} = X_{i+1}  D_{ab}.
\end{split}
\end{align}
Therefore, under these duality transformations 
a model with $\Z^a_2\times \Z^b_2$ symmetry generated by $U_a$
and $U_b$ to another model with the same $\Z^a_2\times \Z^b_2$ symmetry
generated by the same $U_a$ and $U_b$.  In other words, 
the three operators $D_a$, $D_b$, and $D_{ab}$ implement FSD automorphisms.

Comparing the actions of non-invertible duality transformations
\eqref{eq:KW duality transformations} on the fixed-point Hamiltonians 
\eqref{eq:fixed-point Ham for Z2xZ2} and the corresponding Lagrangian algebras
\eqref{eq:LA for Z2xZ2}, we identify the three FSD automorphisms to be
\begin{subequations}
\begin{align}
&
\mathfrak{A}_{D_a}: 
\left(e_a, \, e_b, \, m_a,\, m_b\right)
\mapsto
\left(m_a, \, e_b, \, e_a,\, m_b\right),
\\
&
\mathfrak{A}_{D_b}: 
\left(e_a, \, e_b, \, m_a,\, m_b\right)
\mapsto
\left(e_a, \, m_b, \, m_a,\, e_b\right),
\\
&
\mathfrak{A}_{D_{ab}}: 
\left(e_a, \, e_b, \, m_a,\, m_b\right)
\mapsto
\nonumber\\
&
\qquad\qquad\qquad
\left(e_a\,e_b\,m_a\,m_b, \, m_a\,m_b,  \, m_a,\, m_a\,e_b\right).
\end{align}
\end{subequations}
Note that each of these automorphisms are (by definition) invertible in the
bulk symTO while they are non-invertible at the  1+1D boundary.

Using these automorphisms one can verify that all Lagrangian condensable
algebras \eqref{eq:LA for Z2xZ2} can be connected to one another. In other
words, distinct pSPT classes are related to each other under gauging distinct
$\Z_2$ subgroups of $\Z^a_2\times\Z^b_2$ symmetry~\footnote{This is only
possible for Abelian internal symmetries whereas for non-Abelian internal
symmetries not every pSPT class is connected by an FSD automorphisms, see Table
\ref{Phases}. This is because gauging non-normal subgroups changes the symmetry
$\cR_\text{def}$ at the top boundary and cannot be implemented by an FSD
automorphism.}.  We then identify a single aSPT-class for the global
$\Z^a_2\times\Z^b_2$ symmetry:
\begin{align*}
\Big\{
[(\cA_{1,1})]_4,\, 
[(\cA_{1,2}), \, (\cA_{1,3}), \, (\cA_{1,5})]_2, \, 
[(\cA_{1,4},\, \cA_{1,6})]_1
\Big\}
\end{align*}
We use the short hand notation to describe 
aSPT-classes,
pSPT-classes, and
SPT-classes:
\begin{align*}
\left\{
[\textbf{1}]_4,\, 
[3\times \textbf{1}]_2, \, 
[\textbf{2}]_1
\right\},
\end{align*}
for the organization of gapped phases. Here, the bold-face number denotes the
number of SPT orders related by FCD automorphisms, while each bracket denotes
the organization of such SSB-SPT orders within each pSPT-class.  For examples,
$3\times \textbf{1}$ refers to the three SSB patterns related by FAD
automorphisms. As is our convention, the subscripts denote the ground state
degeneracies in the same pSPT class. 

\subsection{SPT phases with $\Z_N\times \Z_M$ symmetry}

We close this section by discussing the SPT orders protected by $\Z_N\times \Z_M$ internal symmetry. 
In this case, the corresponding quantum double $\cD(\Z_N\times\Z_M)$ has $N^2\times M^2$ anyons. 
Just as in the case of $\Z_2\times\Z_2$ symmetry, we choose the symmetry boundary to be
\begin{align}
\label{eq:ZNxZM symm boundary}
\cA_\text{char}
=
\bigoplus_{\al=1}^{N}
\bigoplus_{\bt=1}^{M}
e^{\al}_a\,
e^{\bt}_b,
\end{align}{\tiny }
where all charges are condensed. Then, the
SPT orders protected by $\Z_N\times \Z_M$ symmetry are obtained 
by choosing a Lagrangian condensable algebra on the physical boundary
that has no overlap with the symmetry boundary \eqref{eq:ZNxZM symm boundary}.
One such Lagrangian algebra is 
\begin{align}
\label{eq:ZNxZM triv SPT boundary}
\cA_\text{def}
=
\sum_{\al=1}^{N}
\sum_{\bt=1}^{M}
m^{\al}_a\,
m^{\bt}_b,
\end{align}
which corresponds to the condensation of all symmetry fluxes. 
Any other SPT order can be obtained by acting on this boundary 
with the FCD automorphisms that act trivially on the anyons condensed 
at the symmetry boundary. By inspection one finds that 
there are $p=\mathrm{gcd}(N,M)$ such automorphisms which are generated by 
successive applications of the automorphism 
\begin{align}
\mathfrak{A}_\text{FCD}: 
\begin{pmatrix}
e_a, & e_b, & m_a, & m_b
\end{pmatrix}
\mapsto
\begin{pmatrix}
e_a, & e_b, & m_a\,e^{\frac{N}{p}}_b, & m_b\,e^{\frac{M}{p}}_b.
\end{pmatrix}
\end{align} 
One verifies that $\mathfrak{A}_\text{FCD}^{p}$ is the identity automorphism.
Hence, the automorphisms that do not act on the symmetry boundary
$\cA_\text{char}$ \eqref{eq:ZNxZM symm boundary} form the Abelian group $\Z_p$
with its generator being $\mathfrak{A}_\text{FCD}$.  We can declare the
symmetry flux boundary \eqref{eq:ZNxZM triv SPT boundary} to be the trivial SPT
order.  The group $\Z_p$ of the automorphisms then corresponds to the group of
SPT phases under stacking operation. This recovers the well-known
classification of 1+1D SPT orders by the second  group cohomology
\begin{align}
H^{2}\left(\Z_N\times\Z_M, \mathrm{U}(1)\right) = \Z_{\mathrm{gcd}(N,M)}.
\end{align}
This argument naturlly extends to any finite Abelaian symmetry for which the
group of FCD automorphisms are on one to one correspondence with the group of
SPT orders under stacking.

\section{$\eD(S_3)$-symTO}

\subsection{Phases in 1+1D systems with $S_3$-symmetry}

\begin{table*}[t] 
\caption{
The fusion rule of $\cD(S_3)$, which contains three sub fusion categories 
$\cRep_{\Z_2}=\{\onebb, \one\}$,
$\cRep_{S_3}= \{\onebb, \one, \two\}$,
and $\cRep'_{S_3}= \{\onebb, \one, b\}$.
} \label{S3Nijk} \centering
\begin{tabular}{ |c||c|c|c|c|c|c|c|c|}
\hline 
8  & $\onebb$  & $\one$  & $\two$  & $b$  & $b'$  & $b''$  & $c$  & $c'$ \\ 
\hline 
\hline 
$\onebb$  & $ \onebb$  & $ \one$  & $ \two$  & $ b$  & $ b'$  & $ b''$  & $ c$  & $ c'$  \\ 
\hline 
$\one$  & $ \one$  & $ \onebb$  & $ \two$  & $ b$  & $ b'$  & $ b''$  & $ c'$  & $ c$  \\ 
\hline 
$\two$  & $ \two$  & $ \two$  & $ \onebb \oplus \one \oplus \two$  & $ b' \oplus b''$  & $ b \oplus b''$  & $ b \oplus b'$  & $ c \oplus c'$  & $ c \oplus c'$  \\ 
\hline 
$b$  & $ b$  & $ b$  & $ b' \oplus b''$  & $ \onebb \oplus \one \oplus b$  & $ \two \oplus b''$  & $ \two \oplus b'$  & $ c \oplus c'$  & $ c \oplus c'$  \\ 
\hline 
$b'$  & $ b'$  & $ b'$  & $ b \oplus b''$  & $ \two \oplus b''$  & $ \onebb \oplus \one \oplus b'$  & $ \two \oplus b$  & $ c \oplus c'$  & $ c \oplus c'$  \\ 
\hline 
$b''$  & $ b''$  & $ b''$  & $ b \oplus b'$  & $ \two \oplus b'$  & $ \two \oplus b$  & $ \onebb \oplus \one \oplus b''$  & $ c \oplus c'$  & $ c \oplus c'$  \\ 
\hline 
$c$  & $ c$  & $ c'$  & $ c \oplus c'$  & $ c \oplus c'$  & $ c \oplus c'$  & $ c \oplus c'$  & $ \onebb \oplus \two \oplus b \oplus b' \oplus b''$  & $ \one \oplus \two \oplus b \oplus b' \oplus b''$  \\ 
\hline 
$c'$  & $ c'$  & $ c$  & $ c \oplus c'$  & $ c \oplus c'$  & $ c \oplus c'$  & $ c \oplus c'$  & $ \one \oplus \two \oplus b \oplus b' \oplus b''$  & $ \onebb \oplus \two \oplus b \oplus b' \oplus b''$  \\ 
\hline 
\end{tabular}
\end{table*}

\subsubsection{The symTO of 1+1D $S_3$-symmetry}
\label{S3sym}

Let us use a simple example of $S_3$-symmetry in 1+1D to illustrate the notion
of symTO, and its application in determining gapped and gapless phases.  The
symTO of 1+1D $S_3$ is a 2+1D topological order, whose exactions are described
by non-degenerate braided fusion 1-category (also known as modular tensor
category (MTC)) $\eD(S_3)$ -- the quantum double of $S_3$ group.  Physically,
such a topological order is described by the $S_3$ gauge theory.  The
$\eD(S_3)$ quantum double has eight anyons (include the trivial one $\onebb$),
whose topological spin $s$ and quantum dimension $d$ are given below:\\[2mm]
\centerline{
\begin{tabular}{|c|c|c|c|c|c|c|c|c|}
\hline
$\eD(S_3)$ anyon& $\onebb$ & $\one$ & $\two$ & $b$ & $b'$ & $b''$ & $c$ & $c'$ \\ 
\hline
topological spin $s$: & $0$ & $0$ & $0$ & $0$ & $\frac{1}{3}$ & $\frac{2}{3}$ & $0$ & $\frac{1}{2}$ \\ 
\hline
quantum dimension $d$: & $1$ & $1$ & $2$ & $2$ & $2$ & $2$ & $3$ & $3$ \\ 
\hline
\end{tabular}
}
\vskip 2mm
\noindent
The fusion rule of those anyons is given in Table \ref{S3Nijk}.  The anyons
$(\onebb, \one, \two)$ in $\eD(S_3)$ correspond to the three irreducible
representations (irreps) of $S_3$ of dimension $(1,1,2)$.

The $\eD(S_3)$-symTO and its $S_3$ and $\cRep_{S_3}$ symmetries have been
discussed extensively in \Rf{CW220506244,BT240505302,CW240505331}.  Here, we
add some discussions of automorphisms, and its associated SPT orders.


An automorphism of a symTO  is defined as a permutation of anyons that does not
change the quantum dimensions $d_i$, the topological spins $s_i$, and the
fusion rule $N^{ij}_k$ of the anyons.  We conjecture that the automorphisms of
symTO correspond to the FSD automorphisms discussed in Section \ref{autos}.  So
we will also refer the automorphisms of symTO as FSD automorphisms.  The
$\eD(S_3)$ quantum double has an automorphism group generated exchanging two
anyons $(\two,b)$. Those automorphisms correspond to symmetry-preserving
duality transformations, whose quantum circuit realizations on lattice are
given in \Rf{CW240505331}.

The $\eD(S_3)$ quantum double has the following four
Lagrangian condensable algebras\footnote{Here $\cA  =  \onebb \oplus \one
\oplus 2b$ means $\cA  = \onebb \oplus \one \oplus b \oplus b$.}
\begin{align}
\label{S3Lcas}
& \cA_{1,1} =  \onebb \oplus \one \oplus 2\two
\nonumber \\
& \cA_{1,2} =  \onebb \oplus \one \oplus 2b
\nonumber \\
& \cA_{2,1} =  \onebb \oplus c \oplus \two
\nonumber \\
& \cA_{2,2} =  \onebb \oplus c \oplus b
.
\end{align}
In the above, we group the four Lagrangian condensable algebras into two
\emph{aSPT-classes}.  The Lagrangian condensable algebras in one
aSPT-class are connected by the FSD automorphisms and the Lagrangian
condensable algebras in different aSPT-class are not connected by the
FSD automorphisms.  The first subscript of $\cA$ is the index of
aSPT-classes, and the second subscript is the index of the Lagrangian
condensable algebras in the aSPT-classes.

Thus the $\eD(S_3)$ topological order has four gapped boundary phases, induced
by the condensation of the four Lagrangian condensable algebras.  However, the
Lagrangian condensable algebras in the same aSPT-class will produce the
same fusion category $\tl\cR_\text{def}$ that describes the boundary
excitations on their induced boundaries.  The aSPT-class-1 induces a
fusion category $\tl\cR_\text{def} = \cVec_{S_3}$.  The aSPT-class-2
induces a fusion category $\tl\cR_\text{def} = \cRep_{S_3}$. 

According to sym/TO correspondence, a symTO describes several symmetries.
Those symmetries form a so called Morita-equivalent class \cite{KZ200514178}.
From the above discussion, we see that Morita-equivalent class described by the
$\eD(S_3)$ quantum double contains two symmetries.  One is the
$\cVec_{S_3}$-symmetry -- the usual $S_3$ group-like symmetry.  The other is
the $\cRep_{S_3}$-symmetry -- a non-invertible symmetry
\cite{TW191202817,KZ200514178}.  

As Morita-equivalent symmetries, the phase diagram for $S_3$-symmetric systems
and for $\cRep_{S_3}$-symmetric systems are identical, and the corresponding
critical points of phase transitions are also identical \cite{KZ200514178}.

The symTO $\eD(S_3)$ also has 4 non-Lagrangian condensable algebras,
grouped into three aSPT-classes by the FSD automorphisms:
\begin{align}
\label{S3nLcas}
& \cA'_{1,1} =  \onebb \oplus \two
,\ \ \
\cA'_{1,2} =  \onebb \oplus b,
\nonumber \\
& \cA'_{2,1} =  \onebb \oplus \one,
\nonumber \\
& \cA'_{3,1} =  \onebb .
\end{align}
Thus the $\eD(S_3)$ symTO has 4 classes of gapless phases.  Those gapless
phases are described by symTO resolved multi-component partition functions
\cite{CW220506244}.  Those resolved multi-component partition functions are
given in Appendix \ref{S3Zs}.  

The two gapless phases, $\cA'_{1,1} $-phase and $ \cA'_{1,2} $-phase, are
connected by FSD automorphisms, and belong to the same aSPT-class.  As a
result, they have identical spectrum in the symmetric sub-Hilbert space, \ie
the $Z_\onebb$ component of the partition functions for the two gapless phases
are identical (see \eqref{Zm4S3A11} and \eqref{Zm4S3A12}, as well as
\eqref{Zm4S3A11a} and \eqref{Zm4S3A12a}).

\subsubsection{$\cVec_{S_3}$-symmetry =  $S_3$-symmetry}

The Morita-equivalent class of $\eD(S_3)$-symTO contain two symmetries:
$\cVec_{S_3}$-symmetry and $\cRep_{S_3}$-symmetry.  In this section, we will
consider the $\cVec_{S_3}$-symmetry.

First, we like to use the sym/TO approach to show that the
$\cVec_{S_3}$-symmetry is anomaly-free. 
To do this, we consider the isomorphic holographic decomposition of
$\cVec_{S_3}$-symmetric systems as described in Fig.\ \ref{CCmorph}, by choosing
the top gapped boundary $\tl\cR_\text{def} = \cVec_{S_3}$.  We use such a
isomorphic holographic decomposition to describe the $\cVec_{S_3}$-symmetry.

The gapped phases of  $\cVec_{S_3}$-symmetric systems are obtained by choosing
the bottom boundary $\cC$ to be one of the four gapped boundaries induced by
the four Lagrangian condensable algebras in \eqref{S3Lcas}. Thus the
$\cVec_{S_3}$-symmetric systems has four different gapped phases.

If one of the gapped phases has a non-degenerate ground state, then the
$\cVec_{S_3}$-symmetry is anomaly-free, by definition.  To compute the ground
state degeneracy, we first note that the top boundary $\tl\cR_\text{def}$ in
Fig \ref{CCmorph} is one of the four gapped phases described by $\cVec_{S_3}$,
which is induced by the Lagrangian condensable algebra $\cA_{1,1}$.  If the
bottom  boundary $\cC$ is induced by the Lagrangian condensable algebra
$\cA_i$, then the ground state degeneracy is given by the inner product
$(\cA_{1,1}, \cA_i)$ \cite{CW221214432}:
\begin{align}
\label{GSDS3A11}
\text{GSD}^{\cVec_{S_3}}_{\cA_i} = (\cA_{1,1}, \cA_i).
\end{align}
Here the ``inner product'' between two Lagrangian condensable algebras is
defined as follows: we write a condensable algebra $\cA$ as $\cA = A_a a \oplus
A_b b \oplus \cdots$.  In other words, $\cA$ can be viewed as a vector in a
vector space with basis vectors given by the anyons.  The expansion
coefficients $A_a$ of $\cA$ on such an anyon basis are non-negative integers.
Then $(\cA_{1,1}, \cA_i)$ is the inner product of such integral vectors, where
the anyon basis is regarded as an orthonormal basis.  We find the ground state
degeneracies of four gapped phases of $\cVec_{S_3}$-symmetric systems are given
by
\begin{align}
& \text{GSD}^{\cVec_{S_3}}_{\cA_{1,1}} = 6,\ \ \ 
\text{GSD}^{\cVec_{S_3}}_{\cA_{1,2}} = 2,
\nonumber \\
& \text{GSD}^{\cVec_{S_3}}_{\cA_{2,1}} = 3,\ \ \
\text{GSD}^{\cVec_{S_3}}_{\cA_{2,2}} = 1
\end{align}

Since $ \text{GSD}^{\cVec_{S_3}}_{\cA_{2,2}} = 1$, the $\cVec_{S_3}$-symmetry
is anomaly-free. This is a well known result: $\cVec_{S_3}$-symmetry is nothing
but the $S_3$-symmetry, and the $S_3$ group-like symmetry is anomaly-free.  The
four gapped phases are the four spontaneous symmetry breaking (SSB) phases of
the $S_3$-symmetry:\cite{CW220506244}
\begin{align}
\cA_{1,1}\text{-phase} \big |_{\text{GSD}=6} &= (S_3\to \Z_1)\text{-SSB phase},
\nonumber\\
\cA_{1,2}\text{-phase} \big |_{\text{GSD}=2} &= (S_3\to \Z_3)\text{-SSB phase},
\nonumber\\
\cA_{2,1}\text{-phase} \big |_{\text{GSD}=3} &= (S_3\to \Z_2)\text{-SSB phase} ,
\nonumber\\
\cA_{2,2}\text{-phase} \big |_{\text{GSD}=1} &= (S_3\to S_3)\text{-symmetric-phase}.
\end{align}
$\cA_{1,1}$-, $\cA_{1,2}$-, and $\cA_{2,1}$-phases are SSB phase since their
condensable algebras $\cA_{1,1}$, $\cA_{1,2}$, and $\cA_{2,1}$ contain
$S_3$-charges.  $\cA_{2,2}$-phase is a $S_3$-symmetric phase since its
condensable algebra $\cA_{2,2}$ contains no $S_3$-charges.
The above four gapped phases can be divided into aSPT-classes (grouped by $\{
\cdots\}$), pSPT-classes (grouped by $[ \cdots]_\text{GSD}$), and SPT-classes
(grouped by $( \cdots)$), unsing the duality automorphism
$\two \leftrightarrow b)$:
\begin{align}
&\Big\{ \big[ ( \cA_{1,2})\big]^\text{SSB}_{2},
\big[ ( \cA_{1,1})\big]^\text{SSB}_{6}\Big\},
\Big\{ \big[ ( \cA_{2,2})\big]^\text{Sym}_{1},
\big[ ( \cA_{2,1})\big]^\text{SSB}_{3}\Big\} .
\end{align}

We see that the $S_3$-symmetry has only one gapped phase with
non-degenerate ground state.  Thus, the $\cVec_{S_3}$-symmetry has no
non-trivial SPT phase.  This is again a well known result: $S_3$-symmetry has
no non-trivial SPT phase since its second group cohomology class
$H^2(S_3, \mathrm{U}(1))=0$.  We remark that the sym/TO approach for SPT phases are more
general than the group cohomology approach, since the sym/TO approach also
apply to generalized symmetries.

$S_3$-symmetry has 4 classes of gapless phases, described by 4 non-Lagrangian
condensable algebras \eqref{S3nLcas}.  We can use the same formula
\eqref{GSDS3A11} to compute the GSD on a ring for those gapless phases:
\begin{align}
& \text{GSD}^{\cVec_{S_3}}_{\cA'_{1,1}} = 3, 
&
& \text{GSD}^{\cVec_{S_3}}_{\cA'_{1,2}} = 1,
\nonumber \\
& \text{GSD}^{\cVec_{S_3}}_{\cA'_{2,1}} = 2, 
&
& \text{GSD}^{\cVec_{S_3}}_{\cA'_{3,1}} = 1.
\end{align}
For a gapless state with a linear dispersion relation on a ring, GSD has the
following meaning.  The usual gapless excitations have an energy gap of order
$1/L$ above the ground state, where $L$ is the size of the ring.  Then the GSD
is the number of ground states, whose energy separations are of order $o(1/L)$.  
$\text{GSD}^{\cVec_{S_3}}_{\cA'_{1,1}} = 3$ indicates that the gapless phases
in the $\cA'_{1,1}$-class of gapless phases have a $(S_3 \to \Z_2)$-SSB.  So
the $\cA'_{1,1}$ gapless phases are gapless $\Z_2$-phases.  We note that
$\cA'_{1,1}$ is the intersection of $\cA_{1,1}$ and $\cA_{2,1}$.  Thus the
$\cA'_{1,1}$ gapless states describe the critical points at the transition
between two gapped phases: $\cA_{1,1}$-phase ($\Z_1$-phase) and
$\cA_{2,1}$-phase ($\Z_2$-phase) \cite{CW220506244}.  We see that the
$\cA'_{1,1}$ gapless states have a $\Z_2$ symmetry, agreeing with the GSD
analysis.

Similarly, $\text{GSD}^{\cVec_{S_3}}_{\cA'_{2,1}} = 2$ indicates that the
$\cA'_{2,1}$-class of gapless phases have a $(S_3 \to \Z_3)$-SSB.  So the
$\cA'_{2,1}$ gapless phases are gapless $\Z_3$-phases.  Now $\cA'_{2,1}$ is the
intersection of $\cA_{1,1}$ and $\cA_{1,2}$.  Thus the $\cA'_{2,1}$ gapless
states describe the critical points at the transition between two gapped
phases: $\cA_{1,1}$-phase ($\Z_1$-phase) and $\cA_{1,2}$-phase ($\Z_3$-phase).
We see that the $\cA'_{2,1}$ gapless states have a $\Z_3$ symmetry, agreeing
with the GSD analysis.

On the other hand, the $\cA'_{1,2}$ and $\cA'_{3,1}$ classes of gapless phases,
having GSD $=1$, are $S_3$ symmetric gapless phases.  The above results are
summarized below:
\begin{align}
\cA'_{3,1}\text{-phases}  \big |_{\text{GSD}=1}
&= S_3\text{-symmetric phases},
\nonumber \\
\cA'_{1,2}\text{-phases}  \big |_{\text{GSD}=1}
&= S_3\text{-symmetric phases},
\nonumber \\
\cA'_{2,1}\text{-phases}  \big |_{\text{GSD}=2}
&= \Z_3\text{-phases (SSB)},
\nonumber \\
\cA'_{1,1}\text{-phases}  \big |_{\text{GSD}=3}
&= \Z_2\text{-phases (SSB)}.
\end{align}

We like to point out that each condensation $\cA'_i$ may correspond to many
gapless phases.  Their symTO resolved multi-component partition functions are
given in Appendix \ref{S3Zs}.  The modular invariant partition function of
those gapless phases are give by \cite{CW221214432}
\begin{align}
\label{ZS3A11}
Z^{\cVec_{S_3}}_\text{inv} = (\cA_{1,1}, \vc Z^{\cVec_{S_3}}_{\cA_i}).
\end{align}

For example, one of the  $\cA'_{1,1}$ gapless phases described by
\eqref{Zm4S3A11} has the following modular invariant partition function
\begin{align}
Z^{\cVec_{S_3}}_{\text{inv},\cA'_{1,1}} &=
Z_{\onebb}^{\eD(S_3)} +
Z_{\one}^{\eD(S_3)} +
2Z_{\two}^{\eD(S_3)} 
\\
&=
\Big( {\color{blue} \chi^{m4 \times \overline{m4}}_{\onebb,0; \onebb,0}} 
+  {\color{red} \chi^{m4 \times \overline{m4}}_{\mathbf{b},\frac{1}{2}; \mathbf{b},-\frac{1}{2}} }
\Big)
+  \chi^{m4 \times \overline{m4}}_{\mathbf{a},\frac{1}{16}; \mathbf{a},-\frac{1}{16}} 
\nonumber \\ 
&\ \ \ \
+ 2\Big({\color{blue} \chi^{m4 \times \overline{m4}}_{\onebb,0; \onebb,0}} 
+  \chi^{m4 \times \overline{m4}}_{\mathbf{a},\frac{1}{16}; \mathbf{a},-\frac{1}{16}} 
+  \chi^{m4 \times \overline{m4}}_{\mathbf{b},\frac{1}{2}; \mathbf{b},-\frac{1}{2}} 
\Big)
\nonumber\\
& =
3\Big( {\color{blue} \chi^{m4 \times \overline{m4}}_{\onebb,0; \onebb,0}} 
+  {\color{red} \chi^{m4 \times \overline{m4}}_{\mathbf{b},\frac{1}{2}; \mathbf{b},-\frac{1}{2}} }
+  \chi^{m4 \times \overline{m4}}_{\mathbf{a},\frac{1}{16}; \mathbf{a},-\frac{1}{16}} 
\Big)
\nonumber 
\end{align}
The total partition function $Z^{\cVec_{S_3}}_{\text{inv},\cA'_{1,1}}$ encodes
the energy spectrum of the $\cA'_{1,1}$ gapless phase.  The term $3 \chi^{m4
\times \overline{m4}}_{\onebb,0; \onebb,0}$ in $ Z^{\cVec_{S_3}}_\text{inv}$
indicates that the GSD = 3, and the  $\cA'_{1,1}$ gapless phase has a $S_3\to
\Z_2$ SSB.  On the other hand, the total  modular invariant partition function
for  $\cA'_{1,2}$ gapless phase is given by 
\begin{align}
Z^{\cVec_{S_3}}_{\text{inv},\cA'_{1,2}} &=
Z_{\onebb}^{\eD(S_3)} +
Z_{\one}^{\eD(S_3)} +
2Z_{\two}^{\eD(S_3)} 
\\
&=
\Big( {\color{blue} \chi^{m4 \times \overline{m4}}_{\onebb,0; \onebb,0}} 
+  {\color{red} \chi^{m4 \times \overline{m4}}_{\mathbf{b},\frac{1}{2}; \mathbf{b},-\frac{1}{2}} }
\Big)
+  \chi^{m4 \times \overline{m4}}_{\mathbf{a},\frac{1}{16}; \mathbf{a},-\frac{1}{16}} 
\nonumber \\ 
& =
{\color{blue} \chi^{m4 \times \overline{m4}}_{\onebb,0; \onebb,0}} 
+  {\color{red} \chi^{m4 \times \overline{m4}}_{\mathbf{b},\frac{1}{2}; \mathbf{b},-\frac{1}{2}} }
+  \chi^{m4 \times \overline{m4}}_{\mathbf{a},\frac{1}{16}; \mathbf{a},-\frac{1}{16}} 
\nonumber 
\end{align}
The term $ \chi^{m4 \times \overline{m4}}_{\onebb,0; \onebb,0}$ in $
Z^{\cVec_{S_3}}_{\text{inv},\cA'_{1,2}}$ indicates that the GSD = 1, and the
$\cA'_{1,2}$ gapless phase is $S_3$ symmetric.

The above four gapless phases can be divided into aSPT-classes, 
pSPT-classes, and SPT-classes:
\begin{align}
&\Big\{ \big[ ( \cA'_{1,2})\big]^\text{Sym}_{1},
\big[ ( \cA'_{1,1})\big]^\text{SSB}_{3}\Big\},
\nonumber\\
&\Big\{ \big[ ( \cA'_{2,1})\big]^\text{SSB}_{2}\Big\},
\Big\{ \big[ ( \cA'_{3,1})\big]^\text{Sym}_{1}\Big\},
\end{align}
We note that there is a $\cA'_{1,2}$ gapless phase that is in the same
aSPT-class as the above $\cA'_{1,1}$ gapless phase.  As a result, their
partition function $Z_\onebb$ in the symmetric sub-Hilbert space are identical
(see \eqref{Zm4S3A11} and \eqref{Zm4S3A12}).  This is surprising since
$\cA'_{1,1}$ and $\cA'_{1,2}$ gapless phases even have different SSB patterns.

It is also interesting to note that we have two classes of gapless phases that
have the full $S_3$ symmetry.  However, these two classes have very different
condensations.  The $\cA'_{3,1}$ class has no non-trivial condensation since
$\cA'_{3,1}=\onebb$, while the $\cA'_{1,2}$ class has a condensation of
symmetry defect $b$ since $\cA'_{1,2}=\onebb \oplus b$.  
As a result, the two classes have different unbroken symTOs.  The unbroken
symTO for $\cA'_{3,1}$ gapless phases is $\eD(S_3)_{/\cA'_{3,1}}=\eD(S_3)$,
while the unbroken symTO for $\cA'_{1,2}$ gapless phases is
$\eD(S_3)_{/\cA'_{1,2}}=\eD(\Z_2)$ \cite{CW220506244}.  This is an example that
two different fully symmetric phases have different unbroken symTO.  We see
that the sym/TO correspondence provides a more refined description of symmetry
breaking patterns, than the usual symmetry approach where we can only say that
the $\cA'_{1,2}$-phases and $\cA'_{3,1}$-phases are fully symmetric.

\subsection{Phases in 1+1D systems with $\cRep_{S_3}$-symmetry}

Now let us discuss the non-invertible $\cRep_{S_3}$-symmetry in the
Morita-equivalence class of $\eD(S_3)$-symTO.  We choose the top gapped boundary
in Fig.\ \ref{CCmorph}  to be $\tl\cR_\text{def} = \cRep_{S_3}$, which is
induced by Lagrangian condensable algebra $\cA_{2,2}$ in \eqref{S3Lcas}.  Just
as its Morita-equivalent $\cVec_{S_3}$-symmetry, the $\cRep_{S_3}$-symmetry also
has four gapped phases described by the four Lagrangian condensable algebras in
\eqref{S3Lcas}.  The ground state degeneracies of those gapped phases are given
by $ \text{GSD}^{\cRep_{S_3}}_{\cA_i} = (\cA_{2,2}, \cA_i)$:
\begin{align}
\text{GSD}^{\cRep_{S_3}}_{\cA_{1,1}} &= 1,\ \ 
\text{GSD}^{\cRep_{S_3}}_{\cA_{1,2}} = 3,
\nonumber\\
\text{GSD}^{\cRep_{S_3}}_{\cA_{2,1}} &= 2,\ \ 
\text{GSD}^{\cRep_{S_3}}_{\cA_{2,2}} = 3.
\end{align}
Since $ \text{GSD}^{\cRep_{S_3}}_{\cA_{1,1}} = 1$, the $\cRep_{S_3}$-symmetry is
anomaly-free.  We also see that the $\cRep_{S_3}$-symmetry has no non-trivial
SPT phase. The four gapped phases of $\cRep_{S_3}$ symmetric systems are
given by the $\cA_i$ condensations (see \eqref{S3Lcas}) on the lower boundary in Fig.\ \ref{CCmorph}
\begin{align}
\cA_{1,1}\text{-phase} \big |_{\text{GSD}=1} &= \cRep_{S_3}\text{-symmetric-phase},
\nonumber\\
\cA_{1,2}\text{-phase} \big |_{\text{GSD}=3} &= \cRep_{S_3}\text{-SSB phase},
\nonumber\\
\cA_{2,1}\text{-phase} \big |_{\text{GSD}=2} &= \cRep_{S_3}\text{-SSB phase},
\nonumber\\
\cA_{2,1}\text{-phase} \big |_{\text{GSD}=3} &= \cRep_{S_3}\text{-SSB phase} .
\end{align}
The above four gapped phases can be divided into aSPT-classes, 
pSPT-classes, and SPT-classes:
\begin{align}
\Big\{ \big[ ( \cA_{1,1})\big]^\text{Sym}_{1},
\big[ ( \cA_{1,2})\big]^\text{SSB}_{3}\Big\},
\Big\{ \big[ ( \cA_{2,1})\big]^\text{SSB}_{2},
\big[ ( \cA_{2,2})\big]^\text{SSB}_{3}\Big\}.
\end{align}

Similarly, the four classes of gapless phases of $\cRep_{S_3}$ symmetric
systems are given by the $\cA'_i$ condensations (see \eqref{S3nLcas}) on the
lower boundary in Fig.\ \ref{CCmorph}
\begin{align}
\cA'_{3,1}\text{-phases} \big |_{\text{GSD}=1} 
&= \cRep_{S_3}\text{-symmetric phases},
\nonumber \\
\cA'_{1,1}\text{-phases} \big |_{\text{GSD}=1} 
&= \cRep_{S_3}\text{-symmetric phases},
\nonumber \\
\cA'_{2,1}\text{-phases} \big |_{\text{GSD}=1} 
&= \cRep_{S_3}\text{-symmetric phases},
\nonumber \\
\cA'_{1,2}\text{-phases} \big |_{\text{GSD}=2} 
&= \cRep_{S_3}\text{-SSB phases},
\end{align}
We see three different condensations all have GSD = 1, which give rise to
phases that does not breaking any part of $\cRep_{S_3}$-symmetry.  The
$\cA'_{1,2}$-phases is a $\cRep_{S_3}$-SSB phases since the condensable algebra
$\cA'_{1,2}$ contains a $\cRep_{S_3}$-charge $b$.

\section{$\eD(S_3) \times \eD(\Z_3)$-symTO}
\label{sec:S3xZ3 SymTO}

In the section, we like to study gapped phases for systems with $S_3\times
\Z_3$ symmetry, as well as $\cRep_{S_3\times \Z_3}$ non-invertible symmetry.
Both symmetries have the same symTO $\eD(S_3) \times \eD(\Z_3)$ (which is
$\eD(S_3) \times \eD(\Z_3)$ quantum double).  

The first part of the quantum double, $\eD(S_3)$, has been described in Section 
\ref{S3sym}. The $\eD(\Z_3)$ quantum double is described by \\[2mm]
\centerline{
\begin{tabular}{|c|c|c|c|c|c|c|c|c|c|}
\hline
$\eD(\Z_3)$ anyon& $\onebb$ & $e$ & $e^2$ & $m^2$ & $m$ & $em^2$ & $e^2m$ & $e^2m^2$ & $em$ \\ 
\hline
$s$: & $0$ & $0$ & $0$ & $0$ & $0$ & $\frac{1}{3}$ & $\frac{1}{3}$ & $\frac{2}{3}$ & $\frac{2}{3}$ \\ 
\hline
$d$: & $1$ & $1$ & $1$ & $1$ & $1$ & $1$ & $1$ & $1$ & $1$ \\ 
\hline
\end{tabular}
}

\

\noindent
The FSD automorphism group of the $\eD(S_3) \times \eD(\Z_3)$ quantum double is $\Z_2\times
\Z_2 \times \Z_2$, generated by the $e\leftrightarrow m$ exchange and the
$(e,m)\leftrightarrow (e^2,m^2)$ exchange in $\eD(\Z_3)$, as well as the
$\two\leftrightarrow b$ exchange in $\eD(S_3)$.
The quantum double has 10
Lagrangian condensable algebras, divided into three aSPT-classes:
\begin{widetext}
\begingroup
\allowdisplaybreaks
\begin{align}
\label{eq:LCA S3xZ3}
& \cA_{1,1} =  (\onebb,\onebb) \oplus (\one,\onebb) \oplus 2(\two,m^2) \oplus 2(\two,m) \oplus (\one,m) \oplus (\onebb,m) \oplus (\onebb,m^2) \oplus 2(\two,\onebb) \oplus (\one,m^2)
\nonumber \\
& \cA_{1,2} =  (\onebb,\onebb) \oplus (\one,\onebb) \oplus (\one,e) \oplus (\one,e^2) \oplus 2(b,e) \oplus (\onebb,e^2) \oplus (\onebb,e) \oplus 2(b,\onebb) \oplus 2(b,e^2)
\nonumber \\
& \cA_{1,3} =  (\onebb,\onebb) \oplus (\one,\onebb) \oplus (\one,e) \oplus (\one,e^2) \oplus (\onebb,e^2) \oplus (\onebb,e) \oplus 2(\two,e^2) \oplus 2(\two,e) \oplus 2(\two,\onebb)
\nonumber \\
& \cA_{1,4} =  (\onebb,\onebb) \oplus (\one,\onebb) \oplus (\one,m) \oplus 2(b,m^2) \oplus (\onebb,m) \oplus 2(b,\onebb) \oplus 2(b,m) \oplus (\onebb,m^2) \oplus (\one,m^2)
\nonumber \\
& \cA_{2,1} =  (\onebb,\onebb) \oplus (c,m^2) \oplus (c,\onebb) \oplus (c,m) \oplus (\two,m^2) \oplus (\two,m) \oplus (\onebb,m) \oplus (\onebb,m^2) \oplus (\two,\onebb)
\nonumber \\
& \cA_{2,2} =  (\onebb,\onebb) \oplus (c,m^2) \oplus (c,\onebb) \oplus (c,m) \oplus (b,m^2) \oplus (\onebb,m) \oplus (b,\onebb) \oplus (b,m) \oplus (\onebb,m^2)
\nonumber \\
& \cA_{2,3} =  (\onebb,\onebb) \oplus (c,e^2) \oplus (c,e) \oplus (c,\onebb) \oplus (b,e) \oplus (\onebb,e^2) \oplus (\onebb,e) \oplus (b,\onebb) \oplus (b,e^2)
\nonumber \\
& \cA_{2,4} =  (\onebb,\onebb) \oplus (c,e^2) \oplus (c,e) \oplus (c,\onebb) \oplus (\onebb,e^2) \oplus (\onebb,e) \oplus (\two,e^2) \oplus (\two,e) \oplus (\two,\onebb)
\nonumber \\
& \cA_{3,1} =  (\onebb,\onebb) \oplus (b'',e^2m) \oplus (b'',em^2) \oplus (b',em) \oplus (b',e^2m^2) \oplus (\one,\onebb) \oplus (\two,m^2) \oplus (\two,m) \oplus (b,e) \oplus (b,e^2)
\nonumber \\
& \cA_{3,2} =  (\onebb,\onebb) \oplus (b'',e^2m) \oplus (b'',em^2) \oplus (b',em) \oplus (b',e^2m^2) \oplus (\one,\onebb) \oplus (b,m^2) \oplus (b,m) \oplus (\two,e^2) \oplus (\two,e)
\end{align}
The quantum double also has 22 non-Lagrangian condensable algebras, divided into eleven aSPT-classes:
\begin{align}
& \cA'_{1,1} =  (\onebb,\onebb) \oplus (\two,m^2) \oplus (\two,m) \oplus (\onebb,m) \oplus (\onebb,m^2) \oplus (\two,\onebb)
\nonumber \\
& \cA'_{1,2} =  (\onebb,\onebb) \oplus (b,e) \oplus (\onebb,e^2) \oplus (\onebb,e) \oplus (b,\onebb) \oplus (b,e^2)
\nonumber \\
& \cA'_{1,3} =  (\onebb,\onebb) \oplus (b,m^2) \oplus (\onebb,m) \oplus (b,\onebb) \oplus (b,m) \oplus (\onebb,m^2)
\nonumber \\
& \cA'_{1,4} =  (\onebb,\onebb) \oplus (\onebb,e^2) \oplus (\onebb,e) \oplus (\two,e^2) \oplus (\two,e) \oplus (\two,\onebb)
\nonumber \\
& \cA'_{2,1} =  (\onebb,\onebb) \oplus (\one,\onebb) \oplus 2(b,\onebb)
\nonumber \\
& \cA'_{2,2} =  (\onebb,\onebb) \oplus (\one,\onebb) \oplus 2(\two,\onebb)
\nonumber \\
& \cA'_{3,1} =  (\onebb,\onebb) \oplus (\one,\onebb) \oplus (\one,e) \oplus (\one,e^2) \oplus (\onebb,e^2) \oplus (\onebb,e)
\nonumber \\
& \cA'_{3,2} =  (\onebb,\onebb) \oplus (\one,\onebb) \oplus (\one,m) \oplus (\onebb,m) \oplus (\onebb,m^2) \oplus (\one,m^2)
\nonumber \\
& \cA'_{4,1} =  (\onebb,\onebb) \oplus (\one,\onebb) \oplus (\two,m^2) \oplus (\two,m)
\nonumber \\
& \cA'_{4,2} =  (\onebb,\onebb) \oplus (\one,\onebb) \oplus (b,e) \oplus (b,e^2)
\nonumber \\
& \cA'_{4,3} =  (\onebb,\onebb) \oplus (\one,\onebb) \oplus (b,m^2) \oplus (b,m)
\nonumber \\
& \cA'_{4,4} =  (\onebb,\onebb) \oplus (\one,\onebb) \oplus (\two,e^2) \oplus (\two,e)
\nonumber \\
& \cA'_{5,1} =  (\onebb,\onebb) \oplus (c,\onebb) \oplus (b,\onebb)
\nonumber \\
& \cA'_{5,2} =  (\onebb,\onebb) \oplus (c,\onebb) \oplus (\two,\onebb)
\nonumber \\
& \cA'_{6,1} =  (\onebb,\onebb) \oplus (b',em) \oplus (b',e^2m^2) \oplus (\one,\onebb)
\nonumber \\
& \cA'_{7,1} =  (\onebb,\onebb) \oplus (b'',e^2m) \oplus (b'',em^2) \oplus (\one,\onebb)
\nonumber \\
& \cA'_{8,1} =  (\onebb,\onebb) \oplus (b,\onebb)
\nonumber \\
& \cA'_{8,2} =  (\onebb,\onebb) \oplus (\two,\onebb)
\nonumber \\
& \cA'_{9,1} =  (\onebb,\onebb) \oplus (\onebb,e^2) \oplus (\onebb,e)
\nonumber \\
& \cA'_{9,2} =  (\onebb,\onebb) \oplus (\onebb,m) \oplus (\onebb,m^2)
\nonumber \\
& \cA'_{10,1} =  (\onebb,\onebb) \oplus (\one,\onebb)
\nonumber \\
& \cA'_{11,1} =  (\onebb,\onebb)
\nonumber \\
\end{align}
\endgroup

\end{widetext}

\subsection{$S_3\times \Z_3$-symmetry }

The Morita-equivalence class of $\eD(S_3) \times \eD(\Z_3)$-symTO contains 3
symmetries.  One of them is the $S_3\times \Z_3$-symmetry induced by the
condensation of $\cA_{1,3}$.  

The FCD automorphism group for $S_3\times \Z_3$-symmetry is trivial.  
The FAD automorphism group is $\Z_2$ generated by  
$(e,m)\leftrightarrow (e^2,m^2)$ exchange.

The $S_3\times \Z_3$-symmetric systems has ten gapped phases, which
are divided into aSPT-classes, pSPT-classes, and SPT-classes:
\begin{align}
&\Big\{ \big[( \cA_{1,4})\big]^\text{SSB}_{2},
\big[( \cA_{1,1})\big]^\text{SSB}_{6},
\big[( \cA_{1,2})\big]^\text{SSB}_{6},
\big[( \cA_{1,3})\big]^\text{SSB}_{18}\Big\},
\nonumber\\
&\Big\{ \big[( \cA_{2,2})\big]^\text{Sym}_{1},
\big[( \cA_{2,1})\big]^\text{SSB}_{3},
\big[( \cA_{2,3})\big]^\text{SSB}_{3},
\big[( \cA_{2,4})\big]^\text{SSB}_{9}\Big\},
\nonumber\\
&\Big\{ \big[( \cA_{3,1})\big]^\text{SSB}_{2},
\big[( \cA_{3,2})\big]^\text{SSB}_{6}\Big\}.
\end{align}
We see that there is only one symmetric phase, and the rest are SSB phases.

$S_3\times \Z_3$ group has nine conjugacy classes of subgroups, labeled by
unbroken symmetry groups: $\Z_1,\ \Z_2,\ \Z^{(1)}_3,\ \Z^{(2)}_3,\ \Z^{(3)}_3,\
\Z_6,\ S_3,\ \Z_3\times \Z_3,\ S_3\times \Z_3 $.  To understand the 3 conjugacy
classes of subgroups $ \Z^{(1)}_3,\ \Z^{(2)}_3,\ \Z^{(3)}_3$, we may break
$S_3\times \Z_3 $ down to $\Z_3\times \Z_3$ first, where the first $\Z_3$ comes
from $S_3$.  Then there are four way to break $\Z_3\times \Z_3$  down to
$\Z_3$.  The unbroken $\Z_3$ is generated by $(1,0)$, $(0,1)$, $(1,1)$, or
$(-1,1)$ in $\Z_3\times \Z_3$.  But the  two $\Z_3$'s generated by $(1,1)$ and
$(-1,1)$ can be mapped into each other by the conjugation of $\Z_2\subset S_3$,
and belong to one conjugacy class.  So there are nine SSB patterns, which fail
to give rise to the ten gapped phases.  

To understand the extra gapped phases, we note that the $(S_3\times \Z_3  \to
\Z_3\times \Z_3 )$-SSB phase has unbroken $\Z_3\times \Z_3$-symmetry, which
give rise to three SPT-phases labeled by $\om_0$, $\om_1$, $\om_{-1}$, since
$H^2(\Z_3\times \Z_3,\mathrm{U}(1))=\Z_3= \{\om_0, \om_1, \om_{-1}\}$.  So one
may naively expect to have eleven gapped phases (with the two extra phases from
$\om_1, \om_{-1}$).

The reason that we have only ten gapped phases is the following: the $(S_3\times
\Z_3  \to \Z_3\times \Z_3 )$-SSB phase has two degenerate ground states.
Combining with the $\Z_3\times \Z_3$-SPT orders, we only get two gapped phases,
whose two degenerate ground states have the following SPT orders:
$(\om_0,\om_0)$, $(\om_1,\om_{-1})$.  Note that the two ground states are
related by a $\Z_2 \subset S_3$ symmetry transformation, which maps $\om_0\to
\om_0$ and $\om_1\leftrightarrow \om_{-1}$.  Therefore, the patterns like
$(\om_1,\om_0)$, $(\om_1,\om_1)$, \etc\ are not allowed.  To summarize, we
obtain the following ten gapped phases
\begingroup
\allowdisplaybreaks
\begin{align}
\big(  \cA_{2,2}\text{-phase} \big)_{1} 
&= S_3\times \Z_3\text{-symmetric phase}
\nonumber \\
\big(  \cA_{1,4}\text{-phase} \big)_{2} 
&= (S_3\times \Z_3 \to \Z_3\times \Z_3)\text{-SSB phase}
\nonumber \\
\big(  \cA_{3,1}\text{-phase} \big)_{2} 
&= (S_3\times \Z_3 \to \Z_3\times \Z_3)\text{-SSB phase}
\nonumber \\
\big(  \cA_{2,3}\text{-phase} \big)_{3} 
&= (S_3\times \Z_3 \to S_3)\text{-SSB phase}
\nonumber \\
\big(  \cA_{2,1}\text{-phase} \big)_{3} 
&= (S_3\times \Z_3 \to \Z_6)\text{-SSB phase}
\nonumber \\
\big(  \cA_{1,2}\text{-phase} \big)_{6} 
&= (S_3\times \Z_3 \to \Z^{(1)}_3)\text{-SSB phase}
\nonumber \\
\big(  \cA_{1,1}\text{-phase} \big)_{6} 
&= (S_3\times \Z_3 \to \Z^{(2)}_3)\text{-SSB phase}
\nonumber \\
\big(  \cA_{3,2}\text{-phase} \big)_{6} 
&= (S_3\times \Z_3 \to \Z^{(3)}_3)\text{-SSB phase}
\nonumber \\
\big(  \cA_{2,4}\text{-phase} \big)_{9} 
&= (S_3\times \Z_3 \to \Z_2)\text{-SSB phase}
\nonumber \\
\big(  \cA_{1,3}\text{-phase} \big)_{18} 
&= (S_3\times \Z_3 \to \Z_1)\text{-SSB phase}
\end{align}
\endgroup
All the phases belong to different SPT-classes, since the FCD automorphisms are trivial.  However, $\cA_{1,4}$-phase and $\cA_{3,1}$-phase
have the same $S_3\times \Z_3 \to \Z_3\times \Z_3$ SSB pattern.  But the 2
phases do not differ by an SPT-order of the unbroken $\Z_3\times
\Z_3$-symmetry, \ie the two phase are not connected by a $S_3\times
\Z_3$-FCD automorphism.  This is a new result: \emph{We find two 1+1D
gapped phases with the same SSB pattern, but the two phases do not differ by an
SPT-order of the unbroken symmetry.}

The $S_3\times \Z_3$ symmetry has 22 classes of gpaless phases, which
are divided into aSPT-classes, pSPT-classes, and SPT-classes:
\begin{align}
&\Big\{ \big[( \cA'_{1,3})\big]^\text{Sym}_{1},
\big[( \cA'_{1,1})\big]^\text{SSB}_{3},
\big[( \cA'_{1,2})\big]^\text{SSB}_{3},
\big[( \cA'_{1,4})\big]^\text{SSB}_{9}\Big\},
\nonumber\\
&\Big\{ \big[( \cA'_{2,1})\big]^\text{SSB}_{2},
\big[( \cA'_{2,2})\big]^\text{SSB}_{6}\Big\},
\ \ 
\Big\{ \big[( \cA'_{3,2})\big]^\text{SSB}_{2},
\big[( \cA'_{3,1})\big]^\text{SSB}_{6}\Big\},
\nonumber\\
&\Big\{ \big[( \cA'_{4,1})\big]^\text{SSB}_{2},
\big[( \cA'_{4,2})\big]^\text{SSB}_{2},
\big[( \cA'_{4,3})\big]^\text{SSB}_{2},
\big[( \cA'_{4,4})\big]^\text{SSB}_{6}\Big\},
\nonumber\\
&\Big\{ \big[( \cA'_{5,1})\big]^\text{Sym}_{1},
\big[( \cA'_{5,2})\big]^\text{SSB}_{3}\Big\},
\ \
\Big\{ \big[( \cA'_{6,1})\big]^\text{SSB}_{2}\Big\},
\nonumber\\
&\Big\{ \big[( \cA'_{7,1})\big]^\text{SSB}_{2}\Big\},
\ \
\Big\{ \big[( \cA'_{8,1})\big]^\text{Sym}_{1},
\big[( \cA'_{8,2})\big]^\text{SSB}_{3}\Big\},
\nonumber\\
&\Big\{ \big[( \cA'_{9,2})\big]^\text{Sym}_{1},
\big[( \cA'_{9,1})\big]^\text{SSB}_{3}\Big\},
\nonumber\\
&\Big\{ \big[( \cA'_{10,1})\big]^\text{SSB}_{2}\Big\},
\ \
\Big\{ \big[( \cA'_{11,1})\big]^\text{Sym}_{1}\Big\}.
\end{align}
We see that there are five classes of $S_3\times \Z_3$ symmetric gapless phases
with GSD = 1.  But the differences between those phases are beyond $S_3\times
\Z_3$ SPT orders, since those phases are not connected via
FCD automorphisms.

\subsection{$\cRep_{S_3\times \Z_3}$-symmetry -- symmetric gapped phase with beyond-SPT order}

The second symmetry in the Morita-equivalence class of $\eD(S_3) \times
\eD(\Z_3)$-symTO is a non-invertible symmetry described by fusion category
$\cRep_{S_3\times \Z_3}$.  We call such a symmetry as $\cRep_{S_3\times
\Z_3}$-symmetry.  The fusion category $\cRep_{S_3\times \Z_3}$ describes
excitations on the boundary induced by $\cA_{2,2}$ condensation.  Thus the
$\cRep_{S_3\times \Z_3}$-symmetry comes from $\cA_{2,2}$ condensation. 

The FCD automorphism group for $\cRep_{S_3\times \Z_3}$-symmetry is trivial.
The FAD automorphism group is $\Z_2$ generated by  $(e,m)\leftrightarrow
(e^2,m^2)$ exchange.

The $\cRep_{S_3\times \Z_3}$-symmetric systems has ten gapped phases, which
are divided into aSPT-classes, pSPT-classes, and SPT-classes:
\begin{align}
&\Big\{ \big[( \cA_{1,3})\big]^\text{Sym}_{1},
\big[( \cA_{1,1})\big]^\text{SSB}_{3},
\big[( \cA_{1,2})\big]^\text{SSB}_{3},
\big[( \cA_{1,4})\big]^\text{SSB}_{9}\Big\},
\nonumber\\
&\Big\{ \big[( \cA_{2,4})\big]^\text{SSB}_{2},
\big[( \cA_{2,3})\big]^\text{SSB}_{3},
\big[( \cA_{2,1})\big]^\text{SSB}_{6},
\big[( \cA_{2,2})\big]^\text{SSB}_{9}\Big\},
\nonumber\\
&\Big\{ \big[( \cA_{3,1})\big]^\text{Sym}_{1},
\big[( \cA_{3,2})\big]^\text{SSB}_{3}\Big\}.
\end{align}
Since there are gapped phases with non degenerate ground state, the
$\cRep_{S_3\times \Z_3}$-symmetry is anomaly-free.  This is expected since both
$\cRep_{S_3}$- and $\cRep_{\Z_3}$-symmetries are anomaly-free.

We see that there are two symmetric gapped phases with non-degenerate ground
state.  One may think those two phases differ by an SPT order.  But the two
phases are not connected by FCD automorphisms.  So their difference is beyond
an SPT order.  We find a surprising result: \frmbox{two symmetric gapped phases
with trivial topological order may not differ by an SPT order.} One of the
symmetric phase, the $\cA_{1,3}$-phase, is the trivial symmetric product state.
The other symmetric phase, the $\cA_{3,1}$-phase, is a new type of symmetric
state, which is not described by a SPT order.  

In fact, the two phases are not even connected by FSD automorphisms. We
expect that the two symmetric phases, $\cA_{1,3}$-phase and
$\cA_{3,1}$-phase, have different bulk properties (such as different bulk
spectra even in symmetric sub-Hilbert space), and can be distinguished by bulk
measurements.  

The $\cRep_{S_3\times \Z_3}$-symmetry has 22 classes of gpaless phases, which
are divided into aSPT-classes, pSPT-classes, and SPT-classes:
\begin{align}
&\Big\{ \big[( \cA'_{1,4})\big]^\text{Sym}_{1},
\big[( \cA'_{1,2})\big]^\text{SSB}_{2},
\big[( \cA'_{1,1})\big]^\text{SSB}_{3},
\big[( \cA'_{1,3})\big]^\text{SSB}_{6}\Big\},
\nonumber\\
&\Big\{ \big[( \cA'_{2,2})\big]^\text{Sym}_{1},
\big[( \cA'_{2,1})\big]^\text{SSB}_{3}\Big\},
\ \
\Big\{ \big[( \cA'_{3,1})\big]^\text{Sym}_{1},
\big[( \cA'_{3,2})\big]^\text{SSB}_{3}\Big\},
\nonumber\\
&\Big\{ \big[( \cA'_{4,1})\big]^\text{Sym}_{1},
\big[( \cA'_{4,2})\big]^\text{Sym}_{1},
\big[( \cA'_{4,4})\big]^\text{Sym}_{1},
\big[( \cA'_{4,3})\big]^\text{SSB}_{3}\Big\},
\nonumber\\
&\Big\{ \big[( \cA'_{5,2})\big]^\text{SSB}_{2},
\big[( \cA'_{5,1})\big]^\text{SSB}_{3}\Big\},
\ \
\Big\{ \big[( \cA'_{6,1})\big]^\text{Sym}_{1}\Big\},
\nonumber\\
&\Big\{ \big[( \cA'_{7,1})\big]^\text{Sym}_{1}\Big\}, 
\ \
\Big\{ \big[( \cA'_{8,2})\big]^\text{Sym}_{1},
\big[( \cA'_{8,1})\big]^\text{SSB}_{2}\Big\},
\nonumber\\
&\Big\{ \big[( \cA'_{9,1})\big]^\text{Sym}_{1},
\big[( \cA'_{9,2})\big]^\text{SSB}_{3}\Big\},
\nonumber\\
&\Big\{ \big[( \cA'_{10,1})\big]^\text{Sym}_{1}\Big\}, 
\ \
\Big\{ \big[( \cA'_{11,1})\big]^\text{Sym}_{1}\Big\}
\end{align}
We see that $\cRep_{S_3\times \Z_3}$-symmetry has no non-trivial gapless SPT
phases.

\subsection{Anomaly-free $\tl\cR^{\eD(S_3\times \Z_3)}_{\cA_{3,2}}$-symmetry}

The Morita-equivalence class of symTO $\eD(S_3\times \Z_3)$ contains three
symmetries. The first two are $S_3\times \Z_3$ and $\cRep_{S_3\times \Z_3}$
symmetries.  The third one is induced by $\cA_{3,2}$-condensation, and is
called $\tl\cR^{\eD(S_3\times \Z_3)}_{\cA_{3,2}}$-symmetry.  The gapped and
gapless phases, as well as as their SPT-classes, pSPT-classes, aSPT-classes,
are summarized in Table \ref{Phases} and \ref{gaplessPhases}.

\section{Lattice models for $\cRep_{S_3\times \Z_3}$ SPTs}

Building on the previous section, we construct exactly solvable lattice models
for the two SPT phases protected by $\cRep_{S_3\times \Z_3}$ symmetry that are
not connected by any automorphisms. By using a gauging-induced duality,
we show that the absence of automorphisms is related to the fact that the
charged excitations realize distinct fusion categories in the two SPT ground states. 

\subsection{Representations of group  $S_3\times \Z_3$}

We start with briefly reviewing the representation theory of the group $S_3\times \Z_3$. 
The non-Abelian group  $S_3$ has cardinality 6 and two generators which we denote
by $r$ and $s$ such that
\begin{align}
r^3 = s^2 = e,  \quad s\, r = r^2\,s,
\end{align}
where $e$ denotes the identity element.
The Abelian group $\Z_3$ has cardinality 3 and the single generator which we 
denote by $g$, \ie $g^3=e$. Hence, the elements of the group $S_3\times \Z_3$
are labeled by the pair $(s^a\,r^b, g^c)$ where $a=1,2$ and $b,c=1,2,3$.

The groups $S_3$ and $\Z_3$ both have 3 irreps, which we denote as
$\one, \one', \two$ and $\one, \bm{\omega}, \bm{\omega}^*$, respectively, with $\one$
being the trivial irrep.
These irreps satisfy the fusion rules
\begin{align}
\begin{split}
&
\one' \otimes \one' = \one, 
\quad 
\one' \otimes \two = \two,
\quad
\two \otimes \two = \one \oplus \one' \oplus \two,
\\
&
\bm{\omega} \otimes \bm{\omega} = \bm{\omega}^*,
\quad
\bm{\omega} ^*\otimes \bm{\omega}^* = \bm{\omega},
\quad
\bm{\omega} \otimes \bm{\omega}^* = \one.
\end{split}
\end{align}
Hence, the direct product $S_3\times \Z_3$ has 9 irreps that are labeled by 
$(\rho_{S_3}, \rho_{\Z_3})$ where $\rho_{S_3}$ and $\rho_{\Z_3}$ are irreps of 
$S_3$ and $\Z_3$, respectively. The fusion rules of these irreps are then deduced from the 
fusion rules of the irreps of $S_3$ and~$\Z_3$.

\subsection{SPT phases protected by $\cRep(S_{3})\times \Z_{3}$ symmetry }

We focus on two Lagrangian algebras of 
the symTO $\cD(S_3\times\Z_3)$, namely, $\cA_{1,3}$ and $\cA_{3,1}$
defined in Eq.\ \eqref{eq:LCA S3xZ3}. 
When symmetry boundary is chosen to be $\cA_{2,2}$, the symTO 
describes 1+1D systems with $\cRep(S_{3})\times \Z_{3}$ symmetry.
If so these two Lagrangian algebras describes the two SPT phases protected by 
non-invertible $\cRep(S_{3})\times \Z_{3}$ symmetry.

We claim that the two Hamiltonians~\footnote{Here we use the short-hand notation $A^{-\sigma^x_i}_i$ for
$A^{\sigma^{z}_{j}} = \frac{1+\sigma^{z}_{j}}{2}A +  
\frac{1-\sigma^{z}_{j}}{2}A^\dag$.}
\begin{subequations}
\label{eq:RepS3xZ3 SPTs}
\begin{align}
\label{eq:RepS3xZ3 SPTs triv}
&
H_{1,3}\,
=
-
\sum_{i=1}^L
\left\{
X_{i}
+
\tl X_{i}
+
\mathrm{H.c.}
\right\}
-
\sum_{i=1}^L
\sigma^{x}_{i},
\\
\label{eq:RepS3xZ3 SPTs nontriv}
&
H_{3,1}\,
=
-
\sum_{i=1}^L
\left\{
\tl Z_i\,
X^{\dag}_i\,
\tl Z^{\dag}_{i+1}\,
+
Z_{i-1}\,
\tl X_{i}\,
Z^{-\sigma^{x}_{i}}_{i}\,
+
\mathrm{H.c.}
\right\}
\nonumber
\\
&\qquad\quad\,\,\,
-
\sum_{i=1}^L
\sigma^{x}_{i},
\end{align}
\end{subequations}
realize precisely these two SPT phases protected by $\cRep_{S_3\times \Z_3}$ symmetry.
Here,   $\{X_i,\,Z_i\}$, $\{\tl X_i,\,\tl Z_i\}$, and
$\{\sigma^{x}_i,\,\sigma^{z}_i\}$ are two flavors  of qutrit 
and a single flavor of qubit operators acting on a 18 dimensional local Hilbert space
and satisfy the algebra  
\begin{align}
\begin{split}
&
Z_{i}\,X_{j}
=
e^{\mathrm{i}\frac{2\pi}{3}\delta_{i,j}}
X_{j}\,Z_{i},
\quad
Z^{3}_{i}
=
X^{3}_{i}
=
1,
\\
&
\tl Z_{i}\,\tl X_{j}
=
e^{\mathrm{i}\frac{2\pi}{3}\delta_{i,j}}
\tl X_{j}\,\tl Z_{j},
\quad
\tl Z^{3}_{i}
=
\tl X^{3}_{i}
=
1,
\\
&
\sigma^{z}_{i}\,
\sigma^{x}_{j}
=
(-1)^{\delta_{i,j}}\,
\sigma^{x}_{j}\,
\sigma^{z}_{i},
\quad
\left(\sigma^{z}_{i}\right)^{2}
=
\left(\sigma^{x}_{i}\right)^{2}
=
1.
\end{split}
\end{align}

We choose a basis $\ket{\bm{a},\bm{b},\bm{c}}$ where
$\bm{a},\,\bm{b}\in \Z^L_3$ and $\bm{c}\in \Z^L_2$ 
such that
\begin{align}
\begin{split}
&
Z_{j}\ket{\bm{a},\bm{b},\bm{c}}
=
e^{\mathrm{i}\frac{2\pi}{3} a_{i}}\,
\ket{\bm{a},\bm{b},\bm{c}},
\quad
X_{j}\ket{\bm{a},\bm{b},\bm{c}}
=
\ket{\bm{a}+\bm{\delta}_i,\,\bm{b},\bm{c}},
\\
&
\tl Z_{j}\ket{\bm{a},\bm{b},\bm{c}}
=
e^{\mathrm{i}\frac{2\pi}{3} b_{i}}\,
\ket{\bm{a},\bm{b},\bm{c}},
\quad
\tl X_{i}\ket{\bm{a},\bm{b},\bm{c}}
=
\ket{\bm{a},\bm{b}+\bm{\delta}_i\,\bm{c}},
\\
&
\sigma^{z}_{i}\ket{\bm{a},\bm{b},\bm{c}}
=
(-1)^{c_{i}}
\ket{\bm{a},\bm{b},\bm{c}},
\quad
\sigma^{x}_{i}\ket{\bm{a},\bm{b},\bm{c}}
=
\ket{\bm{a},\bm{b},\bm{c}+\bm{\delta}_i},
\end{split}
\end{align}
where $\bm{\delta}_i$ is a vector whose only nonzero entry is 
$1$ at site~$i$. 

Both Hamiltonians
\eqref{eq:RepS3xZ3 SPTs triv}  and  \eqref{eq:RepS3xZ3 SPTs nontriv}
commute' with the the $\cRep(S_{3})\times\Z_3$ symmetry generated by
\begin{align}
\label{eq:RepS3xZ3 generators}
\begin{split}
&
W_{\one'}
=
\prod_j \sigma^{x}_{j},
\\
&
W_{\two}
=\frac{1+W_{\one'}}{2}
\left(
\prod_j
X^{\prod_{k=2}^{j}\sigma^x_k}_j
+
\mathrm{H.c.}
\right),
\\
&
W_g
=
\prod_j \tl X_{j}.
\end{split}
\end{align}
Here, $W_{g}$ generates the $\Z_{3}$ symmetry,
while $W_{\one'}$ and $W_{\two}$ operators
generate $\cRep(S_{3})$ symmetry.
The latter is confirmed by noting the these operators
satisfy $\cRep(S_{3})$ the fusion rules
\begin{align}
W_{\one'}\,
W_{\one'}
=
W_{\one},
\quad
W_{\two}\,W_{\two}
=
W_{\one}
+
W_{\one'}
+
W_{\two},
\end{align}
with $W_{\one}=\onebb$ being the identity operator,
which is the fusion rules of irreps of $S_{3}$.

Both Hamiltonians 
\eqref{eq:RepS3xZ3 SPTs triv}  and  \eqref{eq:RepS3xZ3 SPTs nontriv}
have nondegenerate and gapped ground states. 
More precisely,
Hamiltonian \eqref{eq:RepS3xZ3 SPTs triv} consists of pairwise commuting terms
and has the nondegenerate and gapped ground state 
\begin{align}
\label{eq:RepS3xZ3 SPT triv}
\ket{\psi_{1,3}}
=
\sum_{\left\{\bm{a},\,\bm{b},\,\bm{c}\right\}}
\ket{\bm{a},\,\bm{b},\,\bm{c}}.
\end{align}
which is fully disordered and a trivial SPT.
In contrast, Hamiltonian \eqref{eq:RepS3xZ3 SPTs nontriv}
does not consists of pairwise commuting terms but its ground state can be
obtained exactly since $\sigma^x_i$ is a conserved quantity
and in the $\sigma^x_i=1$ subspace all the terms in the Hamiltonian commute one another. 
We find its non-degenerate and gapped ground state to be
\begin{align}
\label{eq:RepS3xZ3 SPT nontriv}
\ket{\psi_{3,1}}
=
\sum_{\left\{\bm{a},\,\bm{b},\,\bm{c}\right\}}
e^{+\mathrm{i}\frac{2\pi}{3}\sum_j a_{j}(b_{j+1}-b_{j})}
\ket{\bm{a},\,\bm{b},\,\bm{c}},
\end{align}
which is nothing but $\Z_{3}$ cluster state~\cite{GM14101580, S150200066}. While these two SPT states cannot be 
deformed to one another via a locally $\Z_3\times\Z_3$
FDQC, it is not a priori obvious that this is also so 
for the $\cRep_{S_3\times \Z_3}$ symmetry generated by 
operators \eqref{eq:RepS3xZ3 generators}.
In the next section, we argue that these two ground states are 
distinct SPT states by showing that after gauging the entire 
$\cRep_{S_3\times \Z_3}$ symmetry, they map to 
distinct SSB patterns of the dual $S_3\times\Z_3$ symmetry.
More precisely, we are going to show that 
while the fusion category of charged excitations in the SPT state  
\eqref{eq:RepS3xZ3 SPT triv} is $\cVec_{S_3\times \Z_3}$, 
this not so for the nontrivial SPT state \eqref{eq:RepS3xZ3 SPT nontriv}. 

\subsection{Two SSB patterns with $S_3\times \Z_3$ symmetry}

We seek to gauge the global $\cRep_{S_3\times \Z_3}$
symmetry of Hamiltonians \eqref{eq:RepS3xZ3 SPTs}.
This can be achieved in two ways: (i) either in two steps
by gauging only invertilbe group-like, 
first $\Z_2$ subsymmetry and then the remaining $\Z_3\times\Z_3$ subsymmetry~\cite{CW240505331}, (ii)
or gauging the entire non-invertible 
$\cRep_{S_3\times \Z_3}$ symmetry in one go using 
matrix product operator representation~\cite{PA240918113}.
Also see \Rf{SY250302925} for a more general 
prescription of gauging non-invertible symmetries on the 
lattice.

Following \Rf{CW240505331},
we use the gauging map obtained therein from 
the  $\cRep_{S_3\times \Z_3}$-symmetric local operators
and the $S_3\times\Z_3$-symmetric ones~\footnote{With an 18-dimensional local
Hilbert space, there are exactly 6 local generators of algebra of $\cRep_{S_3}$-symmetric
operators and 2 local generators of algebra of $\Z_3$-symmetric operator~\cite{CW240505331}. 
Here we choose the left-hand and right-hand sides of the mapping to be symmetric under 
symmetries, respectively.}
\begin{align}
\begin{split}
X_i + X^\dag_i 
&\mapsto 
Z_i\,Z^\dag_{i+1} + Z^\dag_i\,Z_{i+1},
\\
X_i - X^\dag_i 
&\mapsto 
\sigma^z_{i+1}(Z_i\,Z^\dag_{i+1} - Z^\dag_i\,Z_{i+1}),
\\
Z^{\sigma^x_{i+1}}_i\,Z^\dag_{i+1} + Z^{-\sigma^x_{i+1}}_i\,Z_{i+1}
&\mapsto
X_{i+1} + X^\dag_{i +1},
\\
Z^{\sigma^x_{i+1}}_i\,Z^\dag_{i+1} - Z^{-\sigma^x_{i+1}}_i\,Z_{i+1}
&\mapsto
\sigma^z_{i}(X_{i+1} - X^\dag_{i +1}),
\\
\sigma^x_i 
&\mapsto \sigma^z_{i-1}\,\sigma^z_{i},
\\
\sigma^z_{i}\,\sigma^z_{i+1}
&\mapsto
\sigma^x_{i},
\\
\tl X_{i}
& \mapsto
\tl Z_i\, \tl Z^\dag_{i+1},
\\
\tl Z_i\, \tl Z^\dag_{i+1} 
&\mapsto
\tl X^\dag_{i+1}.
\end{split}
\end{align}
Under this mapping, the two Hamiltonians \eqref{eq:RepS3xZ3 SPTs}
are mapped to
\begin{subequations}
\label{eq:S3xZ3 fixed point Hams}
\begin{align}
\label{eq:S3xZ3 Ham 1}
&
H^\vee_{1,3}
=
-
\sum_{i=1}^L
\left\{
Z_{i}\,
Z^{\dag}_{i+1}
+
\tl Z_{i}\,
\tl Z^{\dag}_{i+1}
+
\mathrm{H.c.}
\right\}
\nonumber\\
&\qquad\quad\,\,
-
\sum_{i=1}^L
\sigma^{z}_{i}\,
\sigma^{z}_{i+1},
\\
\label{eq:S3xZ3 Ham 2}
&
H_{3,1}
=
-
\sum_{i=1}^L
\left\{
\tl Z_{i}
X^{\sigma^{z}_{i}}_{i}
\tl Z^{\dag}_{i+1}
+
Z^{\dag}_{i-1}
\tl X^{\sigma^{z}_{i}}_{i}
Z_{i}
+
\mathrm{H.c.}
\right\}
\nonumber\\
&\qquad\quad\,\,
-
\sum_{i=1}^L
\sigma^{z}_{i}\,
\sigma^{z}_{i+1}.
\end{align}
\end{subequations}
Both Hamiltonians 
\eqref{eq:S3xZ3 Ham 1} and \eqref{eq:S3xZ3 Ham 2}
are symmetric under the $S_{3}\times \Z_{3}$ symmetry 
generated by the unitary operators
\begin{align}
U_{r}
=
\prod_j
X_{j},
\quad
U_{s}
=
\prod_j
C_{j}\,
\sigma^{x}_ {j},
\quad
U_{g}
=
\prod_j
\tl X_{j},
\end{align} 
where the operator $C_{j}$
implements charge conjugation on qutrits operators $Z_{j}$
and $X_{j}$, \ie ${C: (Z, X)\mapsto
(Z^{\dag}, X^{\dag})}$.
The unitary operators $U_{r}$ and $U_{s}$
generate the $S_{3}$ subgroup while $U_{g}$
generates the $\Z_{3}$ subgroup.
Since Hamiltonian \eqref{eq:S3xZ3 Ham 1} consists
of pairwise commuting operators, its
eigenstates can be obtained exactly.
It has 18 degenerate ground states
\begin{align}
\label{eq:GS S3xZ3 SSB}
\ket{\psi_{1,3;\,\al,\bt,\ga}}
=
\bigotimes_j
\ket{a_{j}=\al,\, b_{j}=\bt,\, c_{j}=\ga},
\end{align}
labeled by $\al,\bt=0,1,2$ and $\ga=0,1$.
These ground states correspond to a ferromagnetically ordering 
qutrits and qubits for which $S_{3}\times\Z_{3}$
symmetry is fully broken spontaneously.

Hamiltonian \eqref{eq:S3xZ3 Ham 2} does not consists of pairwise commuting terms.
However, observing that $\sigma^{z}_{j}\,\sigma^{z}_{j+1}$
terms commute with rest of the Hamiltonian, one can first minimize them, 
which ferromagnetically orders the qubits, \ie $\sigma^{z}_{j}$ is 
pinned to $\sigma^{z}_{j}=\pm1$. 
One then finds that Hamiltonian \eqref{eq:S3xZ3 Ham 2} 
has 2 degenerate ground states 
\begin{align}
\label{eq:GS nontriv}
\begin{split}
&
\ket{\psi_{3,1;\,+}}
=
\sum_{\left\{\bm{a},\,\bm{b}\right\}}
e^{+\mathrm{i}\frac{2\pi}{3}\sum_j a_{j}(b_j-b_{j+1})}
\bigotimes_j
\ket{a_j,\,b_{j},\,c_{j}=0},
\\
&
\ket{\psi_{3,1;\,-}}
=
\sum_{\left\{\bm{a},\,\bm{b}\right\}}
e^{-\mathrm{i}\frac{2\pi}{3}\sum_j a_{j}(b_j-b_{j+1})}
\bigotimes_j
\ket{a_j,\,b_{j},\,c_{j}=1}.
\end{split}
\end{align}
While just like before  the qubit degrees of freedom are 
ferromagnetically ordered, the qutrits degrees of freedom are
in the nontrivial cluster SPT state protected by the unbroken 
$\Z_{3}\times \Z_{3}$ symmetry.  In contrast, Hamiltonian \eqref{eq:S3xZ3 Ham 1}
has 18-fold degenerate ground state manifold independent of the boundary conditions.

As discussed in Sec.\ \ref{sec:S3xZ3 SymTO},
while there are three SPT states protected by 
$\Z_{3}\times\Z_{3}$ symmetry~\footnote{This follows from the second cohomology 
group $H^{2}\left(\Z_{3}\times\Z_{3},\mathrm{U}(1)\right)=\Z_{3}$.},
there are only two possible SSB patterns where $S_3\times\Z_3$ is broken down to $\Z_3\times\Z_3$. 
In other words, on the two degenerate ground states 
that break the $\Z_{2}$ subgroup, it possible to paste only one nontrivial SPT. This is because
the nontrivial action of the $\Z_{2}$ generator $U_{s}$ on $U_{r}$
implies that if one degenerate ground state describe a $\Z_{3}$ cluster state the other one must 
describe its inverse SPT state under stacking. 
This means that there are only two possible stacking with SPT: (i) one where both
ground states describe trivial SPT, (ii) and one where the two ground states describe two nontrivial SPT phases that 
are inverse of each other under stacking. Indeed, the former is the case of states  \eqref{eq:GS S3xZ3 SSB}
while the latter is the case for \eqref{eq:GS nontriv}.

As claimed, the two Hamiltonians \eqref{eq:RepS3xZ3 SPTs} with $\cRep_{S_3\times \Z_3}$ symmetry
are dual to the Hamiltonians \eqref{eq:S3xZ3 fixed point Hams} that describe two distinct SSB patterns 
of $S_3\times\Z_3$ symmetry. Therefore, 
the corresponding SPT states are distinct from one another. 
Note that this duality is not captured by an automorphism of the $\cD(S_3)\times\cD(\Z_3)$-symTO. 
This is evident as the two pairs of dual Hamiltonians \eqref{eq:RepS3xZ3 SPTs} and \eqref{eq:S3xZ3 fixed point Hams}
have different symmetries while the FSD automorphisms do not change the global symmetry.

With this dual picture in mind, the fact that there are no FSD
automorphisms between the two SPT states \eqref{eq:RepS3xZ3 SPT triv} and
\eqref{eq:RepS3xZ3 SPT nontriv} can be understood as follows.  Conventionally,
two SPT states with the same protecting symmetries are expected to be
indistingushable from the bulk properties. For  group-like symmetries this is
true as SPT phases are connected by FDQCs, or equivalently FCD automorphisms
from bulk symTO point of view. However, the absence of any FSD
automorphisms between these two states suggests that the two SPT phases can be
distinguished by certain bulk properties. In particular, the properties of
symmetry charges can differ between SPT phases with non-invertible symmetries.
This kind of distinction in SPT phases  with non-invertible symmetries was
suggested in the talk \cite{W2024Talk}. In \Rf{MG241220546} 
it was shown that the charged excitations of non-invertible symmetries can
differ on distinct SPT phases. 

For the particular example of $\cRep_{S_3\times \Z_3}$ symmetry and the two SPT states \eqref{eq:RepS3xZ3 SPT triv}
and \eqref{eq:RepS3xZ3 SPT nontriv}, we use the dual picture where they correspond to distinct SSB patterns described by
states \eqref{eq:GS S3xZ3 SSB} and \eqref{eq:GS nontriv}, respectively. 
The duality map between $\cRep_{S_3\times \Z_3}$ and 
$S_3\times\Z_3$ symmetries dictate that insertion of a point-like symmetry defects of $\cRep_{S_3\times \Z_3}$ corresponds to choosing a different symmetry charge sector of dual $S_3\times\Z_3$ symmetry.
We note that the 18-fold degenerate ground states \eqref{eq:GS S3xZ3 SSB} where $S_3\times\Z_3$ symmetry is completely 
broken form the regular representation $(\one \oplus \one' \oplus \two )\otimes (\one \oplus \bm{\omega} \oplus \bm{\omega}^*)$. 
Hence, these states form the fusion category of irreps of $S_3\times\Z_3$, \ie $\cRep_{S_3\times \Z_3}$, which is precisely the 
protecting non-invertible symmetry. In turn,
this implies that the fusion category of charged excitations in the SPT phase \eqref{eq:RepS3xZ3 SPT triv}
is given by $\cVec_{S_3\times \Z_3}$.

In constrast, the 2-fold degenerate ground states \eqref{eq:GS nontriv} form the two-dimensional
representation $\one\oplus \one'$ 
which implies that the fusion category of charged excitations  in the SPT state \eqref{eq:RepS3xZ3 SPT nontriv}
are distinct from $\cVec_{S_3\times \Z_3}$. 
In fact, it has been shown that in the SPT state \eqref{eq:RepS3xZ3 SPT nontriv},
that the charges  form the $\Z_3\times\Z_3$
Tambara-Yamagami fusion category  $TY(\Z_3\times\Z_3)$~\cite{MG241220546}.
Thus, there cannot be an FSD automorphisms between the two SPT phases as these
preserve the fusion structure  (in the symmetric subspace).

\section{Conclusion \& Outlook}

In this work, we studied gapped and gapless phases for systems with a symmetry
(which can be non-invertible generalized symmetry).  We demonstrated that 
the utility of symTO in one-higher dimension to describe generalized symmetry.
In this holographic description, the gapped and gapless phases will be gapped
and gapless boundaries of the corresponding symTO.  This allows us to classify gapped phases
using the Lagrangian condensable algebras of the symTO.  We can also obtain a
useful characterization of gapless phases using the non-Lagrangian condensable
algebras of the symTO.  In particular, we obtained a comprehensive description of
SSB phases for non-invertible symmetries, despite such symmetries not having
the notion of subgroups.

Using the automorphisms of symTO and their action with the
condensable algebras, we showed that one further organize the gapped/gapless phases into
equivalence classes: the SPT-class, pSPT-class, and aSPT-class, which are defined by equivalence under
FCD, FAD, and FSD automorphisms. The states in an
equivalence class are related by symmetry-preserving duality transformations.
For the states in an  aSPT-class, the symmetry-preserving duality transformations
can permute symmetry charges and symmetry defects.  For the states in an
pSPT-class, the symmetry-preserving duality transformations may only permute
symmetry charges.  For the states in a SPT-class, the symmetry-preserving
duality transformations do not act on the symmetry charges.

The phases in an SPT-class have the same SSB patterns, and differ only by SPT
orders.  This gives us a way to compute SPT orders for non-invertible
symmetries, using sym/TO correspondence.  Such a notion of  SPT orders 
naturally extends to the SSB phases and gapless phases.
Similarly, we may define pSPT orders as the difference between the phases 
related by FAD autormorphisms. Similarly, define aSPT orders as the difference between the phases related
by FSD automorphism. In this sense, the pSPT orders and aSPT orders are generalization of SPT orders.

We then define an SPT phase (a different notion than SPT order) as a gapped phase
with non-degenerate ground state on any closed manifold.  For certain
non-invertible symmetries, we found new SPT phases whose are not related by any automorphisms,
\ie they do not differ by any SPT, pSPT, or aSPT orders.
Our results for varoius SSB phases, SPT phases, SPT orders,
pSPT orders, and aSPT orders with non-invertlble symmetries are summarized in 
Tables \ref{Phases}, \ref{PhasesS3xS3}, and \ref{gaplessPhases}.

While we showcased the usefulness of sym/TO correspondence to obtain
basic structures of gapped/gapless phases of 1+1D systems with a symmetry,
the sym/TO correspondence can be useful  in revealing further properties of
gapped/gapless phases, such as their boundaries and  interfaces between them~\cite{BS240902166}.
The sym/TO correspondence extends to the higher dimensions, 
for which the higher condensation theory becomes important~\cite{KZ240307813,BW250312699}.

\textit{Acknowledgements} ---
We thank Chenqi Meng, Xinping Yang, Tian Lan, and Zhengcheng Gu for sharing
some of their unpublished results with us.  We thank Arkya Chatterjee, Xie
Chen, Ho Tat Lam, Dachuan Lu, Sal Pace, and Xinping Yang for helpful
discussions.  This work was partially supported by NSF grant DMR-2022428 and by
the Simons Collaboration on Ultra-Quantum Matter, which is a grant from the
Simons Foundation (651446, XGW). 

\appendix

\section{Gapless SPT, SSB, and SSB-SPT orders for systems with symmetries}

\begin{table*}[t] 
\caption{ 
} \label{gaplessPhases} \centerline{
\begin{tabular}{ 
|c|l|
}
\hline 
Symmetry & Classes of gapless phases (non-Lagrangian condensable algebras)\\ 
\hline 
$S_3$ & 
\{[\textbf{1}]$_{1}$,[\textbf{1}]$_{3}$\},
\{[\textbf{1}]$_{1}$\},
\{[\textbf{1}]$_{2}$\} \\
\hline 
$\cRep_{S_3}$ & 
2$\times$\{[\textbf{1}]$_{1}$\},
\{[\textbf{1}]$_{1}$,[\textbf{1}]$_{2}$\} \\
\hline 
$A_4$ & 
\{[\textbf{1}]$_{1}$\},
\{[\textbf{2}]$_{1}$,[\textbf{1}]$_{4}$\},
\{[\textbf{1}]$_{3}$\},
\{[\textbf{2}]$_{3}$,[\textbf{1}]$_{6}$\} \\
\hline 
$\cRep_{A_4}$ & 
2$\times$\{[\textbf{1}]$_{1}$\},
2$\times$\{[\textbf{2}]$_{1}$,[\textbf{1}]$_{2}$\} \\
\hline 
$\tl\cR^{\eD(A_4)}_{\cA_{1,1}}$ &
\{[\textbf{1}]$_{1}$\},
\{[3$\times$\textbf{1}]$_{2}$\},
\{[\textbf{1}]$_{3}$\},
\{[3$\times$\textbf{1}]$_{4}$\} \\
\hline 
$S_3\hskip-1mm \times \hskip-1mm \Z_3$ &
\{[\textbf{1}]$_{1}$\},
3$\times$\{[\textbf{1}]$_{1}$,[\textbf{1}]$_{3}$\},
\{[\textbf{1}]$_{1}$,2$\times$[\textbf{1}]$_{3}$,[\textbf{1}]$_{9}$\},
3$\times$\{[\textbf{1}]$_{2}$\},
2$\times$\{[\textbf{1}]$_{2}$,[\textbf{1}]$_{6}$\},
\{3$\times$[\textbf{1}]$_{2}$,[\textbf{1}]$_{6}$\} \\
\hline 
$\cRep_{S_3\times \Z_3}$ &
4$\times$\{[\textbf{1}]$_{1}$\},
3$\times$\{[\textbf{1}]$_{1}$,[\textbf{1}]$_{3}$\},
\{[\textbf{1}]$_{1}$,[\textbf{1}]$_{2}$\},
\{[\textbf{1}]$_{1}$,[\textbf{1}]$_{2}$,[\textbf{1}]$_{3}$,[\textbf{1}]$_{6}$\},
\{3$\times$[\textbf{1}]$_{1}$,[\textbf{1}]$_{3}$\},
\{[\textbf{1}]$_{2}$,[\textbf{1}]$_{3}$\} \\
\hline 
$\tl\cR^{\eD(S_3\times \Z_3)}_{\cA_{3,2}}$ &
3$\times$\{[\textbf{1},\textbf{1}]$_{1}$\},
\{[\textbf{1}]$_{1}$\},
\{[\textbf{1},\textbf{1}]$_{1}$,[\textbf{1},\textbf{1}]$_{3}$\},
2$\times$\{[\textbf{1},\textbf{1}]$_{2}$\},
\{[\textbf{1}]$_{2}$\},
\{[\textbf{1},\textbf{1}]$_{2}$,[\textbf{1},\textbf{1}]$_{4}$\},
2$\times$\{[\textbf{1}]$_{4}$\} \\
\hline 
$D_8$ &
\{[\textbf{1}]$_{1}$\},
\{[\textbf{2}]$_{1}$,[\textbf{1}]$_{2}$\},
\{[\textbf{2}]$_{1}$,[\textbf{1},\textbf{1}]$_{2}$\},
\{[\textbf{2},\textbf{2}]$_{1}$,[\textbf{1}]$_{2}$,[\textbf{1}]$_{4}$\},
\{[\textbf{2}]$_{1}$,[\textbf{2},\textbf{2}]$_{1}$,[\textbf{2},\textbf{2}]$_{2}$,[\textbf{1},\textbf{1}]$_{4}$\},
\{[\textbf{1}]$_{2}$\},
\{[\textbf{1}]$_{2}$,[\textbf{2},\textbf{2}]$_{2}$,[\textbf{1}]$_{4}$\} \\
\hline 
$\cRep_{D_8}$ &
2$\times$\{[\textbf{1}]$_{1}$\},
\{[3$\times$\textbf{1}]$_{1}$\},
2$\times$\{[3$\times$\textbf{1}]$_{1}$,[3$\times$\textbf{1}]$_{2}$\},
\{[3$\times$\textbf{1}]$_{1}$,[\textbf{1}]$_{2}$\},
\{[6$\times$\textbf{1}]$_{1}$,[3$\times$\textbf{1}]$_{2}$,[3$\times$\textbf{1}]$_{3}$\} \\
\hline 
$\tl\cR^{\eD(D_8)}_{\cA_{1,1}}$ &
\{[\textbf{1}]$_{1}$\},
\{[6$\times$\textbf{1}]$_{2}$\},
\{[12$\times$\textbf{1}]$_{2}$\},
\{[3$\times$\textbf{1}]$_{2}$\},
\{[4$\times$\textbf{1}]$_{2}$\},
\{[\textbf{1}]$_{4}$\},
\{[6$\times$\textbf{1}]$_{4}$\} \\
\hline 
$H_8$ &
2$\times$\{[\textbf{1}]$_{1}$\},
\{[\textbf{1},\textbf{1}]$_{1}$,[\textbf{1},\textbf{1}]$_{3}$\},
2$\times$\{[\textbf{1}]$_{1}$,[\textbf{1}]$_{2}$\},
\{[\textbf{1}]$_{1}$,2$\times$[\textbf{1}]$_{2}$,[\textbf{1}]$_{3}$\},
\{[\textbf{1}]$_{2}$\} \\
\hline 
$\tl\cR^{\eM_\mathrm{Haag}}_{\cA_{1,1}}$ &
\{[\textbf{1}]$_{1}$\}, \{[\textbf{1}]$_{2}$\} \\
\hline 
$\tl\cR^{\eM_\mathrm{Haag}}_{\cA_{2,1}}$ &
\{[\textbf{1}]$_{1}$\}, \{[\textbf{1}]$_{2}$\} \\
\hline 
$\tl\cR^{\eM_\mathrm{tHaag}}$ &
\{[\textbf{1}]$_{1}$\}, \{[\textbf{1}]$_{2}$\} \\
\hline 
\end{tabular}
}
\end{table*}

In the main text, we studied gapped phases for 1+1D systems with symmetries.
Those gapped phases can be SPT phases (defined by having unique ground state),
and SSB phases (defined by having degenerate ground states).  We used duality
automorphisms, such as the FCD, FAD, and FSD automorphisms of the symTO, to
further characterize those gapped phases.

For a group-like symmetry, its different SPT phases always have identical bulk
excitations (\ie connected by FCD automorphisms), and those SPT phases can be
distinguished only via their different boundary states.  For some
non-invertible symmetries, such as $\cRep(S_3\times \Z_3)$ symmetry, we for
that there can be different SPT phases that are not connected by by any duality
automorphisms and have different bulk excitations.  We also find that the
states with different symmetries and different symmetry breaking patterns (\ie
different ground state degeneracies) can have identical bulk energy spectrum,
within the symmetric sub-Hilbert space.  Thus those symmetries and their
correspond SSB states are Morita-equivalent, which was called
holo-equivalence in \Rf{KZ200514178}.

In this section, we will summarize the similar results for gapless phases.  We
note that each non-Lagrangian condensable algebra corresponds many gapless
phases, which form a class.  Different non-Lagrangian condensable algebras
correspond to different classes of gapless phases.  In this paper, we will
study gapless phases in terms of those classes.  We will abbreviate  ``a class
of gapless phases'' as ``a gapless phase''.  In table \ref{gaplessPhases}, we
list distinct gapless phases (\ie distinct non-Lagrangian condensable algebras)
for some generalized symmetries.

The gapless phases are grouped into SPT-classes.  The gapless phases in the same
SPT-class, and the corresponding non-Lagrangian condensable algebras, are
connected by the FCD automorphisms.  Each bold-face number in table
\ref{gaplessPhases} corresponds to a SPT-class, and the number indicates the
number of gapless phases in the SPT-class.  

A bold-face number larger than 1 indicates the presence of distinct gapless
symmetric or SSB phases, that have identical bulk excitations.  The
corresponding bulk excitations also carry identical symmetry charges.  Despite
the identical bulk, those gapless phases are separated by phase transitions,
and have distinct boundary properties.

We regard the phases in the same SPT-class as differing by SPT orders. In other
words, two phases differ by a SPT order if the two phases are connected by a
FCD automorphisms.  
If the number phases in a SPT-class is more then one, then
the different symmetric phases in the same SPT-class correspond to different
gapless SPT phases.  Those phases are called gSPT phases in \Rf{BW240300905}.
But different gSPT phases may not differ by SPT orders.

Similarly, the different SSB phases in the same SPT-class correspond to
different gapless SSB phases that have the same symmetry breaking pattern but
differ by SPT orders.  We refer those phases as gapless SSB-SPT phases.  

The SPT-classes are further grouped into pSPT-classes.  Each
$[\cdots]_\text{GSD}$ represents a pSPT-class.  The phases in the same
pSPT-classes have the same ground state degeneracy $GSD$.  If GSD $>$ 1, then
the phases in the pSPT-class all have SSB (but may have different SSB
patterns). If  GSD = 1, then the phases in the pSPT-class are all symmetric.
Also, the phases in the same pSPT-class have identical spectrum for bulk
excitations. But, in contrast to SPT-class, the corresponding bulk excitations
may carry different symmetry charges of the same symmetry.

The SPT-classes are further grouped into aSPT-classes.  Each curly bracket $\{\cdots\}$
represents an aSPT-class.  The phases in the same aSPT-class have identical
spectrum for bulk excitations, in symmetric sub-Hilbert space (see an example
at the end of Section \ref{S3sym}).  

We note that as opposed to their gapped counterparts, automorphisms of the symTO 
do not fully characterize all gapless SPT phases. For example, the so-called canonical gSPT~\cite{BW240300905}
correspondence condensation of identity anyon $\onebb$ which is invariant under any automorphism.
We then refer to the differences between gapless phases by an FCD automorphism as
a gapless SPT order.

In the table \ref{gaplessPhases}, we use the shorthand notation, say, 3$\times$\textbf{1}
to denote $\textbf{1}, \textbf{1}, \textbf{1}$, and $H_8$ is the  
non-invertible symmetry described by a Hopf algebra.  

From the table, we see that for $A_4$ symmetry, there are gapless SPT phases
$[\two]_1$, as well as gapless $(A_4\to \Z_2\times \Z_2)$-SSB-SPT phases
$[\two]_3$.  Those $A_4$-SPT phases and $(A_4\to \Z_2\times \Z_2)$-SSB-SPT
phases are labeled by $\Z_2$ torsor (\ie connected by FCD automorphisms that
form a $\Z_2$ group).  It is also interesting to note that the two gapless
$A_4$-SPT phases are Morita-equivalent to a gapless $(A_4\to \Z_3)$-SSB phase
(\ie those phases have identical spectrum in the $A_4$-symmetric sub-Hilbert
space).  In other words, there is duality automorphism that maps a
$A_4$-symmetric model to another $A_4$-symmetric model, and maps a gapless
$A_4$-SPT phase to a gapless $(A_4\to \Z_3)$-SSB phase.  Since the  gapless
$A_4$-SPT phase and the gapless $(A_4\to \Z_3)$-SSB phase are related by a
duality automorphism, the two phases have identical spectrum in the symmetric
sub-Hilbert space.  Similarly, the two gapless $(A_4\to \Z_2\times
\Z_2)$-SSB-SPT phases are Morita-equivalent to a gapless $(A_4\to \Z_2)$-SSB
phase.

We like to mention that the table \ref{gaplessPhases} is computed using
necessary conditions for non-Lagrangian condensable algebras.  Thus we may get
fake non-Lagrangian condensable algebras, which may lead to fake aSPT-classes.

In the rest of the appendix, we will discuss some symmetries one by one, to
obtain the results for gapped phases described in the main text, as well as the
results for gapless phases described above.

\section{Symmetries with $\eD(A_4)$-symTO }

\begin{table*}[t] 
\caption{
The fusion rule of $\cD(A_4)$.
} \label{A4Nijk} \centering
\begin{tabular}{ |c||c|c|c|p{12mm}|p{12mm}|p{12mm}|p{12mm}|p{12mm}|p{14mm}|p{14mm}|p{14mm}|p{14mm}|p{14mm}|p{14mm}|}
\hline 
14  & $\onebb$  & $u$  & $v$  & $x$  & $y$  & $z$  & $a$  & $b$  & $g$  & $h$  & $i$  & $j$  & $s$  & $t$ \\ 
\hline 
\hline 
$\onebb$  & $ \onebb$  & $ u$  & $ v$  & $ x$  & $ y$  & $ z$  & $ a$  & $ b$  & $ g$  & $ h$  & $ i$  & $ j$  & $ s$  & $ t$  \\ 
\hline 
$u$  & $ u$  & $ v$  & $ \onebb$  & $ x$  & $ y$  & $ z$  & $ a$  & $ b$  & $ s$  & $ j$  & $ g$  & $ t$  & $ i$  & $ h$  \\ 
\hline 
$v$  & $ v$  & $ \onebb$  & $ u$  & $ x$  & $ y$  & $ z$  & $ a$  & $ b$  & $ i$  & $ t$  & $ s$  & $ h$  & $ g$  & $ j$  \\ 
\hline 
$x$  & $ x$  & $ x$  & $ x$  & $ \onebb \oplus u \oplus v \oplus 2x$  & $ z \oplus a \oplus b$  & $ y \oplus a \oplus b$  & $ y \oplus z \oplus b$  & $ y \oplus z \oplus a$  & $ g \oplus i \oplus s$  & $ h \oplus j \oplus t$  & $ g \oplus i \oplus s$  & $ h \oplus j \oplus t$  & $ g \oplus i \oplus s$  & $ h \oplus j \oplus t$  \\ 
\hline 
$y$  & $ y$  & $ y$  & $ y$  & $ z \oplus a \oplus b$  & $ \onebb \oplus u \oplus v \oplus 2y$  & $ x \oplus a \oplus b$  & $ x \oplus z \oplus b$  & $ x \oplus z \oplus a$  & $ g \oplus i \oplus s$  & $ h \oplus j \oplus t$  & $ g \oplus i \oplus s$  & $ h \oplus j \oplus t$  & $ g \oplus i \oplus s$  & $ h \oplus j \oplus t$  \\ 
\hline 
$z$  & $ z$  & $ z$  & $ z$  & $ y \oplus a \oplus b$  & $ x \oplus a \oplus b$  & $ \onebb \oplus u \oplus v \oplus 2z$  & $ x \oplus y \oplus b$  & $ x \oplus y \oplus a$  & $ g \oplus i \oplus s$  & $ h \oplus j \oplus t$  & $ g \oplus i \oplus s$  & $ h \oplus j \oplus t$  & $ g \oplus i \oplus s$  & $ h \oplus j \oplus t$  \\ 
\hline 
$a$  & $ a$  & $ a$  & $ a$  & $ y \oplus z \oplus b$  & $ x \oplus z \oplus b$  & $ x \oplus y \oplus b$  & $ \onebb \oplus u \oplus v \oplus 2a$  & $ x \oplus y \oplus z$  & $ g \oplus i \oplus s$  & $ h \oplus j \oplus t$  & $ g \oplus i \oplus s$  & $ h \oplus j \oplus t$  & $ g \oplus i \oplus s$  & $ h \oplus j \oplus t$  \\ 
\hline 
$b$  & $ b$  & $ b$  & $ b$  & $ y \oplus z \oplus a$  & $ x \oplus z \oplus a$  & $ x \oplus y \oplus a$  & $ x \oplus y \oplus z$  & $ \onebb \oplus u \oplus v \oplus 2b$  & $ g \oplus i \oplus s$  & $ h \oplus j \oplus t$  & $ g \oplus i \oplus s$  & $ h \oplus j \oplus t$  & $ g \oplus i \oplus s$  & $ h \oplus j \oplus t$  \\ 
\hline 
$g$  & $ g$  & $ s$  & $ i$  & $ g \oplus i \oplus s$  & $ g \oplus i \oplus s$  & $ g \oplus i \oplus s$  & $ g \oplus i \oplus s$  & $ g \oplus i \oplus s$  & $ 2h \oplus j \oplus t$  & $ \onebb \oplus x \oplus y \oplus z \oplus a \oplus b$  & $ h \oplus j \oplus 2t$  & $ u \oplus x \oplus y \oplus z \oplus a \oplus b$  & $ h \oplus 2j \oplus t$  & $ v \oplus x \oplus y \oplus z \oplus a \oplus b$  \\ 
\hline 
$h$  & $ h$  & $ j$  & $ t$  & $ h \oplus j \oplus t$  & $ h \oplus j \oplus t$  & $ h \oplus j \oplus t$  & $ h \oplus j \oplus t$  & $ h \oplus j \oplus t$  & $ \onebb \oplus x \oplus y \oplus z \oplus a \oplus b$  & $ 2g \oplus i \oplus s$  & $ v \oplus x \oplus y \oplus z \oplus a \oplus b$  & $ g \oplus i \oplus 2s$  & $ u \oplus x \oplus y \oplus z \oplus a \oplus b$  & $ g \oplus 2i \oplus s$  \\ 
\hline 
$i$  & $ i$  & $ g$  & $ s$  & $ g \oplus i \oplus s$  & $ g \oplus i \oplus s$  & $ g \oplus i \oplus s$  & $ g \oplus i \oplus s$  & $ g \oplus i \oplus s$  & $ h \oplus j \oplus 2t$  & $ v \oplus x \oplus y \oplus z \oplus a \oplus b$  & $ h \oplus 2j \oplus t$  & $ \onebb \oplus x \oplus y \oplus z \oplus a \oplus b$  & $ 2h \oplus j \oplus t$  & $ u \oplus x \oplus y \oplus z \oplus a \oplus b$  \\ 
\hline 
$j$  & $ j$  & $ t$  & $ h$  & $ h \oplus j \oplus t$  & $ h \oplus j \oplus t$  & $ h \oplus j \oplus t$  & $ h \oplus j \oplus t$  & $ h \oplus j \oplus t$  & $ u \oplus x \oplus y \oplus z \oplus a \oplus b$  & $ g \oplus i \oplus 2s$  & $ \onebb \oplus x \oplus y \oplus z \oplus a \oplus b$  & $ g \oplus 2i \oplus s$  & $ v \oplus x \oplus y \oplus z \oplus a \oplus b$  & $ 2g \oplus i \oplus s$  \\ 
\hline 
$s$  & $ s$  & $ i$  & $ g$  & $ g \oplus i \oplus s$  & $ g \oplus i \oplus s$  & $ g \oplus i \oplus s$  & $ g \oplus i \oplus s$  & $ g \oplus i \oplus s$  & $ h \oplus 2j \oplus t$  & $ u \oplus x \oplus y \oplus z \oplus a \oplus b$  & $ 2h \oplus j \oplus t$  & $ v \oplus x \oplus y \oplus z \oplus a \oplus b$  & $ h \oplus j \oplus 2t$  & $ \onebb \oplus x \oplus y \oplus z \oplus a \oplus b$  \\ 
\hline 
$t$  & $ t$  & $ h$  & $ j$  & $ h \oplus j \oplus t$  & $ h \oplus j \oplus t$  & $ h \oplus j \oplus t$  & $ h \oplus j \oplus t$  & $ h \oplus j \oplus t$  & $ v \oplus x \oplus y \oplus z \oplus a \oplus b$  & $ g \oplus 2i \oplus s$  & $ u \oplus x \oplus y \oplus z \oplus a \oplus b$  & $ 2g \oplus i \oplus s$  & $ \onebb \oplus x \oplus y \oplus z \oplus a \oplus b$  & $ g \oplus i \oplus 2s$  \\ 
\hline 
\end{tabular}
\end{table*}

In 1+1D, $A_4$-symmetry and $\cRep_{A_4}$-symmetry are Morita-equivalent.  Their
common symTO is the $\eD(A_4)$ quantum double (or 2+1D $A_4$ gauge theory).
The anyons of the symTO are given below:
\\[2mm]
\centerline{
\begin{tabular}{|c|c|c|c|c|c|c|c|c|c|c|c|c|c|c|}
\hline
$\eD(A_4)$ anyon& $\onebb$ 
& $u$ & $v$ & $x$ & $y$ & $z$ & $a$ & $b$ & $g$ & $h$ & $i$ & $j$ & $s$ & $t$ \\ 
\hline
$s$: & $0$ & $0$ & $0$ & $0$ & $0$ & $0$ & $\frac{1}{2}$ & $\frac{1}{2}$ & $0$ & $0$ & $\frac{1}{3}$ & $\frac{1}{3}$ & $\frac{2}{3}$ & $\frac{2}{3}$ \\ 
\hline
$d$: & $1$ & $1$ & $1$ & $3$ & $3$ & $3$ & $3$ & $3$ & $4$ & $4$ & $4$ & $4$ & $4$ & $4$ \\ 
\hline
\end{tabular}
}
\vskip 2mm
\noindent
The fusion rule of the those anyons are given in Table \ref{A4Nijk}.
The FSD automorphisms of the quantum double $\eD(A_4)$
form a group $S_3\times \Z_2\times \Z_2$, which is generated by
the following permutations of anyons:
\begin{align}
&(x ,y) ;\ (x ,y ,z); \ \
(a ,b) ; \ \ 
(g ,h) (i ,j) (s ,t) (u ,v). 
\end{align}
In this section, we will calculate the SPT phases of the non-invertible
$\cRep_{A_4}$-symmetry, using the sym/TO approach.

The quantum double $\eD(A_4)$ has seven Lagrangian condensable algebras that
form 3 aSPT-classes according to the FSD automorphisms of $\eD(A_4)$:
\begingroup
\allowdisplaybreaks
\begin{align}
\label{A4Lca}
& \cA_{1,1} =  \onebb \oplus u \oplus v \oplus x \oplus y \oplus z
\nonumber \\
& \cA_{2,1} =  \onebb \oplus u \oplus v \oplus 3x
\nonumber \\
& \cA_{2,2} =  \onebb \oplus u \oplus v \oplus 3y
\nonumber \\
& \cA_{2,3} =  \onebb \oplus u \oplus v \oplus 3z
\nonumber \\
& \cA_{3,1} =  \onebb \oplus g \oplus h \oplus x
\nonumber \\
& \cA_{3,2} =  \onebb \oplus g \oplus h \oplus y
\nonumber \\
& \cA_{3,3} =  \onebb \oplus g \oplus h \oplus z
\end{align}
Therefore, for 1+1D systems with a symmetry described by $\eD(A_4)$-symTO,
there are only seven distinct gapped phases.  The excitations in charge neutral
sector in a gapped phase are described by a fusion category.  The seven
different phases give rise to three different fusion categories.  The phases in
the same aSPT-class have equivalent dynamics, \ie their excitations are
described by the same fusion category.

The anyons $(\onebb, u,v,x)$ in $\eD(A_4)$ correspond to the 
4 irreps of $A_4$ of dimension $(1,1,1,3)$.  Thus the Lagrangian
condensable algebra $\cA_{2,1}$ in aSPT-class-2 produces a gapped
boundary whose excitations are described by fusion category $\cVec_{A_4}$.
Other Lagrangian condensable algebras in aSPT-class-2 also give rise to
fusion category $\cVec_{A_4}$.  The anyons $(g,h,z)$ in $\eD(A_4)$ correspond
to the three non-trivial gauge flux quanta in the $A_4$ gauge theory.  Thus the
Lagrangian condensable algebra $\cA_{3,3}$ in aSPT-class-3 produces a
gapped boundary whose excitations are described by fusion category
$\cRep_{A_4}$.

The symTO  $\eD(A_4)$ also has eight non-Lagrangian condensable algebras, which
is divided into 4 aSPT-classes by the FSD automorphism of $\eD(A_4)$:
\begin{align}
& \cA'_{1,1} =  \onebb \oplus u \oplus v \oplus x
\nonumber \\
& \cA'_{1,2} =  \onebb \oplus u \oplus v \oplus y
\nonumber \\
& \cA'_{1,3} =  \onebb \oplus u \oplus v \oplus z
\nonumber \\
& \cA'_{2,1} =  \onebb \oplus x
\nonumber \\
& \cA'_{2,2} =  \onebb \oplus y
\nonumber \\
& \cA'_{2,3} =  \onebb \oplus z
\nonumber \\
& \cA'_{3,1} =  \onebb \oplus u \oplus v
\nonumber \\
& \cA'_{4,1} =  \onebb
\end{align}
\endgroup
Thus, there are eight classes of gapless phases for systems with symTO
$\eD(A_4)$.  But the eight classes of gapless phases belong to four different
aSPT classes. Thus the eight gapless phases only give rise to four types of
low energy dynamics (\ie four different partition functions $Z_\onebb$) when
restricted to symmetric sub-Hilbert space.

\subsection{ $A_4$ anomaly-free symmetry}

The $\cVec_{A_4}$-symmetry (\ie the usual $A_4$-symmetry) is induced by the
Lagrangian condensable algebra $\cA_{2,1}$. The symmetry has seven gapped
phases induced by the seven Lagrangian condensable algebras \eqref{A4Lca}.
Their ground state degeneracies are $\text{GSD}^{\cVec_{A_4}}_{\cA_i} =
(\cA_{2,1}, \cA_i)$:
\begin{align}
\text{GSD}^{\cVec_{A_4}}_{\cA_{2,1}} &= 12,\ \ 
\text{GSD}^{\cVec_{A_4}}_{\cA_{2,2}} = 3,\ \
\text{GSD}^{\cVec_{A_4}}_{\cA_{2,3}} = 3,
\nonumber\\
\text{GSD}^{\cVec_{A_4}}_{\cA_{1,1}} &= 6,\ \ 
\\
\text{GSD}^{\cVec_{A_4}}_{\cA_{3,1}} &= 4,\ \ 
\text{GSD}^{\cVec_{A_4}}_{\cA_{3,2}} = 1,\ \ 
\text{GSD}^{\cVec_{A_4}}_{\cA_{3,3}} = 1.
\nonumber 
\end{align}
Noticing that the order-12 $A_4$ group has five (conjugacy classes of)
sub-groups: trivial group, $\Z_2$ group, $\Z_2\times \Z_2$ group, $\Z_4$ group,
and $A_4$ group.  Therefore, there are five SSB patterns.  

As pointed out in \Rf{KZ200308898,KZ200514178}, the SPT phases can be
(partially) generated by FCD automorphisms of $A_4$, where
\emph{FCD automorphisms} are automorphisms that keep the $A_4$ charges, \ie the
anyons in $\cA_{2,1}$, unchanged. FCD automorphisms of $\cA_{2,1}$ form a
$\Z_2\times \Z_2$ group generated by
\begin{align}
(y ,z ),\ \ \ (a ,b ).
\end{align}

The seven gapped phases can be divided into five equivalence classes
(SPT-classes) using FCD automorphisms as equivalence relation:
\begin{align}
\label{A4phases}
\big(  \cA_{3,2}\text{-},\cA_{3,3}\text{-phases} \big)_{1} 
&=\ A_4\text{-SPT phase},
\nonumber \\
\big(  \cA_{2,2}\text{-},\cA_{2,3}\text{-phases} \big)_{3} 
&=\ (A_4\to \Z_2\times\Z_2)\text{-SSB-SPT},
\nonumber \\
\big(  \cA_{3,1}\text{-phase} \big)_{4} 
&=\ (A_4\to \Z_3)\text{-SSB phase} ,
\nonumber \\
\big(  \cA_{1,1}\text{-phase} \big)_{6} 
&=\ (A_4\to \Z_2)\text{-SSB phase} ,
\nonumber \\
\big(  \cA_{2,1}\text{-phase} \big)_{12} 
&=\ (A_4\to \Z_1)\text{-SSB phase}.
\end{align}
In the above, the phases in $(\cdots)$ are phases that are connected by those
FCD automorphisms, which form a SPT-class.  In general, for a symmetry
described by a fusion category $\tl\cR_\text{def}$, its FCD automorphisms maps
between $\tl\cR_\text{def}$-SPT phases.  Thus FCD automorphisms generate
$\tl\cR_\text{def}$-SPT phases from trivial symmetric product state.  

In \eqref{A4phases}, the subscript of $(\cdots)$ is the ground state
degeneracy.  We see that the $\cA_{3,2}$-phase and the $\cA_{3,3}$-phase have
\text{GSD} = 1 and the full $A_4$-symmetry. They are $A_4$-SPT phases.  One of
them is the trivial product state and the other is a non-trivial SPT state.

On the other hand, the $\cA_{2,1}$-phase and the $\cA_{2,2}$-phase both have
$\Z_2\times \Z_2$ unbroken symmetry. They differ by a $(\Z_2\times \Z_2)$-SPT
order.  Thus we call them SSB-SPT phases.

We also see that the FCD automorphism $(y,z)$ maps between $\cA_{3,2}$ and
$\cA_{3,3}$ (\ie between symmetric $\cA_{3,2}$-phase and $\cA_{3,3}$-phase).
Similarly, the same FCD automorphism $(y,z)$ maps between $\cA_{2,1}$ and
$\cA_{2,2}$ (\ie between symmetry breaking $\cA_{2,1}$-phase and
$\cA_{2,2}$-phase).  Thus we may say that the symmetric $\cA_{3,2}$-phase and
$\cA_{3,3}$-phase differ by the same SPT-order as the symmetry-breaking
$\cA_{2,2}$-phase and $\cA_{2,3}$-phase.

%

The $A_4$ symmetric systems have eight classes of gapless phases, which are
divided into six SPT-classes by FCD automorphisms:
\begin{align}
\big(  \cA'_{4,1}\text{-phases} \big)_{1} 
&= A_4\text{-symmetric phases}
\nonumber \\
\big(  \cA'_{2,2}\text{-},\cA'_{2,3}\text{-phases} \big)_{1} 
&= A_4\text{-SPT phases}
\nonumber \\
\big(  \cA'_{3,1}\text{-phases} \big)_{3} 
&= (A_4\to \Z_2\times \Z_2)\text{-SSB}
\nonumber \\
\big(  \cA'_{1,2}\text{-},\cA'_{1,3}\text{-phases} \big)_{3} 
&= (A_4\to \Z_2\times \Z_2)\text{-SSB-SPT}
\nonumber \\
\big(  \cA'_{2,1}\text{-phases} \big)_{4} 
&= (A_4\to \Z_3)\text{-SSB}
\nonumber \\
\big(  \cA'_{1,1}\text{-phases} \big)_{6} 
&= (A_4\to \Z_2)\text{-SSB} 
\end{align}
The subscript of $(\cdots)$ is the ground state degeneracy.  We see that there
are three classes of $A_4$-symmetric gapless phases.  The $\cA'_{2,2}$-class
and the $\cA'_{2,3}$-class are connected by a FCD automorphism, and thus
they differ by a $A_4$-SPT order.  Two classes of SSB gapless phases,
the $\cA'_{1,2}$-class and the $\cA'_{1,3}$-class are also connected by a
FCD automorphism, and they too differ by a $A_4$-SPT order.

Physically, if two gapped or gapless states are connected by FCD automorphism,
then they will have identical bulk excitations, and the corresponding bulk
excitations carry identical symmetry charges.  In other words, the two states
are indistinguishable if we do not twist the periodic condition, \ie they are
indistinguishable by bulk measurements.

In the above, we have divided the gapped and gapless phases of $A_4$-symmetric
systems according to the FCD automorphisms and the associated
SPT-classes.  We can also divide the gapped and gapless phases differently,
according to the \emph{FAD automorphisms} and the associated
\emph{pSPT-classes}.  FAD automorphisms are FSD automorphisms
that keep the set of $A_4$ charges, \ie the Lagrangian condensable
algebra $\cA_{2,1}$, unchanged.  Note that the FCD automorphisms form a
subgroup of the FAD automorphisms.  The FAD automorphisms of
$\cA_{2,1}$ form a $\Z_2\times \Z_2\times \Z_2$ group generated by
\begin{align}
(y ,z ),\ \ \
(g ,h )(i ,j )(s ,t )(u ,v ),\ \ \
(a ,b ).
\end{align}
We find that, for $A_4$ symmetry, the FAD automorphisms happen to divide
the gapped and gapless phases in the same way as the FCD automorphisms.  In
other words, the pSPT-classes coincide with the SPT-classes.

\subsection{ $\cRep_{A_4}$ anomaly-free symmetry}

Now, we are ready to discuss some new results. The $\cRep_{A_4}$-symmetry
induced by $\cA_{3,3}$ condensation also has seven gapped phases induced by the
same seven Lagrangian condensable algebras \eqref{A4Lca}.  

The FCD automorphisms (new fix the anyons in $\cA_{3,3}$) form a $\Z_2\times
\Z_2$ group generated by $ (x ,y)$, and $ (a ,b )$.  On the other hand, the
FAD automorphisms form a $\Z_2\times \Z_2\times \Z_2$ group generated by $ (x ,y
) $, $ (a ,b ) $, and $ (g ,h )(i ,j )(s ,t )(u ,v ) $.

Using FCD automorphisms, FAD automorphisms, and FSD automorphisms, we can
divide the seven gapped phases into aSPT-classes, pSPT-classes, and SPT-classes:
\begin{align}
&\Big\{ \big[( \cA_{1,1})\big]^\text{SSB}_{2}\Big\},
\nonumber\\
&\Big\{ \big[( \cA_{2,1},\cA_{2,2})\big]^\text{Sym}_{1},
\big[( \cA_{2,3})\big]^\text{SSB}_{4}\Big\},
\nonumber\\
&\Big\{ \big[( \cA_{3,1},\cA_{3,2})\big]^\text{SSB}_{3},
\big[( \cA_{3,3})\big]^\text{SSB}_{4}\Big\} .
\end{align}
Here phases in $\{\cdots\}$ form an aSPT-class.  Phases in
$[\cdots]_\text{GSD}$ form an pSPT-class, where GSD is the ground state
degeneracy for all the phases in the pSPT-class.  If GSD = 1, the phases in the
pSPT-class are all symmetric.  If GSD $>$ 1, the phases in the pSPT-class all
have SSB.  Phases in $(\cdots)$ form a SPT-class.  If the number of phases in a
SPT-class is more than one, then there are non-trivial SPT orders.  The entry
\{[\textbf{1}]$_{2}$\},
\{[\textbf{2}]$_{1}$,[\textbf{1}]$_{4}$\}, 
\{[\textbf{2}]$_{3}$,[\textbf{1}]$_{4}$\} for $\cRep_{A_4}$-symmetry in Table
\ref{Phases} summarizes the above result.

From the above, we see that $\cA_{2,1}$-phase and $\cA_{2,2}$-phase have GSD
=1. Thus $\cRep_{A_4}$-symmetry is anomaly-free.  Since, $\cA_{2,1}$-phase and
$\cA_{2,2}$-phase are in the same SPT-class, they differ by a non-trivial
$\cRep_{A_4}$-SPT order.  Thus, 
\frmbox{anomaly-free $\cRep_{A_4}$-symmetry has a pair
of SPT phases differing by a SPT order.
}  
(a trivial one and a non-trivial one).
$\cRep_{A_4}$-symmetry
has also has a  pair of SSB-SPT phases: $\cA_{3,1}$-phase and $\cA_{3,2}$-phase. 

The automorphism $(x,y)$ maps
between $\cA_{2,1}$ and $\cA_{2,2}$ (\ie   between symmetric $\cA_{2,1}$-phase
and $\cA_{2,2}$-phase), as well as between $\cA_{3,1}$ and $\cA_{3,2}$ (\ie
between symmetry-breaking $\cA_{3,1}$-phase and $\cA_{3,2}$-phase).  Thus both
pairs of phases differ by the same $\cRep_{A_4}$-SPT order.  

The $\cRep_{A_4}$-symmetric systems have eight classes of gapless phases, which
are divided into aSPT-classes, pSPT-classes, and SPT-classes:
\begin{align}
&\Big\{ \big[( \cA'_{1,1},\cA'_{1,2})\big]^\text{Sym}_{1},
\big[( \cA'_{1,3})\big]^\text{SSB}_{2}\Big\},
\nonumber\\
&\Big\{ \big[( \cA'_{2,1},\cA'_{2,2})\big]^\text{Sym}_{1}, 
\big[( \cA'_{2,3})\big]^\text{SSB}_{2}\Big\},
\nonumber\\
&\Big\{ \big[( \cA'_{3,1})\big]^\text{Sym}_{1}\Big\},
\ \ \ \
\Big\{ \big[( \cA'_{4,1})\big]^\text{Sym}_{1}\Big\}.
\end{align}
We see that there are six SPT-classes of $\cRep_{A_4}$-symmetric gapless phases.
The $\cA'_{2,1}$  and $\cA'_{2,2}$  differ by a SPT order.  The $\cA'_{1,1}$
and $\cA'_{1,2}$  also differ by a SPT order.  \frmbox{Anomaly-free
$\cRep_{A_4}$-symmetry has two pairs of gapless symmetric phases, $(
\cA'_{1,1},\cA'_{1,2})$ and $( \cA'_{2,1},\cA'_{2,2})$. The two phases in each
pair differ by $\cRep_{A_4}$-SPT orders (\ie are connected by
FCD automorphisms).} We like to remark that $\cA'_{3,1}$ and $\cA'_{4,1}$ are
also gapless symmetric phases, which are not connected to any other gapless
symmetric phases via FCD automorphisms.  We may not want to call them gapless
SPT phases.  (In contrast, we call all the gapped symmetric phases as gapped
SPT phases.)

\subsection{$\tl\cR^{\eD(A_4)}_{\cA_{1,1}}$ anomalous symmetry}

The $\eD(A_4)$-symTO has three Morita-equivalent symmetries. Two of them are
$A_4$ and $\cRep_{A_4}$ symmetries discussed above.  The third symmetry is
described by a fusion category that describes excitations on the $\cA_{1,1}$
condensed boundary.  So we refer to the third symmetry as
$\tl\cR^{\eD(A_4)}_{\cA_{1,1}}$-symmetry.

The FCD automorphisms form a $\Z_2$ group generated by $ (a ,b )$.  The
FAD automorphisms form a $S_3\times \Z_2\times \Z_2$ group generated
by $ (x ,y ) $, $ (x ,y ,z) $, $ (a ,b ) $, and $ (g ,h )(i ,j )(s ,t )(u ,v )
$.

Using FCD automorphisms, FAD automorphisms, and FSD automorphisms,
we can divide the seven gapped phases into aSPT-classes, pSPT-classes, and SPT-classes:
\begin{align}
&\Big\{ \big[( \cA_{1,1})\big]^\text{SSB}_{6}\Big\},
\nonumber\\
&\Big\{ \big[( \cA_{2,1}),( \cA_{2,2}),( \cA_{2,3})\big]^\text{SSB}_{6}\Big\},
\nonumber\\
&\Big\{ \big[( \cA_{3,1}),( \cA_{3,2}),( \cA_{3,3})\big]^\text{SSB}_{2}\Big\},
\end{align}
Since there are no gapped phases with non-degenerate ground state, the
$\tl\cR^{\eD(A_4)}_{\cA_{1,1}}$-symmetry is anomalous.
There is no SSB-SPT phases for $\tl\cR^{\eD(A_4)}_{\cA_{1,1}}$-symmetry, since
each SPT-class contain only one phase.

The $\tl\cR^{\eD(A_4)}_{\cA_{1,1}}$-symmetric systems have eight classes of gapless phases, which
are divided into aSPT-classes, pSPT-classes, and SPT-classes:
\begin{align}
&\Big\{ \big[( \cA'_{1,1}),( \cA'_{1,2}),( \cA'_{1,3})\big]^\text{SSB}_{4}\Big\},
\nonumber\\
&\Big\{ \big[( \cA'_{2,1}),( \cA'_{2,2}),( \cA'_{2,3})\big]^\text{SSB}_{2}\Big\},
\nonumber\\
&\Big\{ \big[( \cA'_{3,1})\big]^\text{SSB}_{3}\Big\},
\nonumber\\
&\Big\{ \big[( \cA'_{4,1})\big]^\text{Sym}_{1}\Big\}.
\end{align}
We see that there is no gapless SPT-order for
the anomalous $\tl\cR^{\eD(A_4)}_{\cA_{1,1}}$-symmetry.

But we have several phases in the same pSPT-class, which are connected by
FAD automorphism.  Physically, if two gapped or gapless states are related by an
FAD automorphism, then the correlation functions of all corresponding local
operators will be identical on a ring with an untwisted periodic condition.
These corresponding local operators may carry different symmetry charges. (If
the FAD automorphism is also an FCD automorphism, then the corresponding local
operators will carry identical symmetry charges.) In other words, the two
states become indistinguishable, apart from a permutation of symmetry charges,
as long as the periodic condition remains untwisted. However, the two states
can be distinguished if we do not permute the symmetry charges. This highlights
the physical significance of FAD automorphisms.

\section{Symmetries with $\eD(D_8)$-symTO } 

In 1+1D, $D_8$-symmetry and $\cRep_{D_8}$-symmetry have the same
symTO which is the $\eD(D_8)$ quantum double (or 2+1D $D_8$ gauge theory).
The anyons of $\eD(D_8)$-symTO are given below\\[2mm] 
\centerline{
\begin{tabular}{|c|c|c|c|c|c|c|c|c|c|c|c|c|c|c|c|c|c|c|c|c|c|c|}
\hline
& $\onebb$ & $\mathbf{a}$ & $\mathbf{b}$ & $g$ & $\mathbf{c}$ & $e$ & $h$ & $\mathbf{d}$ & $i$ & $m$ & $l$ & $j$ & $n$ & $k$ & $u$ & $q$ & $s$ & $v$ & $r$ & $t$ & $x$ & $y$ \\ 
\hline
$s$: & $0$ & $0$ & $0$ & $0$ & $0$ & $0$ & $0$ & $0$ & $0$ & $0$ & $0$ & $0$ & $0$ & $0$ & $\frac{1}{2}$ & $\frac{1}{2}$ & $\frac{1}{2}$ & $\frac{1}{2}$ & $\frac{1}{2}$ & $\frac{1}{2}$ & $\frac{1}{4}$ & $\frac{3}{4}$ \\ 
\hline
$d$: & $1$ & $1$ & $1$ & $1$ & $1$ & $1$ & $1$ & $1$ & $2$ & $2$ & $2$ & $2$ & $2$ & $2$ & $2$ & $2$ & $2$ & $2$ & $2$ & $2$ & $2$ & $2$ \\ 
\hline
\end{tabular}
}
\vskip 2mm \noindent
The anyons $\vc a,\vc b,g,i$ are the gauge charges of the $D_8$ gauge theory.
In fact, $\vc a,\vc b,g$ correspond to the three 1-dimensional irreps, 
and $i$ the 2-dimensional irreps of
$D_8$.  The FSD automorphisms of $D_8$ quantum double form a $S_4$ group of order
24, as one can see from the permutations of $ \vc a, \vc b, \vc c, \vc d$:
\begingroup
\allowdisplaybreaks
\begin{align}
& ( )
\nonumber \\
& (m ,l )(j ,n )(u ,q )(s ,v )(\mathbf{c} ,\mathbf{d} )(e ,h )
\nonumber \\
& (m ,j )(l ,n )(u ,s )(q ,v )(\mathbf{a} ,\mathbf{b} )(e ,h )
\nonumber \\
& (m ,n )(l ,j )(u ,v )(q ,s )(\mathbf{a} ,\mathbf{b} )(\mathbf{c} ,\mathbf{d} )
\nonumber \\
& (i ,m )(n ,k )(u ,r )(v ,t )(\mathbf{b} ,\mathbf{c} )(g ,e )
\nonumber \\
& (i ,m ,l )(j ,k ,n )(u ,q ,r )(s ,t ,v )(\mathbf{b} ,\mathbf{c} ,\mathbf{d} )(g ,e ,h )
\nonumber \\
& (i ,m ,j )(l ,k ,n )(u ,s ,r )(q ,t ,v )(\mathbf{a} ,\mathbf{c} ,\mathbf{b} )(g ,e ,h )
\nonumber \\
& (i ,m ,k ,n )(l ,j )(u ,t ,v ,r )(q ,s )(\mathbf{a} ,\mathbf{c} ,\mathbf{d} ,\mathbf{b} )(g ,e )
\nonumber \\
& (i ,l ,m )(j ,n ,k )(u ,r ,q )(s ,v ,t )(\mathbf{b} ,\mathbf{d} ,\mathbf{c} )(g ,h ,e )
\nonumber \\
& (i ,l )(j ,k )(q ,r )(s ,t )(\mathbf{b} ,\mathbf{d} )(g ,h )
\nonumber \\
& (i ,l ,k ,j )(m ,n )(u ,v )(q ,t ,s ,r )(\mathbf{a} ,\mathbf{d} ,\mathbf{c} ,\mathbf{b} )(g ,h )
\nonumber \\
& (i ,l ,n )(m ,k ,j )(u ,t ,s )(q ,v ,r )(\mathbf{a} ,\mathbf{d} ,\mathbf{b} )(g ,h ,e )
\nonumber \\
& (i ,j ,m )(l ,n ,k )(u ,r ,s )(q ,v ,t )(\mathbf{a} ,\mathbf{b} ,\mathbf{c} )(g ,h ,e )
\nonumber \\
& (i ,j )(l ,k )(q ,t )(s ,r )(\mathbf{a} ,\mathbf{c} )(g ,h )
\nonumber \\
& (i ,j ,k ,l )(m ,n )(u ,v )(q ,r ,s ,t )(\mathbf{a} ,\mathbf{b} ,\mathbf{c} ,\mathbf{d} )(g ,h )
\nonumber \\
& (i ,j ,n )(m ,k ,l )(u ,t ,q )(s ,v ,r )(\mathbf{a} ,\mathbf{c} ,\mathbf{d} )(g ,h ,e )
\nonumber \\
& (i ,n ,k ,m )(l ,j )(u ,r ,v ,t )(q ,s )(\mathbf{a} ,\mathbf{b} ,\mathbf{d} ,\mathbf{c} )(g ,e )
\nonumber \\
& (i ,n ,j )(m ,l ,k )(u ,q ,t )(s ,r ,v )(\mathbf{a} ,\mathbf{d} ,\mathbf{c} )(g ,e ,h )
\nonumber \\
& (i ,n ,l )(m ,j ,k )(u ,s ,t )(q ,r ,v )(\mathbf{a} ,\mathbf{b} ,\mathbf{d} )(g ,e ,h )
\nonumber \\
& (i ,n )(m ,k )(u ,t )(v ,r )(\mathbf{a} ,\mathbf{d} )(g ,e )
\nonumber \\
& (i ,k )(l ,j )(q ,s )(r ,t )(\mathbf{a} ,\mathbf{c} )(\mathbf{b} ,\mathbf{d} )
\nonumber \\
& (i ,k )(m ,l ,n ,j )(u ,q ,v ,s )(r ,t )(\mathbf{a} ,\mathbf{d} ,\mathbf{b} ,\mathbf{c} )(e ,h )
\nonumber \\
& (i ,k )(m ,j ,n ,l )(u ,s ,v ,q )(r ,t )(\mathbf{a} ,\mathbf{c} ,\mathbf{b} ,\mathbf{d} )(e ,h )
\nonumber \\
& (i ,k )(m ,n )(u ,v )(r ,t )(\mathbf{a} ,\mathbf{d} )(\mathbf{b} ,\mathbf{c} )
\end{align}
\endgroup
$D_8$ quantum double has eleven Lagrangian condensable algebras (in three aSPT-classes):
\begingroup
\allowdisplaybreaks
\begin{align}
& \cA_{1,1} =  \onebb \oplus \mathbf{a} \oplus \mathbf{b} \oplus g \oplus \mathbf{c} \oplus e \oplus h \oplus \mathbf{d} 
\nonumber \\
& \cA_{2,1} =  \onebb \oplus i \oplus m \oplus l \oplus \mathbf{a} 
\nonumber \\
& \cA_{2,2} =  \onebb \oplus i \oplus j \oplus n \oplus \mathbf{b} 
\nonumber \\
& \cA_{2,3} =  \onebb \oplus m \oplus j \oplus k \oplus \mathbf{c} 
\nonumber \\
& \cA_{2,4} =  \onebb \oplus l \oplus n \oplus k \oplus \mathbf{d} 
\nonumber \\
& \cA_{3,1} =  \onebb \oplus 2i \oplus \mathbf{a} \oplus \mathbf{b} \oplus g 
\nonumber \\
& \cA_{3,2} =  \onebb \oplus 2m \oplus \mathbf{a} \oplus \mathbf{c} \oplus e 
\nonumber \\
& \cA_{3,3} =  \onebb \oplus 2l \oplus \mathbf{a} \oplus h \oplus \mathbf{d} 
\nonumber \\
& \cA_{3,4} =  \onebb \oplus 2j \oplus \mathbf{b} \oplus \mathbf{c} \oplus h 
\nonumber \\
& \cA_{3,5} =  \onebb \oplus 2n \oplus \mathbf{b} \oplus e \oplus \mathbf{d} 
\nonumber \\
& \cA_{3,6} =  \onebb \oplus 2k \oplus g \oplus \mathbf{c} \oplus \mathbf{d} 
\end{align}
and 33 non-Lagrangian condensable algebras (in seven aSPT-classes):
\begin{align}
& \cA'_{1,1} =  \onebb \oplus g \oplus e \oplus h 
\nonumber \\
& \cA'_{2,1} =  \onebb \oplus \mathbf{a} \oplus \mathbf{b} \oplus g 
\nonumber \\
& \cA'_{2,2} =  \onebb \oplus \mathbf{a} \oplus \mathbf{c} \oplus e 
\nonumber \\
& \cA'_{2,3} =  \onebb \oplus \mathbf{a} \oplus h \oplus \mathbf{d} 
\nonumber \\
& \cA'_{2,4} =  \onebb \oplus \mathbf{b} \oplus \mathbf{c} \oplus h 
\nonumber \\
& \cA'_{2,5} =  \onebb \oplus \mathbf{b} \oplus e \oplus \mathbf{d} 
\nonumber \\
& \cA'_{2,6} =  \onebb \oplus g \oplus \mathbf{c} \oplus \mathbf{d} 
\nonumber \\
& \cA'_{3,1} =  \onebb \oplus i \oplus g 
\nonumber \\
& \cA'_{3,2} =  \onebb \oplus m \oplus e 
\nonumber \\
& \cA'_{3,3} =  \onebb \oplus l \oplus h 
\nonumber \\
& \cA'_{3,4} =  \onebb \oplus j \oplus h 
\nonumber \\
& \cA'_{3,5} =  \onebb \oplus n \oplus e 
\nonumber \\
& \cA'_{3,6} =  \onebb \oplus k \oplus g 
\nonumber \\
& \cA'_{4,1} =  \onebb \oplus i \oplus \mathbf{a} 
\nonumber \\
& \cA'_{4,2} =  \onebb \oplus i \oplus \mathbf{b} 
\nonumber \\
& \cA'_{4,3} =  \onebb \oplus m \oplus \mathbf{a} 
\nonumber \\
& \cA'_{4,4} =  \onebb \oplus m \oplus \mathbf{c} 
\nonumber \\
& \cA'_{4,5} =  \onebb \oplus l \oplus \mathbf{a} 
\nonumber \\
& \cA'_{4,6} =  \onebb \oplus l \oplus \mathbf{d} 
\nonumber \\
& \cA'_{4,7} =  \onebb \oplus j \oplus \mathbf{b} 
\nonumber \\
& \cA'_{4,8} =  \onebb \oplus j \oplus \mathbf{c} 
\nonumber \\
& \cA'_{4,9} =  \onebb \oplus n \oplus \mathbf{b} 
\nonumber \\
& \cA'_{4,10} =  \onebb \oplus n \oplus \mathbf{d} 
\nonumber \\
& \cA'_{4,11} =  \onebb \oplus k \oplus \mathbf{c} 
\nonumber \\
& \cA'_{4,12} =  \onebb \oplus k \oplus \mathbf{d} 
\nonumber \\
& \cA'_{5,1} =  \onebb \oplus g 
\nonumber \\
& \cA'_{5,2} =  \onebb \oplus e 
\nonumber \\
& \cA'_{5,3} =  \onebb \oplus h 
\nonumber \\
& \cA'_{6,1} =  \onebb \oplus \mathbf{a} 
\nonumber \\
& \cA'_{6,2} =  \onebb \oplus \mathbf{b} 
\nonumber \\
& \cA'_{6,3} =  \onebb \oplus \mathbf{c} 
\nonumber \\
& \cA'_{6,4} =  \onebb \oplus \mathbf{d} 
\nonumber \\
& \cA'_{7,1} =  \onebb 
\end{align}
\endgroup

\subsection{$D_8$-symmetry}

The $D_8$-symmetry in the Morita-equivalent class of $\eD(D_8)$-symTO come from
the condensation of $\cA_{3,1}$.  The FCD automorphisms form a $\Z_2$
group generated by $(m ,l )(j ,n )(u ,q )(s ,v )(\mathbf{c} ,\mathbf{d})(e
,h)$.  The FAD automorphisms form a $\Z_2\times \Z_2$ group generated
by $ (m ,l )(j ,n )(u ,q )(s ,v)(\mathbf{c} ,\mathbf{d} )(e ,h )$ and $ (m ,j
)(l ,n)(u ,s )(q ,v )(\mathbf{a} ,\mathbf{b} )(e ,h) $.

The eleven gapped phases of $D_8$-symmetric systems 
can be divided into aSPT-classes, pSPT-classes, and SPT-classes:
\begingroup
\allowdisplaybreaks
\begin{align}
&\Big\{ \big[( \cA_{1,1})\big]^\text{SSB}_{4}\Big\},
\\
&\Big\{ \big[( \cA_{2,3},\cA_{2,4})\big]^\text{Sym}_{1},
\big[( \cA_{2,1}),( \cA_{2,2})\big]^\text{SSB}_{4}\Big\},
\nonumber\\
&\Big\{ \big[( \cA_{3,2},\cA_{3,3}),( \cA_{3,4},\cA_{3,5})\big]^\text{SSB}_{2},
\big[( \cA_{3,6})\big]^\text{SSB}_{2},
\big[( \cA_{3,1})\big]^\text{SSB}_{8}\Big\}.
\nonumber 
\end{align}
The SSB patterns and SPT orders of the  eleven gapped phases are given below:
\begin{align}
\big(  \cA_{2,3}\text{-},\cA_{2,4}\text{-phases} \big)_{1} 
&= D_8\text{-SPT phases}
\nonumber \\
\big(  \cA_{3,6}\text{-phase} \big)_{2} 
&= (D_8\to \Z_4)\text{-SSB phase}
\nonumber \\
\big(  \cA_{3,4}\text{-},\cA_{3,5}\text{-phases} \big)_{2} 
&= (D_8\to \Z_2\times \Z_2)\text{-SSB-SPT}
\nonumber \\
\big(  \cA_{3,2}\text{-},\cA_{3,3}\text{-phases} \big)_{2} 
&= (D_8\to \Z_2\times \Z_2)\text{-SSB-SPT}
\nonumber \\
\big(  \cA_{1,1}\text{-phase} \big)_{4} 
&= (D_8\to \Z_2)\text{-SSB phase}
\nonumber \\
\big(  \cA_{2,2}\text{-phase} \big)_{4} 
&= (D_8\to \Z_2)\text{-SSB phase}
\nonumber \\
\big(  \cA_{2,1}\text{-phase} \big)_{4} 
&= (D_8\to \Z_2)\text{-SSB phase}
\nonumber \\
\big(  \cA_{3,1}\text{-phase} \big)_{8} 
&= (D_8\to \Z_1)\text{-SSB phase}
\end{align}
The subscript of $(\cdots)$ is the ground state degeneracy.  As expected, there
are two $D_8$-SPT phases (including the trivial one), since $H^2(D_8, \mathrm{U}(1))=\Z_2$.  
There are two SSB patterns that breaks $D_8$-symmetry down to
$\Z_2\times Z_2$-symmetry.  Each SSB pattern has two phases differ by
$\Z_2\times Z_2$-SPT order. Thus those phases are SSB-SPT phases.

$D_8$-symmetric systems also has 33 classes of gapless phases, which are
divided into aSPT-classes, pSPT-classes, and SPT-classes:
\begin{align}
&\Big\{ \big[( \cA'_{1,1})\big]^\text{SSB}_{2}\Big\},
\\
&\Big\{ \big[( \cA'_{2,2},\cA'_{2,3}),( \cA'_{2,4},\cA'_{2,5})\big]^\text{SSB}_{2},
\big[( \cA'_{2,6})\big]^\text{SSB}_{2},
\big[( \cA'_{2,1})\big]^\text{SSB}_{4}\Big\},
\nonumber\\
&\Big\{ \big[( \cA'_{3,2},\cA'_{3,3}),( \cA'_{3,4},\cA'_{3,5})\big]^\text{Sym}_{1},
\big[( \cA'_{3,6})\big]^\text{SSB}_{2},
\big[( \cA'_{3,1})\big]^\text{SSB}_{4}\Big\},
\nonumber\\
&\Big\{ \big[( \cA'_{4,4},\cA'_{4,6}),( \cA'_{4,8},\cA'_{4,10})\big]^\text{Sym}_{1},
\big[( \cA'_{4,11},\cA'_{4,12})\big]^\text{Sym}_{1}, 
\nonumber\\ & \ \ 
\big[( \cA'_{4,3},\cA'_{4,5}),( \cA'_{4,7},\cA'_{4,9})\big]^\text{SSB}_{2},
\big[( \cA'_{4,1}),( \cA'_{4,2})\big]^\text{SSB}_{4}\Big\},
\nonumber\\
&\Big\{ \big[( \cA'_{5,2},\cA'_{5,3})\big]^\text{Sym}_{1},
\big[( \cA'_{5,1})\big]^\text{SSB}_{2}\Big\},
\nonumber\\
&\Big\{ \big[( \cA'_{6,3},\cA'_{6,4})\big]^\text{Sym}_{1},
\big[( \cA'_{6,1}),( \cA'_{6,2})\big]^\text{SSB}_{2}\Big\},
\ \
\Big\{ \big[( \cA'_{7,1})\big]^\text{Sym}_{1}\Big\}.
\nonumber 
\end{align}
\endgroup
Many of the gapless phases are connected by the FSD automorphisms of
$\eD(D_8)$, and are identical upto permutations of symmetry-charge
symmetry-twist sectors.  Thus those gapless phases have only seven types of low
energy properties, since the 33 non-Langrangian condensable algebras form seven
aSPT-classes.

We see that there are 15 classes of $D_8$ symmetric gapless phases.  Seven
pairs of them are connected by $D_8$-FCD automorphism, and the gapless
phases in each pair differ only by a SPT order.  Thus, $D_8$-symmetric systems
have seven pairs of gapless SPT phases.  Similarly, $D_8$-symmetric systems
also have four pairs of gapless SSB-SPT phases.

\subsection{$\cRep_{D_8}$-symmetry}

\begin{table*}[t] 
\caption{
The fusion rule of $\eD^{\frac{002}{224}}(D_8)$.
} \label{D8wNijk} \centerline{
\begin{tabular}{ |c||c|c|c|c|c|c|c|c|p{9mm}|p{9mm}|p{9mm}|p{9mm}|p{9mm}|p{9mm}|p{9mm}|p{9mm}|p{9mm}|p{9mm}|p{9mm}|p{9mm}|p{9mm}|p{9mm}|}
\hline 
& $\onebb$  & $a$  & $b$  & $c$  & $d$  & $e$  & $f$  & $g$  & $i$  & $j$  & $k$  & $l$  & $m$  & $n$  & $u$  & $v$  & $w$  & $x$  & $y$  & $z$  & $A$  & $B$ \\ 
\hline 
\hline 
$\onebb$  & $ \onebb$  & $ a$  & $ b$  & $ c$  & $ d$  & $ e$  & $ f$  & $ g$  & $ i$  & $ j$  & $ k$  & $ l$  & $ m$  & $ n$  & $ u$  & $ v$  & $ w$  & $ x$  & $ y$  & $ z$  & $ A$  & $ B$  \\ 
\hline 
$a$  & $ a$  & $ \onebb$  & $ c$  & $ b$  & $ e$  & $ d$  & $ g$  & $ f$  & $ i$  & $ j$  & $ k$  & $ l$  & $ w$  & $ x$  & $ u$  & $ v$  & $ m$  & $ n$  & $ A$  & $ B$  & $ y$  & $ z$  \\ 
\hline 
$b$  & $ b$  & $ c$  & $ \onebb$  & $ a$  & $ f$  & $ g$  & $ d$  & $ e$  & $ i$  & $ j$  & $ u$  & $ v$  & $ m$  & $ n$  & $ k$  & $ l$  & $ w$  & $ x$  & $ A$  & $ B$  & $ y$  & $ z$  \\ 
\hline 
$c$  & $ c$  & $ b$  & $ a$  & $ \onebb$  & $ g$  & $ f$  & $ e$  & $ d$  & $ i$  & $ j$  & $ u$  & $ v$  & $ w$  & $ x$  & $ k$  & $ l$  & $ m$  & $ n$  & $ y$  & $ z$  & $ A$  & $ B$  \\ 
\hline 
$d$  & $ d$  & $ e$  & $ f$  & $ g$  & $ \onebb$  & $ a$  & $ b$  & $ c$  & $ j$  & $ i$  & $ u$  & $ l$  & $ m$  & $ x$  & $ k$  & $ v$  & $ w$  & $ n$  & $ A$  & $ z$  & $ y$  & $ B$  \\ 
\hline 
$e$  & $ e$  & $ d$  & $ g$  & $ f$  & $ a$  & $ \onebb$  & $ c$  & $ b$  & $ j$  & $ i$  & $ u$  & $ l$  & $ w$  & $ n$  & $ k$  & $ v$  & $ m$  & $ x$  & $ y$  & $ B$  & $ A$  & $ z$  \\ 
\hline 
$f$  & $ f$  & $ g$  & $ d$  & $ e$  & $ b$  & $ c$  & $ \onebb$  & $ a$  & $ j$  & $ i$  & $ k$  & $ v$  & $ m$  & $ x$  & $ u$  & $ l$  & $ w$  & $ n$  & $ y$  & $ B$  & $ A$  & $ z$  \\ 
\hline 
$g$  & $ g$  & $ f$  & $ e$  & $ d$  & $ c$  & $ b$  & $ a$  & $ \onebb$  & $ j$  & $ i$  & $ k$  & $ v$  & $ w$  & $ n$  & $ u$  & $ l$  & $ m$  & $ x$  & $ A$  & $ z$  & $ y$  & $ B$  \\ 
\hline 
$i$  & $ i$  & $ i$  & $ i$  & $ i$  & $ j$  & $ j$  & $ j$  & $ j$  & $ \onebb \oplus a \oplus b \oplus c$  & $ d \oplus e \oplus f \oplus g$  & $ l \oplus v$  & $ k \oplus u$  & $ n \oplus x$  & $ m \oplus w$  & $ l \oplus v$  & $ k \oplus u$  & $ n \oplus x$  & $ m \oplus w$  & $ z \oplus B$  & $ y \oplus A$  & $ z \oplus B$  & $ y \oplus A$  \\ 
\hline 
$j$  & $ j$  & $ j$  & $ j$  & $ j$  & $ i$  & $ i$  & $ i$  & $ i$  & $ d \oplus e \oplus f \oplus g$  & $ \onebb \oplus a \oplus b \oplus c$  & $ l \oplus v$  & $ k \oplus u$  & $ n \oplus x$  & $ m \oplus w$  & $ l \oplus v$  & $ k \oplus u$  & $ n \oplus x$  & $ m \oplus w$  & $ z \oplus B$  & $ y \oplus A$  & $ z \oplus B$  & $ y \oplus A$  \\ 
\hline 
$k$  & $ k$  & $ k$  & $ u$  & $ u$  & $ u$  & $ u$  & $ k$  & $ k$  & $ l \oplus v$  & $ l \oplus v$  & $ \onebb \oplus a \oplus f \oplus g$  & $ i \oplus j$  & $ z \oplus B$  & $ y \oplus A$  & $ b \oplus c \oplus d \oplus e$  & $ i \oplus j$  & $ z \oplus B$  & $ y \oplus A$  & $ n \oplus x$  & $ m \oplus w$  & $ n \oplus x$  & $ m \oplus w$  \\ 
\hline 
$l$  & $ l$  & $ l$  & $ v$  & $ v$  & $ l$  & $ l$  & $ v$  & $ v$  & $ k \oplus u$  & $ k \oplus u$  & $ i \oplus j$  & $ \onebb \oplus a \oplus d \oplus e$  & $ y \oplus A$  & $ z \oplus B$  & $ i \oplus j$  & $ b \oplus c \oplus f \oplus g$  & $ y \oplus A$  & $ z \oplus B$  & $ m \oplus w$  & $ n \oplus x$  & $ m \oplus w$  & $ n \oplus x$  \\ 
\hline 
$m$  & $ m$  & $ w$  & $ m$  & $ w$  & $ m$  & $ w$  & $ m$  & $ w$  & $ n \oplus x$  & $ n \oplus x$  & $ z \oplus B$  & $ y \oplus A$  & $ \onebb \oplus b \oplus d \oplus f$  & $ i \oplus j$  & $ z \oplus B$  & $ y \oplus A$  & $ a \oplus c \oplus e \oplus g$  & $ i \oplus j$  & $ l \oplus v$  & $ k \oplus u$  & $ l \oplus v$  & $ k \oplus u$  \\ 
\hline 
$n$  & $ n$  & $ x$  & $ n$  & $ x$  & $ x$  & $ n$  & $ x$  & $ n$  & $ m \oplus w$  & $ m \oplus w$  & $ y \oplus A$  & $ z \oplus B$  & $ i \oplus j$  & $ \onebb \oplus b \oplus e \oplus g$  & $ y \oplus A$  & $ z \oplus B$  & $ i \oplus j$  & $ a \oplus c \oplus d \oplus f$  & $ k \oplus u$  & $ l \oplus v$  & $ k \oplus u$  & $ l \oplus v$  \\ 
\hline 
$u$  & $ u$  & $ u$  & $ k$  & $ k$  & $ k$  & $ k$  & $ u$  & $ u$  & $ l \oplus v$  & $ l \oplus v$  & $ b \oplus c \oplus d \oplus e$  & $ i \oplus j$  & $ z \oplus B$  & $ y \oplus A$  & $ \onebb \oplus a \oplus f \oplus g$  & $ i \oplus j$  & $ z \oplus B$  & $ y \oplus A$  & $ n \oplus x$  & $ m \oplus w$  & $ n \oplus x$  & $ m \oplus w$  \\ 
\hline 
$v$  & $ v$  & $ v$  & $ l$  & $ l$  & $ v$  & $ v$  & $ l$  & $ l$  & $ k \oplus u$  & $ k \oplus u$  & $ i \oplus j$  & $ b \oplus c \oplus f \oplus g$  & $ y \oplus A$  & $ z \oplus B$  & $ i \oplus j$  & $ \onebb \oplus a \oplus d \oplus e$  & $ y \oplus A$  & $ z \oplus B$  & $ m \oplus w$  & $ n \oplus x$  & $ m \oplus w$  & $ n \oplus x$  \\ 
\hline 
$w$  & $ w$  & $ m$  & $ w$  & $ m$  & $ w$  & $ m$  & $ w$  & $ m$  & $ n \oplus x$  & $ n \oplus x$  & $ z \oplus B$  & $ y \oplus A$  & $ a \oplus c \oplus e \oplus g$  & $ i \oplus j$  & $ z \oplus B$  & $ y \oplus A$  & $ \onebb \oplus b \oplus d \oplus f$  & $ i \oplus j$  & $ l \oplus v$  & $ k \oplus u$  & $ l \oplus v$  & $ k \oplus u$  \\ 
\hline 
$x$  & $ x$  & $ n$  & $ x$  & $ n$  & $ n$  & $ x$  & $ n$  & $ x$  & $ m \oplus w$  & $ m \oplus w$  & $ y \oplus A$  & $ z \oplus B$  & $ i \oplus j$  & $ a \oplus c \oplus d \oplus f$  & $ y \oplus A$  & $ z \oplus B$  & $ i \oplus j$  & $ \onebb \oplus b \oplus e \oplus g$  & $ k \oplus u$  & $ l \oplus v$  & $ k \oplus u$  & $ l \oplus v$  \\ 
\hline 
$y$  & $ y$  & $ A$  & $ A$  & $ y$  & $ A$  & $ y$  & $ y$  & $ A$  & $ z \oplus B$  & $ z \oplus B$  & $ n \oplus x$  & $ m \oplus w$  & $ l \oplus v$  & $ k \oplus u$  & $ n \oplus x$  & $ m \oplus w$  & $ l \oplus v$  & $ k \oplus u$  & $ \onebb \oplus c \oplus e \oplus f$  & $ i \oplus j$  & $ a \oplus b \oplus d \oplus g$  & $ i \oplus j$  \\ 
\hline 
$z$  & $ z$  & $ B$  & $ B$  & $ z$  & $ z$  & $ B$  & $ B$  & $ z$  & $ y \oplus A$  & $ y \oplus A$  & $ m \oplus w$  & $ n \oplus x$  & $ k \oplus u$  & $ l \oplus v$  & $ m \oplus w$  & $ n \oplus x$  & $ k \oplus u$  & $ l \oplus v$  & $ i \oplus j$  & $ \onebb \oplus c \oplus d \oplus g$  & $ i \oplus j$  & $ a \oplus b \oplus e \oplus f$  \\ 
\hline 
$A$  & $ A$  & $ y$  & $ y$  & $ A$  & $ y$  & $ A$  & $ A$  & $ y$  & $ z \oplus B$  & $ z \oplus B$  & $ n \oplus x$  & $ m \oplus w$  & $ l \oplus v$  & $ k \oplus u$  & $ n \oplus x$  & $ m \oplus w$  & $ l \oplus v$  & $ k \oplus u$  & $ a \oplus b \oplus d \oplus g$  & $ i \oplus j$  & $ \onebb \oplus c \oplus e \oplus f$  & $ i \oplus j$  \\ 
\hline 
$B$  & $ B$  & $ z$  & $ z$  & $ B$  & $ B$  & $ z$  & $ z$  & $ B$  & $ y \oplus A$  & $ y \oplus A$  & $ m \oplus w$  & $ n \oplus x$  & $ k \oplus u$  & $ l \oplus v$  & $ m \oplus w$  & $ n \oplus x$  & $ k \oplus u$  & $ l \oplus v$  & $ i \oplus j$  & $ a \oplus b \oplus e \oplus f$  & $ i \oplus j$  & $ \onebb \oplus c \oplus d \oplus g$  \\ 
\hline 
\end{tabular}
}
\end{table*}

$\cRep_{D_8}$-symmetry comes from the condensation of $\cA_{2,4}$, which
has been discussed extensively in
\Rfs{SS240401369,WS241215024,MG241220546}.  Here, we add some results related to
automorphisms and SPT orders.  The FCD automorphisms for $\cRep_{D_8}$-symmetry
form a trivial group.  The FAD automorphisms form a $S_3$ group:  
\begingroup
\allowdisplaybreaks
\begin{align}
& ( )
\nonumber \\
& (m ,j )(l ,n )(u ,s )(q ,v )(\mathbf{a} ,\mathbf{b} )(e ,h )
\nonumber \\
& (i ,m )(n ,k )(u ,r )(v ,t )(\mathbf{b} ,\mathbf{c} )(g ,e )
\nonumber \\
& (i ,m ,j )(l ,k ,n )(u ,s ,r )(q ,t ,v )(\mathbf{a} ,\mathbf{c} ,\mathbf{b} )(g ,e ,h )
\nonumber \\
& (i ,j ,m )(l ,n ,k )(u ,r ,s )(q ,v ,t )(\mathbf{a} ,\mathbf{b} ,\mathbf{c} )(g ,h ,e )
\nonumber \\
& (i ,j )(l ,k )(q ,t )(s ,r )(\mathbf{a} ,\mathbf{c} )(g ,h )
\nonumber \\
\end{align}
We note that the $\cRep_{D_8}$-symmetry charges are given by anyons in
$\cA_{2,4}$, which are $k,l,n, \vc d$.  The corresponding symmetry defects are
given by anyons in $\cA_{3,1}$, which are $\vc a, \vc b, g, i$.  The
FAD automorphisms permute the symmetry-charges $k,l,n$ and the
symmetry-defects $\vc a, \vc b, g, i$.

The eleven gapped phases of  $\cRep_{D_8}$-symmetric systems 
can be divided into aSPT-classes, pSPT-classes, and SPT-classes:
\begin{align}
&\Big\{ \big[( \cA_{1,1})\big]^\text{SSB}_{2}\Big\},
\\
&\Big\{ \big[( \cA_{2,1}),( \cA_{2,2}),( \cA_{2,3})\big]^\text{SSB}_{2},
\big[( \cA_{2,4})\big]^\text{SSB}_{5}\Big\},
\nonumber\\
&\Big\{ \big[( \cA_{3,1}),( \cA_{3,2}),( \cA_{3,4})\big]^\text{Sym}_{1},
\big[( \cA_{3,3}),( \cA_{3,5}),( \cA_{3,6})\big]^\text{SSB}_{4}\Big\}.
\nonumber
\end{align}
We see that three different $\cRep_{D_8}$-symmetric gapped phases,
$\cA_{3,1}$-, $\cA_{3,2}$-, $\cA_{3,4}$-phases, all have GSD = 1.  But the
three symmetric gapped phases are not connected by FCD automorphisms,
and thus their difference are beyond SPT order.  We like to remark that the
three gapped phases (\ie the three Lagrangian condensable algebras $\cA_{3,1}$,
$\cA_{3,2}$, $\cA_{3,4}$) are connected by FAD automorphisms that keep
$\cA_{2,4}$ invariant, where $\cA_{2,4}$ is the condensation that gives rise to
the $\cRep_{D_8}$-symmetry.

We also see that three different $\cRep_{D_8}$-SSB gapped phases, $\cA_{2,1}$-,
$\cA_{2,2}$-, $\cA_{2,3}$-phases, are connected by the
FAD automorphisms. The three phases have the condensation of
$\cRep_{D_8}$-charges $l,n,k$, respectively.  Thus $\cA_{2,1}$-, $\cA_{2,2}$-,
$\cA_{2,3}$-phases are different SSB patterns since they condense different
$\cRep_{D_8}$-charges.  We see that FAD automorphisms connect
different SSB patterns.  In contrast, FCD automorphisms never connect
different SSB patterns.  This is why we use FCD automorphisms, rather
than FAD automorphisms, to define SPT orders.

$\cRep_{D_8}$-symmetric systems also has 33 classes of gapless phases, which are
divided into aSPT-classes, pSPT-classes, and SPT-classes:
\begin{align}
&\Big\{ \big[( \cA'_{1,1})\big]^\text{Sym}_{1}\Big\},
\nonumber\\
&\Big\{ \big[( \cA'_{2,1}),( \cA'_{2,2}),( \cA'_{2,4})\big]^\text{Sym}_{1},
\big[( \cA'_{2,3}),( \cA'_{2,5}),( \cA'_{2,6})\big]^\text{SSB}_{2}\Big\},
\nonumber\\
&\Big\{ \big[( \cA'_{3,1}),( \cA'_{3,2}),( \cA'_{3,4})\big]^\text{Sym}_{1},
\big[( \cA'_{3,3}),( \cA'_{3,5}),( \cA'_{3,6})\big]^\text{SSB}_{2}\Big\},
\nonumber\\
&\Big\{ \big[( \cA'_{4,1}),( \cA'_{4,2}),( \cA'_{4,3}),( \cA'_{4,4}),( \cA'_{4,7}),( \cA'_{4,8})\big]^\text{Sym}_{1},
\nonumber\\&\ \ 
\big[( \cA'_{4,5}),( \cA'_{4,9}),( \cA'_{4,11})\big]^\text{SSB}_{2},
\big[( \cA'_{4,6}),( \cA'_{4,10}),( \cA'_{4,12})\big]^\text{SSB}_{3}\Big\},
\nonumber\\
&\Big\{ \big[( \cA'_{5,1}),( \cA'_{5,2}),( \cA'_{5,3})\big]^\text{Sym}_{1}\Big\},
\ \ 
\Big\{ \big[( \cA'_{7,1})\big]^\text{Sym}_{1}\Big\}.
\nonumber\\
&\Big\{ \big[( \cA'_{6,1}),( \cA'_{6,2}),( \cA'_{6,3})\big]^\text{Sym}_{1},
\big[( \cA'_{6,4})\big]^\text{SSB}_{2}\Big\},
\end{align}
\endgroup
We see that there are 20 classes of $\cRep_{D_8}$ symmetric gapless phases.
But there are no gapless $\cRep_{D_8}$ SPT order.

\subsection{$\tl\cR^{\eD(D_8)}_{\cA_{1,1}}$-symmetry} 

The Morita-equivalent class of $\eD(D_8)$-symTO has three symmetries.  The first
two are $D_8$ and $\cRep_{D_8}$ symmetries discussed above.  The third one
comes from $\cA_{1,1}$ condensation.  We will call this symmetry as
$\tl\cR^{\eD(D_8)}_{\cA_{1,1}}$-symmetry.  It is an anomalous symmetry, whose
gapped and gapless phases are summarized in Table \ref{Phases}.

\section{$H_8$-symmetry} 

In this section, we consider a symTO described by twisted quantum double
$\eD^{\frac{002}{224}}(D_8)$.  Since $H^3(D_8, \mathrm{U}(1))=\Z_2 \times \Z_2 \times
\Z_4$, the cohomology classes in $H^3(D_8, \mathrm{U}(1))$ are labeled by $a\in \Z_2$,
$b\in \Z_2$, and $c\in \Z_2$.  Here we denote a cohomology class in
$H^3(D_8, \mathrm{U}(1))$ by $\frac{abc}{224}$.  The  twisted quantum double
$\eD^{\frac{002}{224}}(D_8)$ is twisted by  cohomology class $\frac{002}{224}
\in H^3(D_8, \mathrm{U}(1))$.
The 22 anyons in $\eD^{\frac{002}{224}}(D_8)$ are given by
\\[2mm]
\centerline{
\begin{tabular}{|c|c|c|c|c|c|c|c|c|c|c|c|c|c|c|c|c|c|c|c|c|c|c|}
\hline
& $\onebb$ & $a$ & $b$ & $c$ & $d$ & $e$ & $f$ & $g$ & $i$ & $j$ & $k$ & $l$ & $m$ & $n$ & $u$ & $v$ & $w$ & $x$ & $y$ & $z$ & $A$ & $B$ 
\\ 
\hline
$s$: & $0$ & $0$ & $0$ & $0$ & $\frac{1}{2}$ & $\frac{1}{2}$ & $\frac{1}{2}$ & $\frac{1}{2}$ & $0$ & $0$ & $0$ & $0$ & $0$ & $0$ & $\frac{1}{2}$ & $\frac{1}{2}$ & $\frac{1}{2}$ & $\frac{1}{2}$ & $\frac{1}{8}$ & $\frac{3}{8}$ & $\frac{5}{8}$ & $\frac{7}{8}$ \\ 
\hline
$d$: & $1$ & $1$ & $1$ & $1$ & $1$ & $1$ & $1$ & $1$ & $2$ & $2$ & $2$ & $2$ & $2$ & $2$ & $2$ & $2$ & $2$ & $2$ & $2$ & $2$ & $2$ & $2$ \\ 
\hline
\end{tabular}
}
\vskip 2mm \noindent
Their fusion role is given in Table \ref{D8wNijk}.  The FSD automorphism
group of $\eD^{\frac{002}{224}}(D_8)$ quantum double is $\Z_2\times\Z_2\times
\Z_2$ generated by:
\begingroup
\allowdisplaybreaks
\begin{align}
\nonumber \\
& (d ,g) (e ,f) (k ,l) (m ,n) (u ,v) (w ,x)
\nonumber \\
& (a ,b) (e ,f) (k ,n) (l ,m) (u ,x) (v ,w)
\nonumber \\
& (i ,j)
\end{align}
$\eD^{\frac{002}{224}}(D_8)$ has six Lagrangian condensable algebras
(in two aSPT-classes)
\begin{align}
& \cA_{1,1} =  \onebb \oplus i \oplus a \oplus k \oplus l
\nonumber \\
& \cA_{1,2} =  \onebb \oplus i \oplus b \oplus m \oplus n
\nonumber \\
& \cA_{1,3} =  \onebb \oplus j \oplus a \oplus k \oplus l
\nonumber \\
& \cA_{1,4} =  \onebb \oplus j \oplus b \oplus m \oplus n
\nonumber \\
& \cA_{2,1} =  \onebb \oplus 2i \oplus a \oplus b \oplus c
\nonumber \\
& \cA_{2,2} =  \onebb \oplus 2j \oplus a \oplus b \oplus c
\end{align}
and 15 non-Lagrangian condensable algebras
(in seven aSPT-classes):
\begin{align}
& \cA'_{1,1} =  \onebb \oplus a \oplus k 
\nonumber \\
& \cA'_{1,2} =  \onebb \oplus a \oplus l 
\nonumber \\
& \cA'_{1,3} =  \onebb \oplus b \oplus m 
\nonumber \\
& \cA'_{1,4} =  \onebb \oplus b \oplus n 
\nonumber \\
& \cA'_{2,1} =  \onebb \oplus a \oplus b \oplus c 
\nonumber \\
& \cA'_{3,1} =  \onebb \oplus i \oplus c 
\nonumber \\
& \cA'_{3,2} =  \onebb \oplus j \oplus c 
\nonumber \\
& \cA'_{4,1} =  \onebb \oplus i \oplus a 
\nonumber \\
& \cA'_{4,2} =  \onebb \oplus i \oplus b 
\nonumber \\
& \cA'_{4,3} =  \onebb \oplus j \oplus a 
\nonumber \\
& \cA'_{4,4} =  \onebb \oplus j \oplus b 
\nonumber \\
& \cA'_{5,1} =  \onebb \oplus c 
\nonumber \\
& \cA'_{6,1} =  \onebb \oplus a 
\nonumber \\
& \cA'_{6,2} =  \onebb \oplus b 
\nonumber \\
& \cA'_{7,1} =  \onebb 
\end{align}

The $H_8$-symmetry is the
$\tl\cR^{\eD^{\frac{002}{224}}(D_8)}_{\cA_{1,1}}$-symmetry, coming from the
$\cA_{1,1}$ condensation.  The FCD automorphism group is trivial.  The
FAD automorphism group is $\Z_2$ generated by $ (d ,g )(e ,f )(k ,l
)(m ,n )(u ,v )(w ,x )$.

The six gapped phases of $H_8$-symmetric systems are divided into
aSPT-classes, pSPT-classes, and SPT-classes:
\begin{align}
&\Big\{ \big[ ( \cA_{1,4})\big]^\text{Sym}_{1},
\big[ ( \cA_{1,2})\big]^\text{SSB}_{2},
\big[ ( \cA_{1,3})\big]^\text{SSB}_{4},
\big[ ( \cA_{1,1})\big]^\text{SSB}_{5}\Big\},
\nonumber\\
&\Big\{ \big[ ( \cA_{2,2})\big]^\text{SSB}_{2},
\big[ ( \cA_{2,1})\big]^\text{SSB}_{4}\Big\}.
\end{align}
Since one of the gapped phases has non-degenerate ground state, $H_8$-symmetry
is anomaly-free.  Is this anomaly-free symmetry described by a group?  Is this
anomaly-free symmetry Morita-equivalent to a symmetry described by a group?  It
turns out that, unlike $\cRep_{S_3}$-symmetry that is Morita-equivalent to
group-like $S_3$-symmetry, $H_8$-symmetry is not Morita-equivalent to any
symmetry described by a group.  It is a symmetry described by Hopf algebra
$H_8$ \cite{CS231019867}.  This is why we call it $H_8$-symmetry.  We also see
that there is no $H_8$-SPT phases nor $H_8$-SSB-SPT phases.

The $H_8$-symmetric systems have 15 classes of gapless phases, which are
divided into aSPT-classes, pSPT-classes, and SPT-classes:
\begin{align}
&\Big\{ \big[ ( \cA'_{1,3}),( \cA'_{1,4})\big]^\text{Sym}_{1},
\big[ ( \cA'_{1,1}),( \cA'_{1,2})\big]^\text{SSB}_{3}\Big\},
\nonumber\\
&\Big\{ \big[ ( \cA'_{2,1})\big]^\text{SSB}_{2}\Big\},
\ \
\Big\{ \big[ ( \cA'_{3,2})\big]^\text{Sym}_{1},
\big[ ( \cA'_{3,1})\big]^\text{SSB}_{2}\Big\},
\nonumber\\
&\Big\{ \big[ ( \cA'_{4,4})\big]^\text{Sym}_{1},
\big[ ( \cA'_{4,2})\big]^\text{SSB}_{2},
\big[ ( \cA'_{4,3})\big]^\text{SSB}_{2},
\big[ ( \cA'_{4,1})\big]^\text{SSB}_{3}\Big\},
\nonumber\\
&\Big\{ \big[ ( \cA'_{5,1})\big]^\text{Sym}_{1}\Big\},
\ \
\Big\{ \big[ ( \cA'_{6,2})\big]^\text{Sym}_{1},
\big[ ( \cA'_{6,1})\big]^\text{SSB}_{2}\Big\},
\nonumber\\
&\Big\{ \big[ ( \cA'_{7,1})\big]^\text{Sym}_{1}\Big\},
\end{align}
\endgroup
There is no gapless $H_8$-SPT order.

\section{Symmetries with $\eD(S_3) \times \eD(S_3)$-symTO} 

The $\eD(S_3) \times \eD(S_3)$-symTO has anyons given in Section \ref{S3sym}.
The symTO has 28 Lagrangian condensable algebras, which are divided into
six aSPT-classes:
\begin{widetext}
\begingroup
\allowdisplaybreaks
\begin{align}
& \cA_{1,1} =  (\onebb,\onebb) \oplus (\onebb,\one) \oplus 2(\two,\two)\oplus (c,\onebb) \oplus (c,\one) \oplus 2(c,\two)\oplus (\two,\one) \oplus 2(\onebb,\two)\oplus (\two,\onebb) 
\nonumber \\
& \cA_{1,2} =  (\onebb,\onebb) \oplus (\onebb,\one) \oplus 2(\two,b)\oplus (c,\onebb) \oplus (c,\one) \oplus 2(c,b)\oplus (\two,\one) \oplus 2(\onebb,b)\oplus (\two,\onebb) 
\nonumber \\
& \cA_{1,3} =  (\onebb,\onebb) \oplus (\onebb,\one) \oplus 2(b,\two)\oplus (c,\onebb) \oplus (c,\one) \oplus 2(c,\two)\oplus (b,\one) \oplus (b,\onebb) \oplus 2(\onebb,\two)
\nonumber \\
& \cA_{1,4} =  (\onebb,\onebb) \oplus (\onebb,\one) \oplus 2(b,b)\oplus (c,\onebb) \oplus (c,\one) \oplus 2(c,b)\oplus (b,\one) \oplus (b,\onebb) \oplus 2(\onebb,b)
\nonumber \\
& \cA_{1,5} =  (\onebb,\onebb) \oplus 2(\two,\two)\oplus 2(\two,c)\oplus (\one,c) \oplus (\one,\onebb) \oplus (\onebb,c) \oplus (\one,\two) \oplus (\onebb,\two) \oplus 2(\two,\onebb)
\nonumber \\
& \cA_{1,6} =  (\onebb,\onebb) \oplus 2(\two,b)\oplus 2(\two,c)\oplus (\one,c) \oplus (\one,\onebb) \oplus (\onebb,c) \oplus (\onebb,b) \oplus 2(\two,\onebb)\oplus (\one,b) 
\nonumber \\
& \cA_{1,7} =  (\onebb,\onebb) \oplus 2(b,\two)\oplus 2(b,c)\oplus (\one,c) \oplus (\one,\onebb) \oplus (\onebb,c) \oplus (\one,\two) \oplus 2(b,\onebb)\oplus (\onebb,\two) 
\nonumber \\
& \cA_{1,8} =  (\onebb,\onebb) \oplus 2(b,b)\oplus 2(b,c)\oplus (\one,c) \oplus (\one,\onebb) \oplus (\onebb,c) \oplus 2(b,\onebb)\oplus (\onebb,b) \oplus (\one,b) 
\nonumber \\
& \cA_{2,1} =  (\onebb,\onebb) \oplus (\onebb,\one) \oplus 2(\two,\two)\oplus 2(b,b)\oplus (\one,\one) \oplus (\one,\onebb) \oplus 2(b'',b')\oplus 2(b',b'')
\nonumber \\
& \cA_{2,2} =  (\onebb,\onebb) \oplus (\onebb,\one) \oplus 2(\two,b)\oplus 2(b,\two)\oplus (\one,\one) \oplus (\one,\onebb) \oplus 2(b'',b')\oplus 2(b',b'')
\nonumber \\
& \cA_{3,1} =  (\onebb,\onebb) \oplus (\onebb,\one) \oplus 4(\two,\two)\oplus (\one,\one) \oplus (\one,\onebb) \oplus 2(\one,\two)\oplus 2(\two,\one)\oplus 2(\onebb,\two)\oplus 2(\two,\onebb)
\nonumber \\
& \cA_{3,2} =  (\onebb,\onebb) \oplus (\onebb,\one) \oplus 4(\two,b)\oplus (\one,\one) \oplus (\one,\onebb) \oplus 2(\two,\one)\oplus 2(\onebb,b)\oplus 2(\two,\onebb)\oplus 2(\one,b)
\nonumber \\
& \cA_{3,3} =  (\onebb,\onebb) \oplus (\onebb,\one) \oplus 4(b,\two)\oplus (\one,\one) \oplus (\one,\onebb) \oplus 2(\one,\two)\oplus 2(b,\one)\oplus 2(b,\onebb)\oplus 2(\onebb,\two)
\nonumber \\
& \cA_{3,4} =  (\onebb,\onebb) \oplus (\onebb,\one) \oplus 4(b,b)\oplus (\one,\one) \oplus (\one,\onebb) \oplus 2(b,\one)\oplus 2(b,\onebb)\oplus 2(\onebb,b)\oplus 2(\one,b)
\nonumber \\
& \cA_{4,1} =  (\onebb,\onebb) \oplus (c,c) \oplus (\two,\two) \oplus (\two,c) \oplus (c,\onebb) \oplus (c,\two) \oplus (\onebb,c) \oplus (\onebb,\two) \oplus (\two,\onebb) 
\nonumber \\
& \cA_{4,2} =  (\onebb,\onebb) \oplus (c,c) \oplus (\two,b) \oplus (\two,c) \oplus (c,\onebb) \oplus (c,b) \oplus (\onebb,c) \oplus (\onebb,b) \oplus (\two,\onebb) 
\nonumber \\
& \cA_{4,3} =  (\onebb,\onebb) \oplus (c,c) \oplus (b,\two) \oplus (b,c) \oplus (c,\onebb) \oplus (c,\two) \oplus (\onebb,c) \oplus (b,\onebb) \oplus (\onebb,\two) 
\nonumber \\
& \cA_{4,4} =  (\onebb,\onebb) \oplus (c,c) \oplus (b,b) \oplus (b,c) \oplus (c,\onebb) \oplus (c,b) \oplus (\onebb,c) \oplus (b,\onebb) \oplus (\onebb,b) 
\nonumber \\
& \cA_{4,5} =  (\onebb,\onebb) \oplus (c',c') \oplus (\two,\two) \oplus (\two,c) \oplus (\one,c) \oplus (c,\one) \oplus (c,\two) \oplus (\onebb,\two) \oplus (\two,\onebb) 
\nonumber \\
& \cA_{4,6} =  (\onebb,\onebb) \oplus (c',c') \oplus (\two,b) \oplus (\two,c) \oplus (\one,c) \oplus (c,\one) \oplus (c,b) \oplus (\onebb,b) \oplus (\two,\onebb) 
\nonumber \\
& \cA_{4,7} =  (\onebb,\onebb) \oplus (c',c') \oplus (b,\two) \oplus (b,c) \oplus (\one,c) \oplus (c,\one) \oplus (c,\two) \oplus (b,\onebb) \oplus (\onebb,\two) 
\nonumber \\
& \cA_{4,8} =  (\onebb,\onebb) \oplus (c',c') \oplus (b,b) \oplus (b,c) \oplus (\one,c) \oplus (c,\one) \oplus (c,b) \oplus (b,\onebb) \oplus (\onebb,b) 
\nonumber \\
& \cA_{5,1} =  (\onebb,\onebb) \oplus (c,c) \oplus (c',c') \oplus (\two,\two) \oplus (b,b) \oplus (\one,\one) \oplus (b'',b') \oplus (b',b'') 
\nonumber \\
& \cA_{5,2} =  (\onebb,\onebb) \oplus (c,c) \oplus (c',c') \oplus (\two,b) \oplus (b,\two) \oplus (\one,\one) \oplus (b'',b') \oplus (b',b'') 
\nonumber \\
& \cA_{6,1} =  (\onebb,\onebb) \oplus (c,c) \oplus (c',c') \oplus 2(\two,\two)\oplus (\one,\one) \oplus (\one,\two) \oplus (\two,\one) \oplus (\onebb,\two) \oplus (\two,\onebb) 
\nonumber \\
& \cA_{6,2} =  (\onebb,\onebb) \oplus (c,c) \oplus (c',c') \oplus 2(\two,b)\oplus (\one,\one) \oplus (\two,\one) \oplus (\onebb,b) \oplus (\two,\onebb) \oplus (\one,b) 
\nonumber \\
& \cA_{6,3} =  (\onebb,\onebb) \oplus (c,c) \oplus (c',c') \oplus 2(b,\two)\oplus (\one,\one) \oplus (\one,\two) \oplus (b,\one) \oplus (b,\onebb) \oplus (\onebb,\two) 
\nonumber \\
& \cA_{6,4} =  (\onebb,\onebb) \oplus (c,c) \oplus (c',c') \oplus 2(b,b)\oplus (\one,\one) \oplus (b,\one) \oplus (b,\onebb) \oplus (\onebb,b) \oplus (\one,b) 
\nonumber \\
\end{align}
\endgroup
\end{widetext}
Thus the Morita-equivalent class of $\eD(S_3) \times \eD(S_3)$-symTO has six
symmetries: 
\begin{enumerate}
\item
$S_3\times S_3$-symmetry from $\cA_{3,1}$ condensation.
\item
$\cRep_{S_3}\times \cRep_{S_3}$-symmetry from $\cA_{4,4}$ condensation.
\item
$\cRep_{S_3}\times S_3$-symmetry from $\cA_{1,3}$ condensation. 
\item
$\tl
\cR^{\eD(S_3\times S_3)}_{\cA_{2,1}}$-symmetry from $\cA_{2,1}$ condensation.
\item
$\tl \cR^{\eD(S_3\times S_3)}_{\cA_{5,1}}$-symmetry from $\cA_{5,1}$
condensation. 
\item
$\tl \cR^{\eD(S_3\times S_3)}_{\cA_{6,1}}$-symmetry from
$\cA_{6,1}$ condensation.  
\end{enumerate}
We find all the six symmetries to be anomaly-free.
Their gapped phases are summarized in Table \ref{PhasesS3xS3}.

\section{Haagerup symTO} 

In this section, we consider a symTO described by Haagerup MTC
$\eM_\text{Haag}$. For an earlier detailed discussion, see \Rf{BS241019040B}.
Here, we add some results related to automorphisms and SPT orders.  The 12
anyons in $\eM_\text{Haag}$ are given by
\begin{widetext}
\begin{tabular}{|c|c|c|c|c|c|c|c|c|c|c|c|c|}
\hline
$\eM_{1}$ anyon& $\onebb$ 
& $a$ 
& $b$ 
& $c$ 
& $d$ 
& $e$ 
& $f$ 
& $g$ 
& $h$ 
& $i$ 
& $j$ 
& $k$ 
\\ 
\hline
$s$: & $0$
& $\frac{2}{13}$
& $\frac{5}{13}$
& $\frac{6}{13}$
& $\frac{7}{13}$
& $\frac{8}{13}$
& $\frac{11}{13}$
& $0$
& $0$
& $0$
& $\frac{1}{3}$
& $\frac{2}{3}$
\\ 
\hline
$d$: & $1$
& $\frac{9+3\sqrt{13}}{2}$
& $\frac{9+3\sqrt{13}}{2}$
& $\frac{9+3\sqrt{13}}{2}$
& $\frac{9+3\sqrt{13}}{2}$
& $\frac{9+3\sqrt{13}}{2}$
& $\frac{9+3\sqrt{13}}{2}$
& $\frac{11+3\sqrt{13}}{2}$
& $\frac{13+3\sqrt{13}}{2}$
& $\frac{13+3\sqrt{13}}{2}$
& $\frac{13+3\sqrt{13}}{2}$
& $\frac{13+3\sqrt{13}}{2}$
\\ 
\hline
\end{tabular}
\end{widetext}
The FSD automorphism
group of $\eM_\text{Haag}$ is 
$\Z_2$, generated by:
\begin{align}
(h ,i).
\end{align}
$\eM_\text{Haag}$ has three Lagrangian condensable algebras
(in two aSPT-classes)
\begin{align}
& \cA_{1,1} =  \onebb \oplus g \oplus h \oplus i 
\nonumber \\
& \cA_{2,1} =  \onebb \oplus g \oplus 2h 
\nonumber \\
& \cA_{2,2} =  \onebb \oplus g \oplus 2i 
\nonumber \\
\end{align}
and two non-Lagrangian condensable algebras
(in two aSPT-classes):
\begin{align}
& \cA'_{1,1} =  \onebb \oplus g 
\nonumber \\
& \cA'_{2,1} =  \onebb 
\nonumber \\
\end{align}

\subsection{$\tl\cR^{\eM_\mathrm{Haag}}_{\cA_{1,1}}$ symmetry}

The $\tl\cR^{\eM_\mathrm{Haag}}_{\cA_{1,1}}$-symmetry comes from the
$\cA_\text{char}=\cA_{1,1}$ condensation.  The FCD automorphism is trivial, and
FAD and FSD automorphisms are $\Z_2$ generated by $(h,i)$.

The two gapped phase of $\tl\cR^{\eM_\mathrm{Haag}}_{\cA_{1,1}}$-symmetric
systems can be divided into aSPT-classes, pSPT-classes, and SPT-classes:
\begin{align}
\Big\{ \big[ ( \cA_{1,1})\big]^\text{SSB}_{4}\Big\},\ \ \ \ 
\Big\{ \big[ ( \cA_{2,1}),( \cA_{2,2})\big]^\text{SSB}_{4}\Big\}.
\end{align}
Since none of the gapped phases has non-degenerate ground state,
$\tl\cR^{\eM_\mathrm{Haag}}_{\cA_{1,1}}$-symmetry is anomalous.  The
$\tl\cR^{\eM_\mathrm{Haag}}_{\cA_{1,1}}$-symmetric systems have two classes of
gapless phases, which are divided into aSPT-classes, pSPT-classes, and
SPT-classes:
\begin{align}
\Big\{ \big[ ( \cA'_{1,1})\big]^\text{SSB}_{2}\Big\},\ \ \ \ 
\Big\{ \big[ ( \cA'_{2,1})\big]^\text{Sym}_{1}\Big\}.
\end{align}
There is no gapless $\tl\cR^{\eM_\mathrm{Haag}}_{\cA_{1,1}}$-SPT order.

\subsection{$\tl\cR^{\eM_\mathrm{Haag}}_{\cA_{2,1}}$ symmetry}

The $\tl\cR^{\eM_\mathrm{Haag}}_{\cA_{2,1}}$-symmetry comes from the
$\cA_\text{char}=\cA_{2,1}$ condensation.  The FCD and FAD automorphisms are
trivial, and FSD automorphism is $\Z_2$ generated by $(h,i)$.

The two gapped phase of $\tl\cR^{\eM_\mathrm{Haag}}_{\cA_{2,1}}$-symmetric
systems can be divided into aSPT-classes, pSPT-classes, and SPT-classes:
\begin{align}
\Big\{ \big[ ( \cA_{1,1})\big]^\text{SSB}_{4}\Big\},\ \ \
\Big\{ \big[ ( \cA_{2,2})\big]^\text{SSB}_{2},
\big[ ( \cA_{2,1})\big]^\text{SSB}_{6}\Big\}.
\end{align}
Since none of the gapped phases has non-degenerate ground state, 
$\tl\cR^{\eM_\mathrm{Haag}}_{\cA_{2,1}}$-symmetry
is anomalous.  
The $\tl\cR^{\eM_\mathrm{Haag}}_{\cA_{1,1}}$-symmetric systems have two classes of gapless phases, which are
divided into aSPT-classes, pSPT-classes, and SPT-classes:
\begin{align}
\Big\{ \big[ ( \cA'_{1,1})\big]^\text{SSB}_{2}\Big\}, \ \ \ \
\Big\{ \big[ ( \cA'_{2,1})\big]^\text{Sym}_{1}\Big\}.
\end{align}
There is no gapless $\tl\cR^{\eM_\mathrm{Haag}}_{\cA_{2,1}}$-SPT orders.

\

\section{Twisted Haagerup symTO} 

The twisted Haagerup MTC $\eM_\text{tHaag}$ also described a symTO.  The 12
anyons in $\eM_\text{tHaag}$ are given by
\begin{widetext}

\begin{tabular}{|c|c|c|c|c|c|c|c|c|c|c|c|c|}
\hline
$\eM_{1}$ anyon& $\onebb$ 
& $a$ 
& $b$ 
& $c$ 
& $d$ 
& $e$ 
& $f$ 
& $g$ 
& $h$ 
& $i$ 
& $j$ 
& $k$ 
\\ 
\hline
$s$: & $0$
& $\frac{2}{13}$
& $\frac{5}{13}$
& $\frac{6}{13}$
& $\frac{7}{13}$
& $\frac{8}{13}$
& $\frac{11}{13}$
& $0$
& $0$
& $\frac{1}{9}$
& $\frac{4}{9}$
& $\frac{7}{9}$
\\ 
\hline
$d$: & $1$
& $\frac{9+3\sqrt{13}}{2}$
& $\frac{9+3\sqrt{13}}{2}$
& $\frac{9+3\sqrt{13}}{2}$
& $\frac{9+3\sqrt{13}}{2}$
& $\frac{9+3\sqrt{13}}{2}$
& $\frac{9+3\sqrt{13}}{2}$
& $\frac{11+3\sqrt{13}}{2}$
& $\frac{13+3\sqrt{13}}{2}$
& $\frac{13+3\sqrt{13}}{2}$
& $\frac{13+3\sqrt{13}}{2}$
& $\frac{13+3\sqrt{13}}{2}$
\\ 
\hline
\end{tabular}
\end{widetext}
The FSD automorphism
group of $\eM_\text{tHaag}$ is 
trivial.
$\eM_\text{Haag}$ has one Lagrangian condensable algebras
\begin{align}
 \cA_{1,1} =  \onebb \oplus 2h \oplus g .
\end{align}
and two non-Lagrangian condensable algebras
(in two aSPT-classes):
\begin{align}
& \cA'_{1,1} =  \onebb \oplus g ,
\nonumber \\
& \cA'_{2,1} =  \onebb .
\end{align}

The $\tl\cR^{\eM_\mathrm{tHaag}}$-symmetry comes from the
$\cA_\text{char}=\cA_{1,1}$ condensation.  
The only gapped phase of $\tl\cR^{\eM_\mathrm{tHaag}}$-symmetric systems 
is
\begin{align}
\Big\{ \big[ ( \cA_{1,1})\big]^\text{SSB}_{6}\Big\}.
\end{align}
Since none of the gapped phases has non-degenerate ground state, $\tl\cR^{\eM_\mathrm{tHaag}}$-symmetry
is anomalous.  

The $\tl\cR^{\eM_\mathrm{tHaag}}$-symmetric systems have two classes of gapless phases, which are
divided into aSPT-classes, pSPT-classes, and SPT-classes:
\begin{align}
\Big\{ \big[ ( \cA'_{1,1})\big]^\text{SSB}_{2}\Big\},\ \ \ \ \
\Big\{ \big[ ( \cA'_{2,1})\big]^\text{Sym}_{1}\Big\}.
\end{align}
There is no gapless $\tl\cR^{\eM_\mathrm{tHaag}}$ SPT order.

\begingroup
\allowdisplaybreaks

\section{Gaplesss phases of $\eD(S_3)$-symTO}
\label{S3Zs}

In this section, we list gapless phases of $\eD(S_3)$-symTO.  We note that in
the presence of symmetry, we can consider the partition function on a ring in a
sector with a particular symmetry charge and with a particular symmetry-twisted
periodic condition.  Thus we can have several partition functions labeled by a
pair of indices: the symmetry charge and the symmetry twist.  Such pairs of
indices happen to label the anyons in the bulk symTO. Thus a gapless phase is
described by symTO resolved multi-component partition functions, labeled by the
anyons in the symTO \cite{JW190513279,CW220506244}.

We will list those symTO resolved multi-component partition
functions for $\eD(S_3)$-symTO.  We will group the gaplesss phases by their
associated non-Lagrangian condensable algebras.  For each non-Lagrangian
condensable algebra, there can many gapless boundary phases.  Here, we list all
the gapless boundary phases with small central charges, for each
non-Lagrangian condensable algebra.  The gapless boundary phases with larger
central charges are not included.

The various terms in each component of the partition function are conformal
characters of the minimal models.  For example, the superscript $m4 \times
\overline{m4}$ indicates that both the left and right moving chiral conformal
characters are picked from the same (4,3) minimal model.  $m5$ corresponds to
(5,4) minimal model, \etc.  The expression $\chi^{m4\times
\overline{m4}}_{a,h_a;\, b, -h_b}$ is a short-hand notation for the product of
the left moving chiral conformal character associated with the primary operator
labeled by $a$ (an arbitrary indexing convention) with conformal weight
$(h_a,0)$, and the right moving chiral conformal character associated with the
primary operator labeled $b$ with conformal weight $(0,h_b)$. 

The appearance of the  chiral conformal character of identity primary field
(colored blue), ${\color{blue} \chi^{m4 \times \overline{m4}}_{\onebb,0;
\onebb,0}}$, in the partition function $Z_i$ in indicates the condensation of
the $i^\text{th}$-anyon.  This allows us to identify the non-Lagrangian condensable algebra that characterized
the gapless phase.

For a fixed non-Lagrangian condensable algebra, the gapless phases with the
least number of symmetric relevant operators is most stable.  The  symmetric
relevant operators (colored red) can be read off from the conformal character
in the $Z_\onebb$ component.

\subsection{Condensation $\cA'_{1,1} = \onebb \oplus \two $}

The  gapless boundary phases with central charge $(c,\bar
c) \leq (\frac{7}{10}, \frac{7}{10})$.

\begin{align}
\label{Zm4S3A11}
 Z_{\onebb}^{\eD(S_3)} &={\color{blue} \chi^{m4 \times \overline{m4}}_{\onebb,0; \onebb,0}} 
+  {\color{red} \chi^{m4 \times \overline{m4}}_{\mathbf{b},\frac{1}{2}; \mathbf{b},-\frac{1}{2}} }
 \nonumber \\ 
Z_{\one}^{\eD(S_3)} &= \chi^{m4 \times \overline{m4}}_{\mathbf{a},\frac{1}{16}; \mathbf{a},-\frac{1}{16}} 
 \nonumber \\ 
Z_{\mathbf{2}}^{\eD(S_3)} &={\color{blue} \chi^{m4 \times \overline{m4}}_{\onebb,0; \onebb,0}} 
+  \chi^{m4 \times \overline{m4}}_{\mathbf{a},\frac{1}{16}; \mathbf{a},-\frac{1}{16}} 
+  \chi^{m4 \times \overline{m4}}_{\mathbf{b},\frac{1}{2}; \mathbf{b},-\frac{1}{2}} 
 \nonumber \\ 
Z_{{b}}^{\eD(S_3)} &=0
 \nonumber \\ 
Z_{{b'}}^{\eD(S_3)} &=0
 \nonumber \\ 
Z_{{b''}}^{\eD(S_3)} &=0
 \nonumber \\ 
Z_{{c}}^{\eD(S_3)} &= \chi^{m4 \times \overline{m4}}_{\mathbf{a},\frac{1}{16}; \mathbf{a},-\frac{1}{16}} 
 \nonumber \\ 
Z_{{c'}}^{\eD(S_3)} &= \chi^{m4 \times \overline{m4}}_{\onebb,0; \mathbf{b},-\frac{1}{2}} 
+  \chi^{m4 \times \overline{m4}}_{\mathbf{b},\frac{1}{2}; \onebb,0} 
 \end{align}

\begin{widetext}
\begin{align}
\label{Zm4S3A11a}
 Z_{\onebb}^{\eD(S_3)} &={\color{blue} \chi^{m5 \times \overline{m5}}_{\onebb,0; \onebb,0}} 
+  {\color{red} \chi^{m5 \times \overline{m5}}_{\mathbf{a},\frac{1}{10}; \mathbf{a},-\frac{1}{10}} } 
+  {\color{red} \chi^{m5 \times \overline{m5}}_{\mathbf{b},\frac{3}{5}; \mathbf{b},-\frac{3}{5}} }
+  \chi^{m5 \times \overline{m5}}_{\mathbf{c},\frac{3}{2}; \mathbf{c},-\frac{3}{2}} 
 \nonumber \\ 
Z_{\one}^{\eD(S_3)} &= \chi^{m5 \times \overline{m5}}_{\mathbf{d},\frac{7}{16}; \mathbf{d},-\frac{7}{16}} 
+  \chi^{m5 \times \overline{m5}}_{\mathbf{e},\frac{3}{80}; \mathbf{e},-\frac{3}{80}} 
 \nonumber \\ 
Z_{\mathbf{2}}^{\eD(S_3)} &={\color{blue} \chi^{m5 \times \overline{m5}}_{\onebb,0; \onebb,0}} 
+  \chi^{m5 \times \overline{m5}}_{\mathbf{a},\frac{1}{10}; \mathbf{a},-\frac{1}{10}} 
+  \chi^{m5 \times \overline{m5}}_{\mathbf{b},\frac{3}{5}; \mathbf{b},-\frac{3}{5}} 
+  \chi^{m5 \times \overline{m5}}_{\mathbf{c},\frac{3}{2}; \mathbf{c},-\frac{3}{2}} 
+  \chi^{m5 \times \overline{m5}}_{\mathbf{d},\frac{7}{16}; \mathbf{d},-\frac{7}{16}} 
+  \chi^{m5 \times \overline{m5}}_{\mathbf{e},\frac{3}{80}; \mathbf{e},-\frac{3}{80}} 
 \nonumber \\ 
Z_{{b}}^{\eD(S_3)} &=0
 \nonumber \\ 
Z_{{b'}}^{\eD(S_3)} &=0
 \nonumber \\ 
Z_{{b''}}^{\eD(S_3)} &=0
 \nonumber \\ 
Z_{{c}}^{\eD(S_3)} &= \chi^{m5 \times \overline{m5}}_{\mathbf{d},\frac{7}{16}; \mathbf{d},-\frac{7}{16}} 
+  \chi^{m5 \times \overline{m5}}_{\mathbf{e},\frac{3}{80}; \mathbf{e},-\frac{3}{80}} 
 \nonumber \\ 
Z_{{c'}}^{\eD(S_3)} &= \chi^{m5 \times \overline{m5}}_{\onebb,0; \mathbf{c},-\frac{3}{2}} 
+  \chi^{m5 \times \overline{m5}}_{\mathbf{a},\frac{1}{10}; \mathbf{b},-\frac{3}{5}} 
+  \chi^{m5 \times \overline{m5}}_{\mathbf{b},\frac{3}{5}; \mathbf{a},-\frac{1}{10}} 
+  \chi^{m5 \times \overline{m5}}_{\mathbf{c},\frac{3}{2}; \onebb,0} 
 \end{align}

\subsection{Condensation $\cA'_{1,2} = \onebb \oplus b $}

The  gapless boundary phases with central charge $(c,\bar
c) \leq (\frac{7}{10}, \frac{7}{10})$.

\begin{align}
\label{Zm4S3A12}
 Z_{\onebb}^{\eD(S_3)} &={\color{blue} \chi^{m4 \times \overline{m4}}_{\onebb,0; \onebb,0}} 
+  {\color{red} \chi^{m4 \times \overline{m4}}_{\mathbf{b},\frac{1}{2}; \mathbf{b},-\frac{1}{2}} }
 \nonumber \\ 
Z_{\one}^{\eD(S_3)} &= \chi^{m4 \times \overline{m4}}_{\mathbf{a},\frac{1}{16}; \mathbf{a},-\frac{1}{16}} 
 \nonumber \\ 
Z_{\mathbf{2}}^{\eD(S_3)} &=0
 \nonumber \\ 
Z_{{b}}^{\eD(S_3)} &={\color{blue} \chi^{m4 \times \overline{m4}}_{\onebb,0; \onebb,0}} 
+  \chi^{m4 \times \overline{m4}}_{\mathbf{a},\frac{1}{16}; \mathbf{a},-\frac{1}{16}} 
+  \chi^{m4 \times \overline{m4}}_{\mathbf{b},\frac{1}{2}; \mathbf{b},-\frac{1}{2}} 
 \nonumber \\ 
Z_{{b'}}^{\eD(S_3)} &=0
 \nonumber \\ 
Z_{{b''}}^{\eD(S_3)} &=0
 \nonumber \\ 
Z_{{c}}^{\eD(S_3)} &= \chi^{m4 \times \overline{m4}}_{\mathbf{a},\frac{1}{16}; \mathbf{a},-\frac{1}{16}} 
 \nonumber \\ 
Z_{{c'}}^{\eD(S_3)} &= \chi^{m4 \times \overline{m4}}_{\onebb,0; \mathbf{b},-\frac{1}{2}} 
+  \chi^{m4 \times \overline{m4}}_{\mathbf{b},\frac{1}{2}; \onebb,0} 
 \end{align}

\begin{align}
\label{Zm4S3A12a}
 Z_{\onebb}^{\eD(S_3)} &={\color{blue} \chi^{m5 \times \overline{m5}}_{\onebb,0; \onebb,0}} 
+  {\color{red} \chi^{m5 \times \overline{m5}}_{\mathbf{a},\frac{1}{10}; \mathbf{a},-\frac{1}{10}} }
+  {\color{red} \chi^{m5 \times \overline{m5}}_{\mathbf{b},\frac{3}{5}; \mathbf{b},-\frac{3}{5}} }
+  \chi^{m5 \times \overline{m5}}_{\mathbf{c},\frac{3}{2}; \mathbf{c},-\frac{3}{2}} 
 \nonumber \\ 
Z_{\one}^{\eD(S_3)} &= \chi^{m5 \times \overline{m5}}_{\mathbf{d},\frac{7}{16}; \mathbf{d},-\frac{7}{16}} 
+  \chi^{m5 \times \overline{m5}}_{\mathbf{e},\frac{3}{80}; \mathbf{e},-\frac{3}{80}} 
 \nonumber \\ 
Z_{\mathbf{2}}^{\eD(S_3)} &=0
 \nonumber \\ 
Z_{{b}}^{\eD(S_3)} &={\color{blue} \chi^{m5 \times \overline{m5}}_{\onebb,0; \onebb,0}} 
+  \chi^{m5 \times \overline{m5}}_{\mathbf{a},\frac{1}{10}; \mathbf{a},-\frac{1}{10}} 
+  \chi^{m5 \times \overline{m5}}_{\mathbf{b},\frac{3}{5}; \mathbf{b},-\frac{3}{5}} 
+  \chi^{m5 \times \overline{m5}}_{\mathbf{c},\frac{3}{2}; \mathbf{c},-\frac{3}{2}} 
+  \chi^{m5 \times \overline{m5}}_{\mathbf{d},\frac{7}{16}; \mathbf{d},-\frac{7}{16}} 
+  \chi^{m5 \times \overline{m5}}_{\mathbf{e},\frac{3}{80}; \mathbf{e},-\frac{3}{80}} 
 \nonumber \\ 
Z_{{b'}}^{\eD(S_3)} &=0
 \nonumber \\ 
Z_{{b''}}^{\eD(S_3)} &=0
 \nonumber \\ 
Z_{{c}}^{\eD(S_3)} &= \chi^{m5 \times \overline{m5}}_{\mathbf{d},\frac{7}{16}; \mathbf{d},-\frac{7}{16}} 
+  \chi^{m5 \times \overline{m5}}_{\mathbf{e},\frac{3}{80}; \mathbf{e},-\frac{3}{80}} 
 \nonumber \\ 
Z_{{c'}}^{\eD(S_3)} &= \chi^{m5 \times \overline{m5}}_{\onebb,0; \mathbf{c},-\frac{3}{2}} 
+  \chi^{m5 \times \overline{m5}}_{\mathbf{a},\frac{1}{10}; \mathbf{b},-\frac{3}{5}} 
+  \chi^{m5 \times \overline{m5}}_{\mathbf{b},\frac{3}{5}; \mathbf{a},-\frac{1}{10}} 
+  \chi^{m5 \times \overline{m5}}_{\mathbf{c},\frac{3}{2}; \onebb,0} 
 \end{align}

\subsection{Condensation $\cA'_{2,1} = \onebb \oplus \one $}

The  gapless boundary phases with central charge $(c,\bar
c) \leq (\frac45,\frac45)$.

\begin{align}
Z_{\onebb}^{\eD(S_3)} &={\color{blue} \chi^{m6 \times \overline{m6}}_{\onebb,0; \onebb,0}} 
+  \chi^{m6 \times \overline{m6}}_{\onebb,0; \mathbf{d},-3} 
+  \chi^{m6 \times \overline{m6}}_{\mathbf{d},3; \onebb,0} 
+  \chi^{m6 \times \overline{m6}}_{\mathbf{d},3; \mathbf{d},-3} 
+  {\color{red} \chi^{m6 \times \overline{m6}}_{\mathbf{e},\frac{2}{5}; \mathbf{e},-\frac{2}{5}} }
+  \chi^{m6 \times \overline{m6}}_{\mathbf{e},\frac{2}{5}; \mathbf{i},-\frac{7}{5}} 
+  \chi^{m6 \times \overline{m6}}_{\mathbf{i},\frac{7}{5}; \mathbf{e},-\frac{2}{5}} 
+  \chi^{m6 \times \overline{m6}}_{\mathbf{i},\frac{7}{5}; \mathbf{i},-\frac{7}{5}} 
 \nonumber \\ 
Z_{\one}^{\eD(S_3)} &={\color{blue} \chi^{m6 \times \overline{m6}}_{\onebb,0; \onebb,0}} 
+  \chi^{m6 \times \overline{m6}}_{\onebb,0; \mathbf{d},-3} 
+  \chi^{m6 \times \overline{m6}}_{\mathbf{d},3; \onebb,0} 
+  \chi^{m6 \times \overline{m6}}_{\mathbf{d},3; \mathbf{d},-3} 
+  \chi^{m6 \times \overline{m6}}_{\mathbf{e},\frac{2}{5}; \mathbf{e},-\frac{2}{5}} 
+  \chi^{m6 \times \overline{m6}}_{\mathbf{e},\frac{2}{5}; \mathbf{i},-\frac{7}{5}} 
+  \chi^{m6 \times \overline{m6}}_{\mathbf{i},\frac{7}{5}; \mathbf{e},-\frac{2}{5}} 
+  \chi^{m6 \times \overline{m6}}_{\mathbf{i},\frac{7}{5}; \mathbf{i},-\frac{7}{5}} 
 \nonumber \\ 
Z_{\mathbf{2}}^{\eD(S_3)} &= 2\chi^{m6 \times \overline{m6}}_{\mathbf{b},\frac{2}{3}; \mathbf{b},-\frac{2}{3}} 
+  2\chi^{m6 \times \overline{m6}}_{\mathbf{g},\frac{1}{15}; \mathbf{g},-\frac{1}{15}} 
 \nonumber \\ 
Z_{{b}}^{\eD(S_3)} &= 2\chi^{m6 \times \overline{m6}}_{\mathbf{b},\frac{2}{3}; \mathbf{b},-\frac{2}{3}} 
+  2\chi^{m6 \times \overline{m6}}_{\mathbf{g},\frac{1}{15}; \mathbf{g},-\frac{1}{15}} 
 \nonumber \\ 
Z_{{b'}}^{\eD(S_3)} &= 2\chi^{m6 \times \overline{m6}}_{\onebb,0; \mathbf{b},-\frac{2}{3}} 
+  2\chi^{m6 \times \overline{m6}}_{\mathbf{d},3; \mathbf{b},-\frac{2}{3}} 
+  2\chi^{m6 \times \overline{m6}}_{\mathbf{e},\frac{2}{5}; \mathbf{g},-\frac{1}{15}} 
+  2\chi^{m6 \times \overline{m6}}_{\mathbf{i},\frac{7}{5}; \mathbf{g},-\frac{1}{15}} 
 \nonumber \\ 
Z_{{b''}}^{\eD(S_3)} &= 2\chi^{m6 \times \overline{m6}}_{\mathbf{b},\frac{2}{3}; \onebb,0} 
+  2\chi^{m6 \times \overline{m6}}_{\mathbf{b},\frac{2}{3}; \mathbf{d},-3} 
+  2\chi^{m6 \times \overline{m6}}_{\mathbf{g},\frac{1}{15}; \mathbf{e},-\frac{2}{5}} 
+  2\chi^{m6 \times \overline{m6}}_{\mathbf{g},\frac{1}{15}; \mathbf{i},-\frac{7}{5}} 
 \nonumber \\ 
Z_{{c}}^{\eD(S_3)} &=0
 \nonumber \\ 
Z_{{c'}}^{\eD(S_3)} &=0
 \end{align}

\subsection{Condensation $\cA'_{3,1} = \onebb $}

The  gapless boundary phases with central charge $(c,\bar
c) \leq (\frac45,\frac45)$.

\begin{align}
 Z_{\onebb}^{\eD(S_3)} &={\color{blue} \chi^{m6 \times \overline{m6}}_{\onebb,0; \onebb,0}} 
+  \chi^{m6 \times \overline{m6}}_{\mathbf{d},3; \mathbf{d},-3} 
+  {\color{red} \chi^{m6 \times \overline{m6}}_{\mathbf{e},\frac{2}{5}; \mathbf{e},-\frac{2}{5}} }
+  \chi^{m6 \times \overline{m6}}_{\mathbf{i},\frac{7}{5}; \mathbf{i},-\frac{7}{5}} 
 \nonumber \\ 
Z_{\one}^{\eD(S_3)} &= \chi^{m6 \times \overline{m6}}_{\onebb,0; \mathbf{d},-3} 
+  \chi^{m6 \times \overline{m6}}_{\mathbf{d},3; \onebb,0} 
+  \chi^{m6 \times \overline{m6}}_{\mathbf{e},\frac{2}{5}; \mathbf{i},-\frac{7}{5}} 
+  \chi^{m6 \times \overline{m6}}_{\mathbf{i},\frac{7}{5}; \mathbf{e},-\frac{2}{5}} 
 \nonumber \\ 
Z_{\mathbf{2}}^{\eD(S_3)} &= \chi^{m6 \times \overline{m6}}_{\mathbf{b},\frac{2}{3}; \mathbf{b},-\frac{2}{3}} 
+  \chi^{m6 \times \overline{m6}}_{\mathbf{g},\frac{1}{15}; \mathbf{g},-\frac{1}{15}} 
 \nonumber \\ 
Z_{{b}}^{\eD(S_3)} &= \chi^{m6 \times \overline{m6}}_{\mathbf{b},\frac{2}{3}; \mathbf{b},-\frac{2}{3}} 
+  \chi^{m6 \times \overline{m6}}_{\mathbf{g},\frac{1}{15}; \mathbf{g},-\frac{1}{15}} 
 \nonumber \\ 
Z_{{b'}}^{\eD(S_3)} &= \chi^{m6 \times \overline{m6}}_{\onebb,0; \mathbf{b},-\frac{2}{3}} 
+  \chi^{m6 \times \overline{m6}}_{\mathbf{d},3; \mathbf{b},-\frac{2}{3}} 
+  \chi^{m6 \times \overline{m6}}_{\mathbf{e},\frac{2}{5}; \mathbf{g},-\frac{1}{15}} 
+  \chi^{m6 \times \overline{m6}}_{\mathbf{i},\frac{7}{5}; \mathbf{g},-\frac{1}{15}} 
 \nonumber \\ 
Z_{{b''}}^{\eD(S_3)} &= \chi^{m6 \times \overline{m6}}_{\mathbf{b},\frac{2}{3}; \onebb,0} 
+  \chi^{m6 \times \overline{m6}}_{\mathbf{b},\frac{2}{3}; \mathbf{d},-3} 
+  \chi^{m6 \times \overline{m6}}_{\mathbf{g},\frac{1}{15}; \mathbf{e},-\frac{2}{5}} 
+  \chi^{m6 \times \overline{m6}}_{\mathbf{g},\frac{1}{15}; \mathbf{i},-\frac{7}{5}} 
 \nonumber \\ 
Z_{{c}}^{\eD(S_3)} &= \chi^{m6 \times \overline{m6}}_{\mathbf{a},\frac{1}{8}; \mathbf{a},-\frac{1}{8}} 
+  \chi^{m6 \times \overline{m6}}_{\mathbf{c},\frac{13}{8}; \mathbf{c},-\frac{13}{8}} 
+  \chi^{m6 \times \overline{m6}}_{\mathbf{f},\frac{1}{40}; \mathbf{f},-\frac{1}{40}} 
+  \chi^{m6 \times \overline{m6}}_{\mathbf{h},\frac{21}{40}; \mathbf{h},-\frac{21}{40}} 
 \nonumber \\ 
Z_{{c'}}^{\eD(S_3)} &= \chi^{m6 \times \overline{m6}}_{\mathbf{a},\frac{1}{8}; \mathbf{c},-\frac{13}{8}} 
+  \chi^{m6 \times \overline{m6}}_{\mathbf{c},\frac{13}{8}; \mathbf{a},-\frac{1}{8}} 
+  \chi^{m6 \times \overline{m6}}_{\mathbf{f},\frac{1}{40}; \mathbf{h},-\frac{21}{40}} 
+  \chi^{m6 \times \overline{m6}}_{\mathbf{h},\frac{21}{40}; \mathbf{f},-\frac{1}{40}} 
 \end{align}

\end{widetext}
\endgroup

\bibliography{all,publst, local}

\end{document}

%% file: defs.tex
\usepackage{graphicx}
\usepackage{graphbox}

\usepackage{dcolumn}
\usepackage{bm}
\usepackage{multirow}

\usepackage[normalem]{ulem}

\usepackage{amsmath}
\usepackage{amsthm}
\usepackage{amstext}
\usepackage{amssymb}
\usepackage{mathrsfs}
\usepackage{amsfonts}
\usepackage{amsbsy} 
\usepackage[all,knot,color,matrix,cmtip]{xy}

\usepackage{csquotes}
\MakeOuterQuote{"}

\usepackage{color}

\definecolor{red}{rgb}{1,0,0}
\definecolor{blue}{rgb}{0,0,1}
\definecolor{dblue}{rgb}{0,0,0.4}
\definecolor{green}{rgb}{0,1,0}
\definecolor{black}{rgb}{0,0,0}
\definecolor{white}{rgb}{1,1,1}

\definecolor{brn}{rgb}{.8,.4,.0}
\definecolor{redo}{rgb}{1,.5,.0}
\definecolor{ddgrn}{rgb}{0,0.4,0}
\definecolor{dgrn}{rgb}{0,0.55,0}
\definecolor{dbl}{rgb}{0,0,0.5}

\usepackage[bbgreekl]{mathbbol}
\newcommand{\one}{\mathbf{1}}
\newcommand{\two}{\mathbf{2}}
\newcommand{\onebb}{\mathbb{1}}

\newcommand{\Z}{\mathbb{Z}}

\newcommand{\vc}[1]{\boldsymbol{#1}} 
\newcommand{\tl}[1]{\widetilde{#1}} 
\newcommand{\ii}{\hspace{1pt}\mathrm{i}\hspace{1pt}}
\newcommand{\ee}{\hspace{1pt}\mathrm{e}}

\newcommand{\Rf}[1]{Ref.~\onlinecite{#1}}
\newcommand{\Rfs}[1]{Refs.~\onlinecite{#1}}

\newcommand{\ie}{{\it i.e.~}} 
 
\newcommand{\etc}{{\it etc.}}

\newcommand{\bpm}{\begin{pmatrix}}
\newcommand{\epm}{\end{pmatrix}}
\newcommand{\bmm}{\begin{matrix}}
\newcommand{\emm}{\end{matrix}}

\newcommand{\cA}{ {\cal A} } 

\newcommand{\cC}{ {\cal C} } 
\newcommand{\cD}{ {\cal D} }

\newcommand{\cR}{ {\cal R} }

\newcommand{\cV}{ {\cal V} }

\usepackage{euscript}

\newcommand\eD           {\EuScript{D}}

\newcommand\eM          {\EuScript{M}}

\newcommand\eV        {\EuScript{V}}

\newcommand\eZ         {\EuScript{Z}}

\newcommand{\al}{\alpha} 
\newcommand{\bt}{\beta}

\newcommand{\eps}{\epsilon} 
 
\newcommand{\ga}{\gamma}

\newcommand{\om}{\omega}

\newcommand{\frmbox}[1]{\begin{center}\fbox{\parbox{3.3in}{\parindent=0pt #1}}\end{center}}